\documentclass[aps,prd,twocolumn,superscriptaddress,nofootinbib]{revtex4-1}
\pdfoutput=1

\usepackage{float}
\usepackage{amsfonts}
\usepackage{psfrag}
\usepackage{graphicx}
\usepackage{grffile}
\usepackage{cancel}
\usepackage{amsmath}
\usepackage{natbib}
\usepackage{xr}
\usepackage{hyperref}
\usepackage{verbatim}
\usepackage{epstopdf}
\usepackage{subcaption}
\usepackage{caption}
\usepackage{multirow}
\usepackage{float}
\usepackage{commath}
\usepackage{empheq}
\usepackage{braket}

\usepackage{booktabs,dcolumn,caption}
\hypersetup{
    bookmarks=true,         
    unicode=false,          
    pdftoolbar=true,        
    pdfmenubar=true,        
    pdffitwindow=false,     
    pdfstartview={FitH},    
    pdfauthor={Nathan Rutherford},     
    colorlinks=true,       
    linkcolor=blue,          
    citecolor=blue,        
}
\newcolumntype{d}[1]{D{.}{.}{#1}}

\RequirePackage[normalem]{ulem} 
\RequirePackage{color}\definecolor{RED}{rgb}{1,0,0}\definecolor{BLUE}{rgb}{0,0,1} 
\providecommand{\DIFaddbegin}{} 
\providecommand{\DIFaddend}{} 
\providecommand{\DIFdelbegin}{} 
\providecommand{\DIFdelend}{} 
\providecommand{\DIFaddbeginFL}{} 
\providecommand{\DIFaddendFL}{} 
\providecommand{\DIFdelbeginFL}{} 
\providecommand{\DIFdelendFL}{} 
\newcommand{\DIFscaledelfig}{0.5}
\RequirePackage{settobox} 
\RequirePackage{letltxmacro} 
\newsavebox{\DIFdelgraphicsbox} 
\newlength{\DIFdelgraphicswidth} 
\newlength{\DIFdelgraphicsheight} 
\LetLtxMacro{\DIFOincludegraphics}{\includegraphics} 
\newcommand{\DIFaddincludegraphics}[2][]{{\color{blue}\fbox{\DIFOincludegraphics[#1]{#2}}}} 
\newcommand{\DIFdelincludegraphics}[2][]{
\sbox{\DIFdelgraphicsbox}{\DIFOincludegraphics[#1]{#2}}
\settoboxwidth{\DIFdelgraphicswidth}{\DIFdelgraphicsbox} 
\settoboxtotalheight{\DIFdelgraphicsheight}{\DIFdelgraphicsbox} 
\scalebox{\DIFscaledelfig}{
\parbox[b]{\DIFdelgraphicswidth}{\usebox{\DIFdelgraphicsbox}\\[-\baselineskip] \rule{\DIFdelgraphicswidth}{0em}}\llap{\resizebox{\DIFdelgraphicswidth}{\DIFdelgraphicsheight}{
\setlength{\unitlength}{\DIFdelgraphicswidth}
\begin{picture}(1,1)
\thicklines\linethickness{2pt} 
{\color[rgb]{1,0,0}\put(0,0){\framebox(1,1){}}}
{\color[rgb]{1,0,0}\put(0,0){\line( 1,1){1}}}
{\color[rgb]{1,0,0}\put(0,1){\line(1,-1){1}}}
\end{picture}
}\hspace*{3pt}}} 
} 
\LetLtxMacro{\DIFOaddbegin}{\DIFaddbegin} 
\LetLtxMacro{\DIFOaddend}{\DIFaddend} 
\LetLtxMacro{\DIFOdelbegin}{\DIFdelbegin} 
\LetLtxMacro{\DIFOdelend}{\DIFdelend} 
\DeclareRobustCommand{\DIFaddbegin}{\DIFOaddbegin \let\includegraphics\DIFaddincludegraphics} 
\DeclareRobustCommand{\DIFaddend}{\DIFOaddend \let\includegraphics\DIFOincludegraphics} 
\DeclareRobustCommand{\DIFdelbegin}{\DIFOdelbegin \let\includegraphics\DIFdelincludegraphics} 
\DeclareRobustCommand{\DIFdelend}{\DIFOaddend \let\includegraphics\DIFOincludegraphics} 
\LetLtxMacro{\DIFOaddbeginFL}{\DIFaddbeginFL} 
\LetLtxMacro{\DIFOaddendFL}{\DIFaddendFL} 
\LetLtxMacro{\DIFOdelbeginFL}{\DIFdelbeginFL} 
\LetLtxMacro{\DIFOdelendFL}{\DIFdelendFL} 
\DeclareRobustCommand{\DIFaddbeginFL}{\DIFOaddbeginFL \let\includegraphics\DIFaddincludegraphics} 
\DeclareRobustCommand{\DIFaddendFL}{\DIFOaddendFL \let\includegraphics\DIFOincludegraphics} 
\DeclareRobustCommand{\DIFdelbeginFL}{\DIFOdelbeginFL \let\includegraphics\DIFdelincludegraphics} 
\DeclareRobustCommand{\DIFdelendFL}{\DIFOaddendFL \let\includegraphics\DIFOincludegraphics} 

\newcommand{\Msun}{\mathrm{M}_\odot}
\newcommand{\chibar}{\Bar{\chi}}

\newcommand{\phimu}{\phi_\mu}

\begin{document}

\title{Probing fermionic asymmetric dark matter cores using global neutron star properties}
\author{Nathan Rutherford}
\email[Corresponding author: ]{nathan.rutherford@unh.edu}
\author{Chanda Prescod-Weinstein}
\email{chanda.prescod-weinstein@unh.edu}
\affiliation{Department of Physics and Astronomy, University of New Hampshire, Durham, New Hampshire 03824, USA}

\author{Anna Watts}
\email{A.L.Watts@uva.nl}
\affiliation{Anton Pannekoek Institute for Astronomy, University of Amsterdam,
Science Park 904, 1098XH Amsterdam, The Netherlands}
\bibliographystyle{unsrtnat}
\begin{abstract}
It is possible for asymmetric dark matter (ADM) to accumulate in neutron star interiors and affect their global properties. Considering the effects of this accumulation, neutron star mass-radius measurements can deliver new insights into the cold dense matter equation of state (EoS). In this paper, we employ Bayesian parameter estimation using real and synthetic neutron star mass-radius data to infer constraints on the combined baryonic matter and fermionic ADM EoS, where the fermionic ADM forms a core in the neutron star interior. Using currently available mass-radius data, we find that the lower bound of the ratio between ADM effective self-repulsion strength ($g_\chi/m_\phi$) and particle mass ($m_\chi$) can be constrained at the 68\% (95\%) credible level to $10^{-6.59}$ ($10^{-7.77}$). We also find that, if neutron star mass-radius measurement uncertainties are reduced to the 2\% level, the constraints on the lower bound of the ratio of $g_\chi/m_\phi$ to $m_\chi$ can be improved to $10^{-6.49}$ and $10^{-7.68}$ at the 68\% and 95\% credible levels, respectively. However, all other combinations, of $m_\chi$, $g_\chi$, and the ADM mass-fraction, $F_\chi$, (i.e., the ratio of the gravitational ADM mass to the gravitational mass of the neutron star) are unconstrained. Furthermore, in the pressure-energy density and mass-radius planes, the inferences which include the possibility of fermionic ADM cores are nearly identical with the inferences that neglect fermionic ADM for $F_\chi \leq 1.7\%$ and neutron star mass-radius uncertainties $\geq 2\%$. Therefore, we find that neutron star mass-radius measurements can constrain the ratio of $g_\chi/m_\phi$ to $m_\chi$ and that neutron stars with ADM are indistinguishable from purely baryonic stars. This implies that neutron stars with ADM are equally as consistent with the available mass-radius data as neutron stars without ADM. without ADM.
\end{abstract}
\maketitle
\section{Introduction}\label{intro}
 The extremely compact nature of neutron stars provides a unique environment to probe the behavior of matter at supranuclear densities. Theoretical models of neutron star interiors predict many different types of baryonic matter, such as neutron-rich matter, nuclear pasta, hyperons, and deconfined quarks \cite{Hebeler13,Oertel_2017,Caplan2018,Tolos2020,Burgio_2021,Han2023,Keller2024}. The microphysics of these hypothetical forms of matter are encoded by the equation of state (EoS), which describes the relation between the pressure and energy density maintained throughout the star. The dense matter EoS can be determined from knowing the gravitational masses and radii of neutron stars. In particular, each EoS can be mapped to a unique mass-radius relation, i.e., the numerical expression that relates all possible, stable neutron star mass and radii, through the Tolman-Oppenheimer-Volkoff equations \cite{Lindblom_1992}. Thus, measurements of neutron stars can be used to characterize the mass-radius curve, and thus constrain each hypothetical dense matter EoS.

Analysing the X-ray pulse profile (a rotationally phase and energy-resolved X-ray count spectrum) can be used to infer the masses and radii of neutron stars through the Pulse Profile Modelling (PPM) technique. PPM is a relativistic ray tracing technique that can be used to extract valuable information encoded in the pulse profile, such as mass, radius, and hot spot geometry \cite[for more details on PPM see][]{Watts2019,Bogdanov2019,Bogdanov2021}. So far, data from the NASA's Neutron Star Interior Composition Explorer (NICER) \cite{Gendreau2016}, informed where possible by mass priors from pulsar timing, has been used to infer the masses and radii of three millisecond pulsars (MSPs) using the PPM technique: PSR J0740$+$6620 \cite{Fonseca_2021,Riley0740,Miller2021,Salmi2022,Salmi2024, Dittmann2024}, PSR J0030$+$0451 \cite{Riley0030,Miller2019,Vinciguerra2024}, and most recently, PSR J0437$-$4715 \cite{Choudhury2024,Reardon2024}. Constraints have also been placed on a fourth MSP, PSR~J1231-1411, but these constraints are weaker than the others and will therefore be neglected in this work \cite{Salmi:2024bss}. The mass-radius measurements of these pulsars have been used in many analyses to set constraints on the dense matter EoS \cite[see e.g.,][]{Raaijmakers2019,Miller2019,Raaijmakers2020,Raaijmakers2021,Miller2021,JieLiJ21,Legred21,Pang21,TangSP21,Annala2022,Biswas2022,Rutherford2024,Huang2024}. Most recently, \cite{Rutherford2024} has shown that the inferences with the newest mass-radius results from \textit{NICER} improve the reliability of our understanding on the neutron star EoS via the updated chiral effective field theory ($\chi$EFT) calculations of \cite{Keller2022} and the degree in which the results are data-driven. However, the uncertainty of the cold dense matter EoS remains.

In the near future, improved mass-radius constraints on targets already analyzed by \textit{NICER} are expected. Additionally, the mass-radius measurements of four more sources are anticipated. During the next decade, large area X-ray spectral-timing missions are anticipated to perform PPM on more neutron stars with improved uncertainties. Such missions include the Chinese mission concept \textit{eXTP} (the enhanced X-ray Timing and Polarimetry mission, \cite{extp}), the NASA probe-class mission concept \textit{STROBE-X} (the Spectroscopic Time-Resolving Observatory for Broadband Energy X-rays, \cite{Ray2019}), and the ESA's L-class mission \textit{Athena} (Advanced Telescope for High ENergy Astrophysics, \cite{Athena13}).

 Currently, many of the EoS studies that consider PPM derived mass-radius measurements only account for the presence of baryonic matter and its potential phase transitions. However, there is a growing understanding that dark matter may form part of neutron star structures and therefore affect the observable properties of neutron stars. To fully assess the range of neutron star EoS, we must therefore also study the inclusion of a dark matter component in our models. Dark matter can occupy two spatial regimes: a dark matter halo that extends through and beyond the baryonic radius of the neutron star, and a dark matter core inside the neutron star's interior. Dark matter halos around neutron stars have been shown to increase the gravitational masses and tidal deformabilities of the stars when compared to a neutron star with no ADM with an identical central baryonic energy density, if the dark matter mass distribution is mostly beyond the baryonic surface \cite{Nelson_2018,Sagun:2021oml,Diedrichs2023,Bramante2024, Buras-Stubbs2024,Guha2024,Jockel2024}. However, if most of the dark matter halo distribution resides within the baryonic radii of neutron stars, halos can reduce the gravitational masses and radii of these stars \cite[see e.g.,][]{Shawqi2024}. In addition, the presence of any dark matter halo can significantly impact the exterior space-time around neutron stars, which would alter the interpretation of \textit{NICER}'s mass-radius measurements \cite{Miao_2022,Shawqi2024,Shakeri2024}. Interestingly, dark matter halos could possibly form stable ultra-compact neutron stars with compactness greater than 1/3, which could serve as a black hole mimicker \cite{Pitz2024}. If dark matter forms a core inside neutron star interiors, the gravitational masses, radii, and tidal deformabilities have been shown to decrease when compared to their purely baryonic counterparts with the same central baryonic energy density \cite{Ellis2018,Ivanystkyi2020,Kain2021,Karkevandi2022,Karkevandi2024,Bastero-Gil2024,Scordino2024,Konstantinou2024}. Therefore, it is evident that, because dark matter can affect the measurable properties of neutron stars, the possible presence of it must be accounted for in analyses of the neutron star mass-radius measurements and the EoS.

 There are a variety of proposed methods to constrain the presence of dark matter in and around neutron stars. For instance, by considering one or multiple representative baryonic EoSs, constraints on the dark matter particle mass, mediator mass, and mass-fraction can be made using gravitational wave and neutron star mass-radius measurements \cite{Das_2022,Apran2020,Sen2021,Guha2021,Giangrandi2022,Karkevandi2022,Karkevandi2024,Barbat2024,Miao_2022,Sun2023,Guha2024,P.Thakur2024,P.Thakur2024b,Shirke2024,Pratik-Thankur2024,Pal2024,Mariani2024,Mahapatra2024,Kumar2024,Routaray2023}. More specifically, using one or more baryonic EoSs allows for constraints on the dark matter mass-fraction as a function of particle mass \cite{Karkevandi2022,Giangrandi2022,Karkevandi2024,Barbat2024,Guha2024}, the dark matter Fermi momentum and particle mass \cite{Das_2022,Kumar2024}, several dark matter parameters by imposing hard cut-offs on the maximum neutron star mass and tidal deformability of a 1.4 $\Msun$ neutron star \cite{Sen2021,Guha2021,Pratik-Thankur2024,Pal2024,Mariani2024, Routaray2023}, correlations between dark matter and baryonic matter parameters \cite{Sun2023,P.Thakur2024b,P.Thakur2024,Shirke2024,Mahapatra2024}, and Bayesian inferences to estimate the dark matter parameter space \cite{Apran2020,Miao_2022}. Analytical calculations of the maximum accumulated dark matter mass before gravitational collapse to a black hole inside a neutron star can be used to tightly constrain dark matter models through observations of old neutron stars \cite{Khlopov1985,Bertone2008,Kouvaris2011,Gresham2018}. When an additional dark matter component is considered in simulations of binary neutron star mergers it has been shown that dark matter can leave detectable signatures on the gravitational wave and electromagnetic counterparts, which can constrain the dark matter particle mass and the total accumulated dark matter mass \cite{Bauswein2023,Ruter2023,Emma2022}. Finally, due to their extremely compact natures, neutron stars can efficiently capture dark matter within their interiors and thus can provide constraints on the dark matter--nucleon cross section and particle mass \cite{Tong2024}.

 In our previous work \cite{Rutherford2023}, we investigated constraints on bosonic asymmetric dark matter (ADM) cores inside neutron stars by performing a full Bayesian inference in which all parameters in the neutron star EoS model are allowed to vary. We assumed the dark matter was 100\% comprised of the self-repulsive bosonic ADM of \cite{Nelson_2018}. The main motivation of the ADM paradigm suggests that the cosmic history between dark matter and baryonic is strongly tied together given that the observed dark matter mass density in the universe is only five times greater than that of baryonic matter. That is, similar to the baryon asymmetry in the early universe, dark matter also had an asymmetry between it and anti-dark matter, which produced the relic abundance of dark matter observed today in the universe \cite{Petraki2013,Petraki2014}. A ``dark asymmetry'' between dark matter and anti-dark matter in the early universe would allow for small attractive interactions with baryonic matter and substantial repulsive self-interactions. 
 
 By combining the EoS of the \cite{Nelson_2018} bosonic ADM model with the parametrized piecewise polytropic (PP) model of \cite{Hebeler13} and restricting the study to simulated neutron star mass-radius measurements with and without bosonic ADM cores, in \cite{Rutherford2023} we explored the possible inferred constraints on both of the bosonic ADM and baryonic EoSs. From these inferences, we found that the uncertainties on the baryonic EoS are relaxed when bosonic ADM cores are taken into account. Moreover, in \cite{Rutherford2023} we found that, if the baryonic EoS could be constrained more tightly, constraints on the ADM mass-fraction and the ratio of the effective bosonic ADM self-repulsion strength to the particle mass can be made. Lastly, we concluded that the ADM particle mass and self-repulsion cannot be individually constrained using neutron star mass-radius measurements.

 Several studies, that have also studied dark matter admixed neutron stars using \textit{NICER} mass-radius measurements and/or gravitational wave measurements from LIGO/VIRGO are \cite{Miao_2022, Guha2024, Shakeri2024}. In \cite{Miao_2022}, the authors studied the impact of dark matter halos on the pulse-profile of neutron stars and performed a Bayesian analysis on the \cite{Nelson_2018} fermionic ADM model with a fixed baryonic matter EoS. Additionally, they restricted their analysis to dark matter cores and diffuse halos. Their inference is based on the posterior distributions of two \textit{NICER} pulsars, namely PSR J0740$+$6620 and PSR J0030$+$0451. Although \cite{Miao_2022} did not determine constraints on the ADM self-repulsion strength and mass-fraction, they did infer constraints on the fermionic ADM particle mass. In particular, they determined the ADM particle mass is favored to be less than 1.5 GeV for halos and ADM cores favor masses near 0.6 GeV. In \cite{Shakeri2024}, the authors also studied the affect of dark matter halos on the pulse-profile of neutron stars and studied the constraints on bosonic dark matter using two baryonic EoS models with different stiffness. Moreover, \cite{Shakeri2024} performed a scan over the bosonic dark matter particle mass, self-interaction strength, and mass-fraction using both baryonic matter EoS models and compared the results to the 95\% confidence intervals of the \textit{NICER} and LIGO/VIRGO measurements. Here, they found that the maximum allowed dark matter mass-fraction is around 5\% and 20\% for the relatively soft and stiff baryonic EoSs, respectively. In addition, \cite{Shakeri2024} also found that when the allowed region of particle masses and self-repulsion strengths shrinks for increasing mass-fraction for both baryonic EoSs. Lastly, in \cite{Guha2024}, the authors considered 10 baryonic EoS models of varying stuffiness and a fermionic dark matter model with repulsive and attractive self-interactions. Their analysis consisted of comparing various fermionic dark matter configurations using the 95\% contours of the \textit{NICER} and LIGO/VIRGO measurements as hard-cut offs to rule one configuration in or out. Here, \cite{Guha2024} found that dark matter particle masses in the range of $[0.1,30]$ GeV with fermi momenta in the interval $[0.01,0.07]$ GeV can match the data contours. In order to constrain dark matter in neutron stars, all three studies used at most a handful of baryonic EoSs and used \textit{NICER} and/or LIGO/VIRGO measurements.  

In this work, we take a different approach to those found in the literature and expand on our prior work in \cite{Rutherford2023} by taking into account fermionic ADM cores, considering realistic ADM accumulation mechanisms that apply to all physically allowed ADM particle masses, and including both real and synthetic PPM derived mass-radius data. Modeling the dark matter core as fermionic ADM instead of bosonic ADM is physically interesting because fermionic ADM cores have additional support against gravity through the Fermi degeneracy pressure, thus expanding the allowed ADM parameter space to be studied because the self-repulsion is allowed to be zero. We additionally consider real neutron star mass-radius data for this work because \cite{Konstantinou2024} showed that the presence of ADM cores does not modify the universal relations used to model the oblateness of neutron stars.\footnote{For further details on the universal oblateness relations of neutron stars see \cite{Morsink2007,AlGendy2014}}. By considering both real and synthetic data in addition to ADM accumulation mechanisms that apply to the entire physically allowed parameter space, this work will be able to investigate current and potential future constraints on fermionic ADM cores.

 In this work, we assume the fermionic ADM cores are described by the \cite{Nelson_2018} model and neglect the possibility of ADM halo configurations, because the existence of any halo will alter the exterior space-time and thus modify how PPM is performed. This work will consider the mass-radius measurements of PSR J0740$+$6620 \cite{Riley0740} and PSR J0030$+$0451 \cite{Riley0030}. There have been several updates to PSR J0740$+$6620 \cite{Salmi2022,Salmi2024} and PSR J0030$+$0451 \cite{Vinciguerra2024}, as well as a new mass-radius measurement for PSR J0437$-$4715 \cite{Choudhury2024}, which was released during the completion of this work. However, we still consider the mass-radius posteriors of PSR J0030$+$0451 \cite{Riley0030} and PSR J0740$+$6620 \cite{Riley0740} for two key reasons: the first is so that this work is fully comparable to our previous work on bosonic ADM \cite{Rutherford2023} and the other is because our synthetic scenario best demonstrates what can be achieved with tighter mass-radius uncertainties. 
 
 For the simulated neutron star data, we consider six possible \textit{STROBE-X} sources because \textit{STROBE-X} mission is expected to provide lower uncertainties than \textit{NICER} \cite{Ray2019,STROBEX2}. We call this scenario, \textit{Future-X}, after the original \textit{Future-X} scenario in our previous work. By incorporating fermionic ADM with both real and synthetic data into our Bayesian framework, this work aims to quantify the possible constraints on the fermionic ADM EoS for current missions and future missions, namely \textit{NICER} and \textit{STROBE-X}. Furthermore, the other objective is to determine the effects of including fermionic ADM on the neutron star pressure-energy density and mass-radius posteriors. 
 
 The work presented in this manuscript shows that the current \textit{NICER} and future \textit{STROBE-X} measurements are able to place constraints on the lower bound of the ratio between the ADM particle mass and effective self-repulsion strength. However, under the current uncertainties of the baryonic EoS, neither \textit{NICER} nor \textit{STROBE-X} can constrain the fermionic ADM particle mass, effective self-repulsion strength, or mass-fraction. Finally, we find that the mass-radius ADM admixed neutron star posteriors are equally as consistent with the data as the counterparts that neglect ADM, which implies that \textit{NICER} and \textit{STROBE-X} cannot distinguish between the two cases.

This paper is organized as follows. In Sec.~\ref{tovandeos}, we motivate the two-fluid TOV equations and describe the baryonic matter and fermionic ADM EoSs. Sec.~\ref{method} discusses our Bayesian inference framework for providing constraints on the ADM admixed neutron star EoS, the baryonic matter EoS priors, the constraints on the fermionic ADM EoS parameter space, and the selected neutron star mass-radius measurements. In Sec.~\ref{results}, we study the inferences for both the \textit{NICER} data and \textit{Future-X} scenario. Finally, in Sec.~\ref{conclusion}, we discuss our results. Throughout this work, we use the \textit{diag\big($-$,$+$,$+$,$+$\big)}.

\section{Modeling the structure of ADM admixed neutron stars}\label{tovandeos}
Traditionally, the mass-radius relation of neutron stars is computed by iteratively solving the Tolman-Oppenheimer-Volkoff (TOV) equations \cite{Tolman1939,Oppenheimer1939} for a given baryonic matter equation of state and range of central energy densities \cite{Lindblom_1992}. However, to compute the mass-radius relation of neutron stars with ADM, the single fluid mass-radius relation calculation must be modified. In this section, we describe how to model the structure of ADM admixed neutron stars using the two-fluid formalism, and describe the baryonic matter and ADM equations of state used in our analysis. The combination of the baryonic matter EoS, ADM EoS, and two-fluid formalism, allow for the ADM admixed neutron star mass-radius relation to be computed. 
\subsection{The two-fluid formalism}\label{twofluid}
The global properties, such as mass and radius, of an ADM admixed neutron star can be computed by adopting the two-fluid formalism, which assumes that the interactions between the Standard Model and ADM are solely gravitational \cite[see e.g.,][]{Shakeri2024,Shawqi2024,Konstantinou2024,Karkevandi2024,Ruter2023,Sagun2023,Giangrandi2022,Ellis2018,Sandin2009}. The two-fluid formalism is an appropriate framework to study ADM admixed neutron stars because any non-gravitational interfluid interaction between ADM and the Standard Model is expected to be negligible \cite{Marrodan-Undagoitia2016,Reddy2016}{\footnote{We would like to note that the choice of using the two-fluid formalism is not entirely consistent with the ADM paradigm because it neglects any interaction of dark matter to baryonic matter, which is necessary to generate the dark asymmetry between ADM and anti-ADM. However, since this work assumes that the coupling of baryonic matter to ADM is small compared to the coupling strength of ADM to itself (see Sec.~\ref{EoSs} for further details) and that the dominant interaction between ADM and baryonic matter is gravitational, the two-fluid formalism is a well-motivated treatment to study ADM inside neutron stars. Additionally, since this work considers the \cite{Hebeler13} parametrized PP model for the baryonic EoS, the Standard Model baryon current cannot be computed because there is no associated Lagrangian with the model. Although capturing the uncertainties of the baryonic EoS and the interaction term of baryonic matter with ADM in the ADM EoS need further work, this can be improved in the future, e.g., using the two-fluid interacting system considered in \cite{Hajkarim24}.}. The assumption that ADM and baryonic matter interact only gravitationally implies that ADM and baryonic matter satisfy their own conservation of energy-momentum equation. Thus, both ADM and baryonic matter can be treated as two distinct fluids, which can be expressed in terms of pressure and energy density as
\begin{align}
    p(r) &= p_B(r) + p_{\chi}(r) \label{pB_p_ADM}\\
    \varepsilon(r) &= \varepsilon_B(r) + \varepsilon_\chi(r) \label{epsB_epsADM},
\end{align}
where $p_B$ ($\varepsilon_{B}$) is the baryonic matter pressure (energy density) as a function of radius $r$ and $p_\chi$ ($\varepsilon_\chi$) is the ADM pressure (energy density) as function of $r$. The substitution of Eqs.~\ref{pB_p_ADM} and \ref{epsB_epsADM} into the single-fluid TOV equations yields the two-fluid TOV equations:
\begin{align}
      &\frac{dp_B}{dr} = - \left( \varepsilon_B +p_B \right) \frac{Gc^2M(r) + 4\pi r^3 G p(r)}{c^2 r \left[rc^2-2GM(r)  \right] } \label{dpbdr}\\
       &\frac{dp_\chi}{dr} = - \left( \varepsilon_\chi +p_\chi \right) \frac{Gc^2M(r) + 4\pi r^3 G p(r)}{c^2 r \left[rc^2-2GM(r)  \right] } \label{dpchidr}\\
        &\frac{dM_B(r)}{dr} = 4\pi r^2 \frac{\varepsilon_B(r)}{c^2} \\ 
         &\frac{dM_\chi(r)}{dr} = 4\pi r^2 \frac{\varepsilon_\chi(r)}{c^2},
\end{align}
where $M_\chi(r)$ is the gravitational mass of ADM, $M_B(r)$ is the gravitational mass of baryonic matter, and $M(r) = M_B(r)+M_\chi(r)$. The two-fluid TOV equations can then be simultaneously solved for the gravitational masses, radii, and pressures of each fluid given their respective equations of state and central energy densities. More specifically, the pressures and gravitational masses of ADM and baryonic matter are solved simultaneously until either one of the fluid pressures reaches zero. The integration is stopped and then resumed using the single-fluid TOV equations with the last pressure value of the remaining fluid as the initial condition. Since the two-fluid TOV equations can be solved to obtain the gravitational masses and radii of both ADM and baryonic matter, they can also be used to numerically compute the ADM admixed mass-radius relation. Similar to how the baryonic matter neutron star mass-radius relation is numerically computed, the ADM admixed neutron star mass-radius relation is obtained by iteratively solving the two-fluid TOV equations for a given EoS and range central energy densities for both ADM and baryonic matter, respectively.

The two-fluid TOV equations allow for a clear distinction between the ADM core radius ($R_\chi$) and the baryonic matter radius ($R_B$), which makes it possible to define the ADM mass-fraction, $F_\chi$. The mass-fraction is defined as the ratio of the ADM gravitational mass to the total gravitational mass of the ADM admixed neutron star, and it is given by
\begin{equation}
    F_\chi = \frac{M_\chi(R_\chi)}{M_\chi(R_\chi)+M_B(R_B)} \label{F_chi},
\end{equation}
where $M_\chi(R_\chi)$ is the total accumulated ADM gravitational mass evaluated at $R_\chi$ and $M_B(R_B)$ is the baryonic matter gravitational mass evaluated at $R_B$. The ADM mass-fraction is useful because the total gravitational mass and radius of an ADM admixed neutron star is strongly dependent on the value of $F_\chi$ \cite[see][]{Scordino2024,Collier2022,Karkevandi2022}. Moreover, when specific assumptions about the baryonic matter EoS are made, the ADM mass-fraction can be constrained tightly by neutron star measurements \cite{Karkevandi2024,Rutherford2023,Sagun2023,Karkevandi2022} and can be used to place constraints on the ADM particle mass \cite{Scordino2024,Barbat2024,Sagun2023,Shakeri2024}. The $F_\chi$ is an optimal parameter to consider in our analysis. Finally, to compute the structure of an ADM admixed neutron star for a given $F_\chi$ and central baryonic energy density, we follow the numerical algorithm outlined in \cite{Rutherford2023}. 

\subsection{The baryonic matter and ADM equations of state}\label{EoSs}
In order to model neutron stars that have a baryonic matter and a potential fermionic ADM core component, we need to solve the two-fluid TOV equations. This requires an equation of state for each type of matter. While it is common practice to model the baryonic matter EoS using one or more of the available tabulated EoSs \cite{Konstantinou2024,Shawqi2024,Giangrandi2022,Miao_2022,Ivanystkyi2020,Kain2021,Ellis2018,Nelson_2018}, we model the baryonic matter EoS using the parametrized piecewise polytropic (PP) model used in \cite{Hebeler13,Greif19,Raaijmakers2019,Raaijmakers2020,Raaijmakers2021,Rutherford2024}. We employ the PP model because it, as well as other polytropic models, can fit many of the tabulated EoS \cite{Read2009}. Moreover, when a wide range of PP EoSs are sampled, the PP model is able to capture the uncertainties in the baryonic EoS due to its parametrized nature. This allows for the PP model to span much of the physically viable space in the mass-radius plane. The parametrized PP model is described by three varying polytropes connected at 2 varying transition densities and considers the calculations of $\chi$EFT at low density. In particular, for densities $\lesssim 0.5 n_0$ ($n_0 = 0.16 fm^{-3}$) we consider the neutron star crust to be described by the Baym-Pethick-Sutherland (BPS) crust EoS \cite{Baym71}, which we then connect to a single polytropic fit of the $\chi$EFT band of \cite{Hebeler13} between $0.5 n_0$ and $1.1 n_0$. At densities above $2 n_0$, the $\chi$EFT calculations grow increasingly uncertain and eventually break down \cite{Drischler2021}. Although, $\chi$EFT calculations have been considered up to $1.5-2.0 n_0$ in other EoS analyses \cite[see e.g.,][]{Tews2018,Keller2024,Rutherford2024}, we do not expect the difference in these choices to affect the inferences on the fermionic ADM EoS parameters, as the $\chi$EFT calculations only affect the uncertainties on the baryonic EoS, which is beyond the scope of this work. Additionally, this work seeks to be comparable to our previous work \cite{Rutherford2023}. Thus, for densities $\geq 1.1 n_0$, we connect the \cite{Hebeler13} $\chi$EFT band to the high density piecewise polytropic parameterization. Modeling the baryonic EoS in this way allows us to simultaneously consider the tight constraints delivered by the $\chi$EFT formalism and systemically study the neutron star mass-radius plane \cite[see][]{Read2009,Hebeler10,Hebeler13}.

Our fermionic ADM core is that of \cite{Nelson_2018}, which describes a MeV-GeV mass-scale spin-1/2 ADM particle $\chi$ with repulsive self-interactions mediated by the exchange of an eV-MeV mass-scale vector gauge boson $\phimu$. The ADM vector gauge boson also carries the Standard Model baryon number in order to create the asymmetry between dark matter and anti-dark matter particles in the early universe that is needed to produce the present dark matter mass density of the universe.

In order to obtain the fermionic ADM EoS, \cite{Nelson_2018} assumes that $g_B \ll g_\chi$, where $g_\chi$ is the interacting strength between $\chi$ and $\phimu$ and $g_B$ is the interaction strength of $\phimu$ with the Standard Model baryon current. Although robust calculations of this inequality have not been done yet, assuming that $g_B \ll g_\chi$ is physically reasonable because it is expected that $g_\chi \in [10^{-6},1]$ for $\phimu$ in the eV-MeV mass-scale, which is at least four orders of magnitude greater than the $g_B \leq 10^{-10}$ constraint \cite{Reddy2016,Nelson_2018}\footnote{Note, $g_\chi$ is allowed be zero in the \cite{Nelson_2018} fermionic ADM model because the Fermi degeneracy pressure offers the ADM core support against gravitational collapse.}. Therefore, in order to obey the assumption that $g_B \ll g_\chi$, we approximate the effective self-repulsion strength to a non-zero value that produces ADM cores similar to the physically allowed value of zero in \hyperref[AppendixB]{Appendix A}. Given the assumption that  $g_B \ll g_\chi$, and by assuming that all of the possible accumulated ADM has thermally equilibrated within the neutron star then the ADM EoS can be computed in the zero temperature limit. Thus, with $\hbar$ and $c$ restored, the \cite{Nelson_2018} fermionic ADM EoS is expressed as

\begin{multline}\label{eps_chi}
     \varepsilon_\chi = \frac{c^5 m_\chi^4}{8 \pi^2 \hbar^3} \left[\sqrt{1+z^2} (2z^3 +z) - ln(z + \sqrt{1+z^2})   \right]\\  + \frac{g_\chi^2}{2 m_\phi^2} \frac{c^5 (m_\chi z)^6}{\hbar^3 (3\pi^2)^2}
\end{multline}
\begin{multline}\label{press_chi}
     p_\chi = \frac{c^5 m_\chi^4}{8 \pi^2 \hbar^3} \left[\sqrt{1+z^2} \Big(\frac{2}{3}z^3 -z \Big) + ln(z + \sqrt{1+z^2})   \right]\\ + \frac{g_\chi^2}{2 m_\phi^2} \frac{c^5 (m_\chi z)^6}{\hbar^3 (3\pi^2)^2},
\end{multline}
where $\varepsilon_\chi$ is the fermionic ADM energy density, $p_\chi$ is the fermionic ADM pressure, and $z = \hbar k_\chi/m_\chi c$ is the relativity parameter defined in terms of the ADM Fermi momentum $k_\chi$. By inserting the baryonic matter and fermionic ADM equations of state into the two-fluid TOV equation, the ADM admixed neutron star mass-radius relation can be computed.

\section{Methodology}\label{method}
We will now show how to construct our Bayesian inference framework that considers both baryonic matter and ADM inside neutron star interiors. In this section, we discuss this inference framework, the prior space of the PP parameterization of the baryonic EoS, constraints on the fermionic ADM EoS parameters, and the source selection for both the real and synthetic data analyses.
\subsection{Bayesian framework}\label{method:framework}
We use the inference framework developed in \cite{Rutherford2023}, which adapts the Bayesian analysis of \cite{Raaijmakers2019,Greif19,Raaijmakers2020,Raaijmakers2021,Rutherford2024} to include the possible presence of an ADM EoS. In particular, we use the open source EoS inference code NEoST \texttt{v2.0.0}, which includes ADM functionality \cite{Raaijmakers24} \footnote{\url{https://github.com/xpsi-group/neost}}. A full reproduction package, including the posterior samples and scripts to generate the plots in this work, is available
in a Zenodo repository at \citet{plotdata_fermionicadm}.

We use Bayes's theorem to write the posterior distribution on all ADM and baryonic equations of state as
\begin{align}
    p(\boldsymbol{\theta}, \boldsymbol{\varepsilon_c} |\mathbf{d})\nonumber &\propto p(\boldsymbol{\theta}) p(\boldsymbol{\varepsilon_c}|\boldsymbol{\theta}) ~p(\mathbf{d} |\boldsymbol{\theta}, \boldsymbol{\varepsilon_c}) \\
    & \propto p(\boldsymbol{\theta})  p(\boldsymbol{\varepsilon_c}|\boldsymbol{\theta}) p\boldsymbol{(}\mathbf{d}|\mathbf{M}(\boldsymbol{\theta},\boldsymbol{\varepsilon_c}), \mathbf{R}(\boldsymbol{\theta},\boldsymbol{\varepsilon_c})\boldsymbol{)} \label{eq:Bayes step 1}, 
\end{align}
where $\boldsymbol{\theta}$ is the vector containing all ADM and baryonic EoS parameters, $\boldsymbol{\varepsilon_c}$ is the vector containing the baryonic and ADM central energy densities, $\mathbf{d}$ is the vector containing the masses and radii of the sources from each scenario, $\mathbf{M}(\boldsymbol{\theta},\boldsymbol{\varepsilon_c})$ is the mass of a produced admixed neutron star, and $\mathbf{R}(\boldsymbol{\theta},\boldsymbol{\varepsilon_c})$ is the radius of the admixed neutron star. Moreover, by assuming each of the mass-radius datasets are independent of one another and equating the mass-radius datasets to those derived from PPM, we obtain
\begin{align}
\label{eq:posterior}
    p(\boldsymbol{\theta}, \boldsymbol{\varepsilon_c} |\mathbf{d}) \propto~& p(\boldsymbol{\theta}) p(\boldsymbol{\varepsilon_c}|\boldsymbol{\theta}) \\ & \prod_{i} p(M_i, R_i ~|~d_{PPM,i}) \nonumber, 
\end{align}
where $i$ runs over the number of stars for which PPM delivers the mass and radius and $d_{PPM,i}$ is an element in the $\mathbf{d}$ vector in which PPM was used. Furthermore, since the ADM mass-fraction is a function of baryonic and ADM EoS parameters and central energy densities, we can sample over the ADM mass-fraction instead of the ADM central energy density by introducing $F_{\chi} = F_{\chi}(\boldsymbol{\theta}, \boldsymbol{\varepsilon_{c,B}}, \boldsymbol{\varepsilon_{c,ADM}})$. This implies that the posterior distribution $\boldsymbol{\theta}$ and $\boldsymbol{\varepsilon_c}$ can be rewritten as

\begin{align}
    p(\boldsymbol{\theta}, \boldsymbol{\varepsilon_{c,B}}, \nonumber \boldsymbol{F_{\chi}} |\mathbf{d}) \propto~& p(\boldsymbol{\theta}) p(\boldsymbol{\varepsilon_{c, B}}|\boldsymbol{\theta}) p( \boldsymbol{F_{\chi}} |\boldsymbol{\theta}, \boldsymbol{\varepsilon_c}) \\ & \prod_{i} p(M_i, R_i ~|~d_{PPM,i}),
\end{align}
where $\boldsymbol{\varepsilon_{c,B}}$ and $\boldsymbol{\varepsilon_{c,ADM}}$ are the central energy densities of baryonic matter and ADM, respectively. Sampling over the ADM mass-fraction rather than the ADM central energy density allows for a direct comparison between the potential accumulation methods of ADM in neutron stars and the $F_\chi$ prior space.

\subsection{Baryonic matter priors}
We will now define the priors on the baryonic matter EoS, which we describe using the parametrized PP EoS model used in \cite{Hebeler13,Raaijmakers2019,Raaijmakers2020,Raaijmakers2021, Rutherford2024} from Sec.~\ref{EoSs}. Since the neutron star crust is modeled after the fixed BPS crust EoS \cite{Baym71}, the prior ranges that need to be defined are those on the \cite{Hebeler13} $\chi$EFT band, the three polytropes, and the two varying transition densities between each polytrope. As described in Sec.~\ref{EoSs}, the \cite{Hebeler13} $\chi$EFT band is fitted between $0.5 n_0$ and $1.1 n_0$ using a single polytrope, which is of the form
\begin{equation}\label{ceft_polytrope}
    P_{\chi \mathrm{EFT}}(n_B) = K\big( n_B/n_0\big)^{\Gamma},
\end{equation}
where $P_{\chi EFT}$ is the $\chi$EFT pressure as a function of baryonic number density $n_B$, $K$ is the matching constant to $P_{\chi EFT}$ in units of MeV fm$^{-3}$, and $\Gamma$ is the adiabatic index. Following the fitting procedure in \cite{Raaijmakers2021}, which fits the maximum and minimum pressure bands of a given $\chi$EFT using Eq.~\ref{ceft_polytrope}, the \cite{Hebeler13} band is well reproduced by $K \in [1.676,2.814]$ MeV fm$^{-3}$ and $\Gamma \in [2.486,2.571]$. We take the fit parameters for $K$ and $\Gamma$ to be the prior bounds on the \cite{Hebeler13} $\chi$EFT band. 

In order to produce the three polytropic priors, we consider the allowed ranges described in \cite{Hebeler13,Greif19}, which define the priors on the first polytropic index ($\Gamma_1$), the transition density between the first and second polytrope ($n_1$), the second polytropic index ($\Gamma_2$), the transition density between the second and third polytrope ($n_2$), and the third polytropic index ($\Gamma_3$). The priors on each parameter are given as: $\Gamma_1 \in [1,4.5]$, $\Gamma_2 \in [0.,8.]$, $\Gamma_3 \in [0.5,8]$, and $1.5 n_0 \leq n_1 < n_2 \leq 8.3 n_0$. Here, the priors on $\Gamma_1$ in \citet{Hebeler13} restricted to a smaller range relative to the other two polytropes because it controls the stiffness of the EoS between $1.1 n_0 \leq n \leq 1.5 n_0$, which is still well constrained by the \cite{Hebeler13} $\chi$EFT calculations. Therefore, in order to remain consistent with causality and with the $\chi$EFT band at $1.1 n_0$, the variations in $\Gamma_1$ must be restricted to $[1,4.5]$ \cite{Hebeler13}. The priors of $\Gamma_2$ where chosen to allow for the possibility of a first-order phase transition (i.e., $\Gamma_2 = 0$), such as those in quark matter EoSs, and to capture the maximally stiff EoSs of this PP framework, which corresponds to $\Gamma_{2,max} = 8$, which is determined by causality and a maximum neutron star mass of 3 $\Msun$. For densities beyond $n_2$, the prior of $\Gamma_3$ is chosen similarly to $\Gamma_2$, but 0 is excluded to avoid numerical artifacts associated with a first-order phase transition in the density range of $\Gamma_2$. The transition densities between the three polytropes, $n_1$ and $n_2$, are allowed up to $8.3 n_0$ as this is the maximal central density of the PP model \cite{Greif19,Hebeler13}. Since each polytropic segment in the \cite{Hebeler13} PP model is determined by three parameters, $K_i$, $\Gamma_i$, and the number density $n$, then a prior on $K_i$ must also be defined. However, according to \cite{Read2009}, when an EoS is specified at lower density, continuity of pressure forces $K_i$ to
\begin{equation}
    K_i = \frac{P(n_t)}{n_t^{\Gamma_i}},
\end{equation}
where $n_t$ is the transition density. In this case the $n_t$'s are set by the $\chi$EFT band for the first polytrope, $n_1$ for the second polytrope, and $n_2$ for the third polytrope. Lastly, for each PP EoS parameter, we uniformly sample the considered prior ranges.
 \begin{figure*}
    \centering
    \includegraphics[width = .6\textwidth]{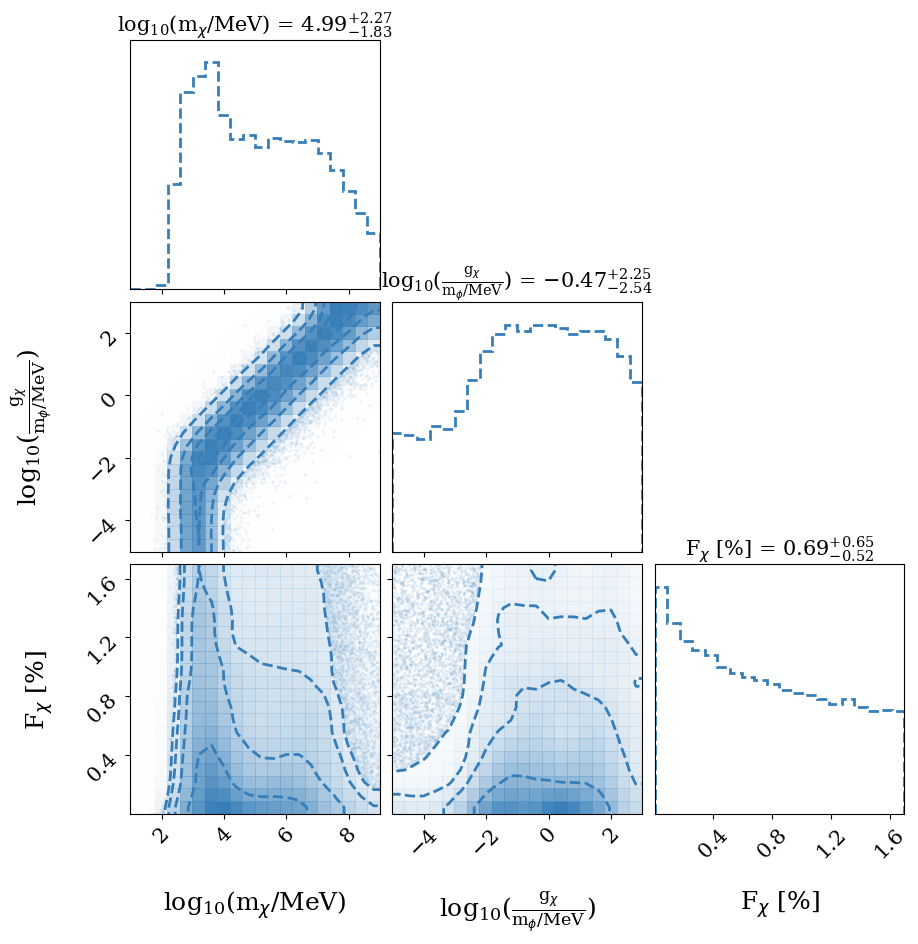}
    \caption{Prior corner plot of the fermionic ADM EoS. Here, the ADM particle mass, effective self-interaction strength, and mass-fraction are plotted against each other, where the dark shaded regions represent a higher prior probability and lighter shaded regions represent a lower prior probability.  The dashed blue lines in the 2-D contour plots represent the 0.5, 1, 1.5, and 2 $\sigma$ contour levels. The top panels in each column show the 1-D prior histogram. The figure titles on the diagonal show the median value with the 0.16 and 0.84 fractional quantiles. In the $\log_{10}(m_\chi/\mathrm{MeV})$$-$$\log_{10}(g_\chi/(m_\phi/\mathrm{MeV}))$ plane we observe that the prior density has two large regions of no shading. }
    \label{fig1}
\end{figure*}

\subsection{Fermionic ADM priors}\label{fermionic priors}
We now define the priors on each of the fermionic ADM EoS parameters. In particular, we use the available literature and physical constraints (if any) to construct the prior spaces on $m_\chi$, $F_\chi$, and $g_\chi/m_\phi$, all three of which completely define the fermionic ADM EoS.

To define the prior space on the fermionic ADM particle mass, we consider the physical constraints on $m_\chi$ from \cite{Kouvaris2011_pt2, Gresham2018}. The considered lower bound on $m_\chi$ was obtained by \cite{Kouvaris2011_pt2}, which showed that the minimum $m_\chi$ such that no ADM particle can exceed the neutron star escape velocity is 
    \begin{equation}
        m_\chi \geq 10^{-2}\, \mathrm{MeV}.
    \end{equation}
On the other hand, \cite{Gresham2018} considered fermionic ADM core collapse to a black hole inside the host neutron star. The authors showed that in order to avoid the formation of a black hole, regardless of whether the ADM is self-interacting,
\begin{equation}
    m_\chi \leq 10^9 \, \mathrm{MeV}.
\end{equation}
Therefore, the prior space on the fermionic ADM particle mass is defined by 
\begin{equation}
    m_\chi \in [10^{-2} \, \mathrm{MeV}, 10^9 \, \mathrm{MeV}].
\end{equation}

While the prior space on the fermionic ADM particle mass is well constrained, the ADM mass-fraction prior space is not. Typically, the fermionic ADM mass-fraction prior space is defined using physically motivated ADM accumulation methods, such as neutron bremmstrahlung \cite{Nelson_2018,Ellis2018}, production of ADM in supernovae \cite{Nelson_2018, Collier2022}, and neutron conversion to ADM \cite{Bastero-Gil2024,Husain2023,Ellis2018}. The neutron Bremsstrahlung reaction of ADM produces the gauge boson $\phimu$ via the conversion of the kinetic energy produced between the scattering of two neutrons ($NN$), i.e., $NN \longrightarrow NN \phimu$. Moreover, since $\phimu$ is strongly coupled to ADM, the reaction of $NN \longrightarrow NN \chibar \chi$ proceeds at a similar rate as that of the neutron bremsstrahlung reaction. 
    
In order to ensure that the ADM does not annihilate with the anti-ADM particles after the neutron bremmstrahlung reaction, \citet{Nelson_2018} assumes that anti-ADM is repulsed by baryonic matter and ADM is attracted to it. By making this assumption, \citet{Nelson_2018} showed that the energy difference of anti-ADM to ADM in the mean-field approximation is given by, in units of $\hbar = c = 1$,
    \begin{equation}
        \Delta E = E_{\Bar{\chi}} - E_\chi = \frac{2 g_\chi g_B}{m_\phi^2} n_B,
    \end{equation}
where $E_{\Bar{\chi}}$ is the energy of the anti-ADM particle, $E_\chi$ is the energy of the ADM particle, and $n_B$ is the baryon number density. Thus, for the mass scales of $m_\phi$ for this model and when $g_B g_\chi \simeq 10^{-16} - 10^{-10}$, $\Delta E$ is comparable to the gravitational binding energy of ADM/anti-ADM, which will preferentially trap the ADM particles and expel the anti-ADM particles. Thus, neutron bremmstrahlung provides a mechanism to produce and trap ADM inside neutron stars without having it diminished by annihilation, which allows for the maximum amount of ADM to be accumulated.

Since young neutron stars have temperatures around 50 MeV and assuming $m_\chi = 100$ MeV, neutron Bremsstrahlung of ADM can produce $\approx 0.02 M_{NS}$, where $M_{NS}$ is the mass of the neutron star \cite{Ellis2018}. Since neutrons inside compact objects can reach Fermi momenta of several hundred MeV, neutrons can decay to ADM for ADM particle masses less than $m_n + \mathcal{O}(k_{\mathrm{F},\chi}^2/2m_n)$, where $m_n$ is the mass of the neutron. This process allows for total ADM masses of $\approx 0.05 M_{NS}$ \cite{Ellis2018,Collier2022}. Lastly, ADM can accumulate inside neutron stars via the production of ADM in supernovae. Since supernovae are very energetic events with luminosities in excess of $\mathcal{O}(10^{52}$ erg/s), they can efficiently produce ADM particles that can then be trapped within the newly born neutron star. As discussed in \cite{Collier2022}, supernovae events can produce total accumulated ADM masses up to about 0.15 $\Msun$ for MeV-scale ADM particles.    

Although neutron Bremsstrahlung of ADM, neutron conversion of ADM, and production of ADM in supernovae each are capable of producing total ADM masses in the range of 0.02-0.15 $\Msun$, all three of these processes are effective for ADM particle masses up to $\mathcal{O}(10^{2-3})$ MeV, which would only apply to a small fraction of the $m_\chi$ prior space. Since the aforementioned physically motivated mechanisms only apply to a few order magnitudes within the $m_\chi$ prior space, we choose not consider these ADM accumulation methods within our $F_\chi$ prior space. Other accumulation methods, such as a neutron star passing through an ADM over-density, accretion of baryonic matter onto a pre-existing ADM core, and a dark star-neutron star merger, could be considered \cite[see][and references therein]{Karkevandi2022, Collier2022}. However, such ADM accumulation mechanisms are highly speculative, thus we also neglect these accumulation methods. Therefore, to define the $F_\chi$ prior space, we will follow the upper bound ADM mass-fraction estimate of \cite{Ivanystkyi2020} using the Nevarro-Frenk-White (NFW) dark matter mass density profile \cite{Nevarro} to compute local dark matter density around our considered sources. The NFW dark matter profile is given by
\begin{equation}
    \rho_\chi(r) = \frac{\rho_0}{\frac{r_s}{r}\left(1+ \frac{r}{r_s} \right)^2},
\end{equation}
where $\rho_\chi(r)$ is the ADM mass density a radius, $r$, from the Galactic center (GC), $\rho_0 = 5.22 \pm 0.46 \cdot 10^7 \Msun/\mathrm{kpc}^3$ is the central density \cite{Lin2019}, $r_s = 8.1 \pm 0.7$ kpc is the scale radius \cite{Lin2019}. By considering the $F_\chi$ approximation of \cite{Ivanystkyi2020}, an upper limit on $F_\chi$ can be determined without a heavily restricted ADM particle mass prior space or having to consider a very hypothetical accumulation scenario.

In order to estimate the upper bound on $F_\chi$, \cite{Ivanystkyi2020} calculated the ratio of the ADM mass density to the combined mass density of baryonic matter and ADM in the vicinity of PSR J0740$+$6620 and PSR J0348$+$0432, which are 8.6 kpc and 9.9 kpc from the GC, respectively. To model the baryonic mass density distribution, \cite{Ivanystkyi2020} used only the contribution of the Milky Way's stellar disc because both pulsars were taken to be sufficiently far away from the Galactic bulge. The shape of Galactic stellar disc profile is
\begin{equation}
    \rho_B(r) = \rho_{s,B} e^{-r/r_{s,B}},
\end{equation}
where $\rho_B(r)$ is baryonic mass density as a function of $r$, from the GC, $\rho_{s,B} = 15\, \Msun/\mathrm{pc}^3$ is the baryonic mass density scale, and $r_{s,B} = 3.0$ kpc is the baryonic scale radius \cite{Sofue2013}. Since this work seeks to constrain fermionic ADM using neutron stars delivered by PPM, we will only consider PSR J0740$+$6620. Using the radial distance of PSR J0740$+$6620 to the GC, \cite{Ivanystkyi2020} found the $F_\chi$ upper bound near PSR J0740$+$6620 to be $F_\chi \leq 1.7\%$. By repeating the $F_\chi$ upper bound estimation for the two other PPM sources from \textit{NICER}, we find that the maximum possible ADM mass-fraction is $\leq 1.69\%$ and $\leq 1.67\%$ for PSR J0030$+$0451 and PSR J0437$-$4715, respectively. Note, we have found the radial distances to the GC to be 8.45 kpc and 8.35 kpc, for PSR J0437$-$4715 and PSR J0030$+$0451, respectively. Therefore, because all three PPM delivered pulsars have similar $F_\chi$ upper estimates and PSR J0740$+$6620 can achieve the highest possible ADM mass-fraction, we adopt the upper bound on the ADM mass-fraction prior space to be 
\begin{equation}
    F_\chi \leq 1.7\%.
\end{equation} 
Lastly, we want to caution that the \cite{Ivanystkyi2020} calculation provides a best case upper estimate on $F_\chi$ in all of the \textit{NICER} targets and that the true ADM mass-fraction in each pulsar due to their respective ADM surroundings is likely smaller than 1.7\%.

Depending on the assumed ADM accumulation mechanism and scenario, the $F_\chi$ prior space can be constrained to be a finite size, but the effective fermionic ADM self-repulsion strength has yet to be physically constrained. In order to ensure that the $g_\chi/m_\phi$ prior space is bounded from above, we adopt the upper bound of $g_\chi/m_\phi \leq 10^3 \, \mathrm{MeV^{-1}}$ to capture the highest self-repulsion strengths used in \cite{Nelson_2018}. From below, $g_\chi/m_\phi$ is physically allowed to be zero because the ADM fermionic degeneracy pressure provides enough support against gravitational collapse to a black hole. However, in Sec.~\ref{EoSs}, we have additionally assumed that $g_B \ll g_\chi$, thus a non-zero approximation to $g_\chi/m_\phi = 0 \, \mathrm{MeV^{-1}}$ is necessary. To accomplish this, we compute the relative radial percent difference (RRPD) between 0 MeV$^{-1}$ and a small non-zero self-repulsion for various baryonic matter EoSs and pairs of ($m_\chi$, $F_\chi$). We find that the RRPDs between $g_\chi/m_\phi = 10^{-5} \, \mathrm{MeV^{-1}}$ and zero self-repulsion do not exceed 4$\cdot 10^{-3}$\%. This shows that $10^{-5} \, \mathrm{MeV^{-1}}$ is an adequate approximation for 0 $ \mathrm{MeV^{-1}}$ down to mass-radius measurements with uncertainties $\mathcal{O}(10^{-3}\%)$ (see \hyperref[AppendixB]{Appendix A} for further details).

In summary, the fermionic ADM EoS prior space is taken to be
\begin{align}
    &\log_{10}(m_\chi/\mathrm{MeV}) \in [-2,9]\\
    &F_\chi \in [0,1.7] \%\\
    &\log_{10}\Big(\frac{g_\chi}{m_\phi/\mathrm{MeV}}\Big) \in [-5,3]. 
\end{align}
Within each interval above, we uniformly sample each ADM parameter. We also assign all halo configurations to have a zero likelihood because the existence of any ADM halo has been shown to modify the pulse profile of neutron stars and thus the interpretation of the \textit{NICER} mass-radius measurements \cite{Shawqi2024}. This results in only ADM cores within the prior space. Moreover, within the remaining ADM core configurations, we also assign any ADM admixed neutron star with a mass $< 1 \Msun$ to have a zero likelihood evaluation. The minimum neutron star mass constraint is motivated by the theoretical description of a newly born neutron star \cite{Strobel_1999}. In addition, the 1 $\Msun$ constraint is compatible with the minimum neutron star remnant masses from core-collapse supernovae simulations \cite[see e.g.,][]{Radice17,Suwa18}\footnote{We also want to note that our imposed 1 $\Msun$ constraint, although well supported, is in tension with the mass-radius measurement of the HESS J1731$-$347 supernova remnant \cite{Doroshenko2022}. However, this measurement is challenged by \cite{Alford2023} because the \cite{Doroshenko2022} analysis relies on several assumptions about the distance to the star, the spectral modeling, and the data set chosen in analysis.}. The consequences of the no-ADM halo and 1 $\Msun$ constraints can be seen as the non-shaded regions above and below the stripe in the $\log_{10}\boldsymbol{(}g_\chi/(m_\phi/\mathrm{MeV})\boldsymbol{)} \, vs. \, \log_{10}(m_\chi/\mathrm{MeV})$ plot of Fig.~\ref{fig1}. In Fig.~\ref{fig1}, we show the prior corner plots of fermionic ADM EoS parameters, which shows a nonuniform distribution for all three fermionic ADM EoS parameters.

\subsection{Source selection: real and synthetic}
\begin{figure}
\centering
\begin{subfigure}[b]{0.5\textwidth}
   \includegraphics[width=\textwidth,scale = 1.5]{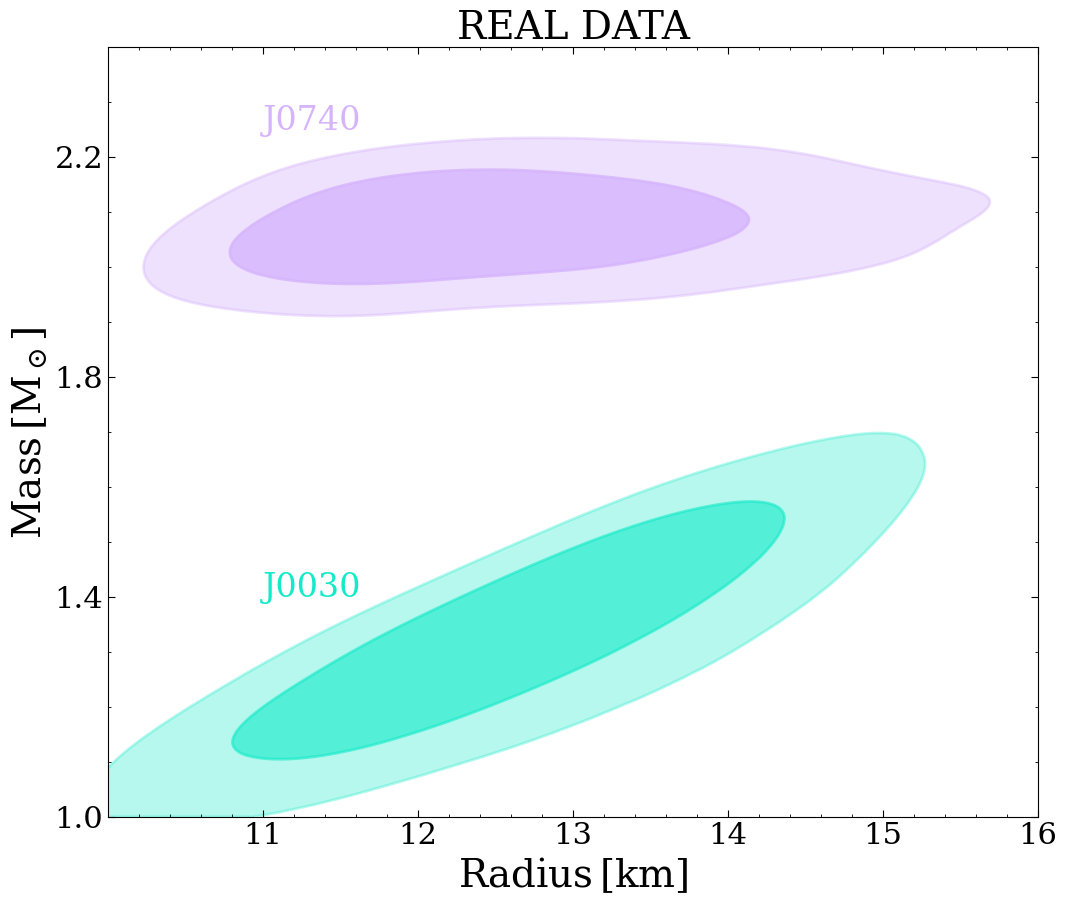}
   \label{fig_ellipses:real} 
\end{subfigure}

\begin{subfigure}[b]{0.5\textwidth}
   \includegraphics[width=\textwidth,scale = 1.5]{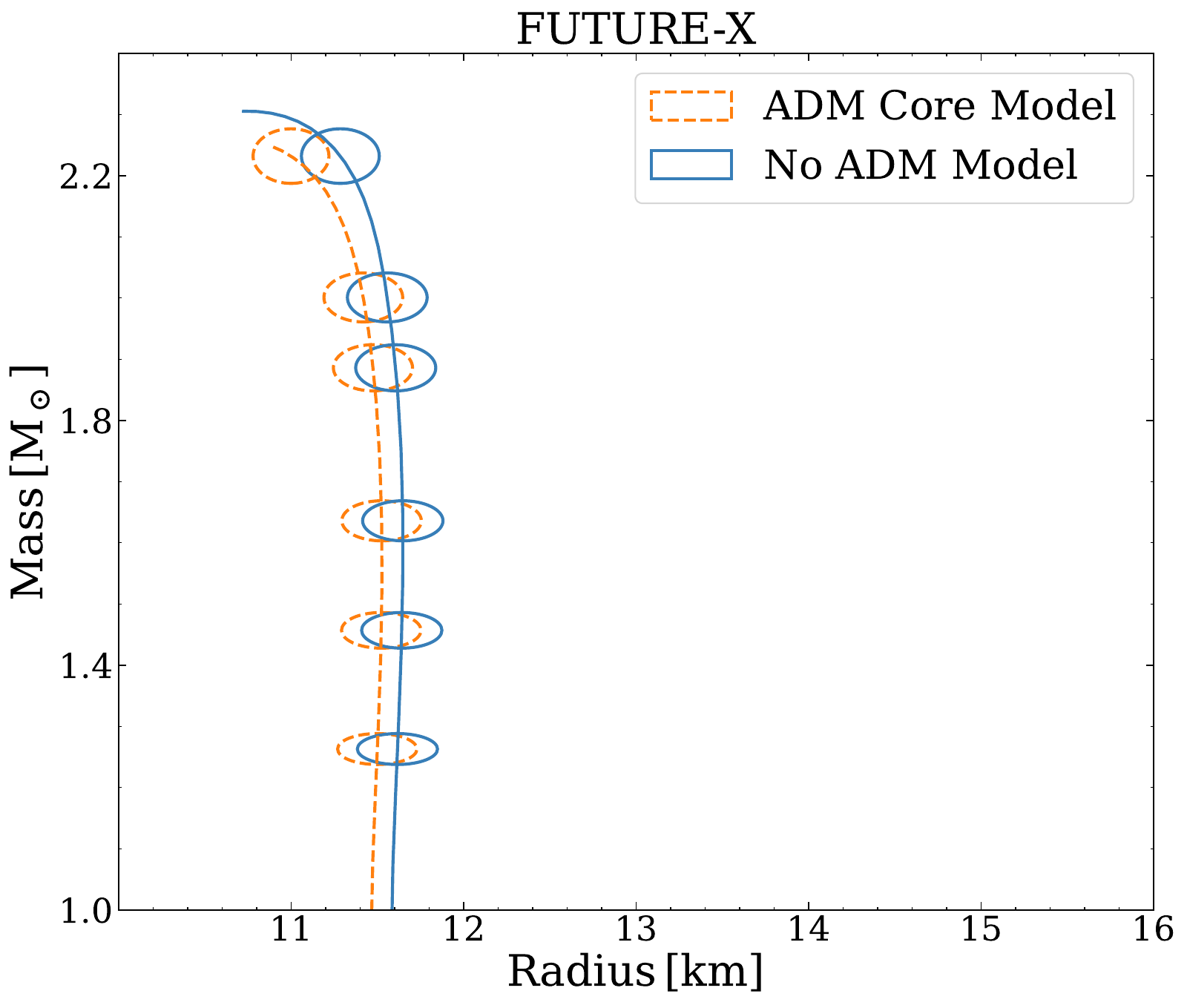}
   \label{fig_ellipses:synthetic}
\end{subfigure}
\caption{Top panel: The 68\% and 95\% level uncertainty ellipses of the mass-radius measurements of PSR J0740$+$6620 from \cite{Riley0740} and PSR J0030$+$0451 from \cite{Riley0030}; Bottom panel: Uncertainty ellipses from the 1$\sigma$ level of the 2-D Gaussian for each of the synthetic \textit{Future-X} sources calculated from both ground truth models defined in Sec.~\ref{syntheticdata}.}
\label{fig_ellipses}
\end{figure}
By considering both real and synthetic data, we will be able to demonstrate the current constraining power of \textit{NICER}, and the potential future constraints of large area X-ray telescopes, like \textit{STROBE-X}.

In order to assess the current capabilities of PPM delivered measurements, we consider the mass-radius posteriors of the \textit{NICER} targets PSR J0740$+$6620 from \cite{Riley0740} and PSR J0030$+$0451 from \cite{Riley0030}. In top panel of Fig.~\ref{fig_ellipses}, we show the mass-radius posteriors of PSR J0740$+$6620 of \cite{Riley0740} and PSR J0030$+$0451 of \cite{Riley0030} for our real data inferences.

Although the current mass-radius uncertainties on the \textit{NICER} targets are at the $\sim$10\% level, it is interesting to consider the impact of future measurements in which more neutron stars will be observed at significantly lower mass-radius uncertainties. For the inferences where we consider synthetic neutron star mass-radius measurements, we model our sources using the \textit{Future-X} scenario of \cite{Rutherford2023}. The \textit{Future-X} scenario assumes six sources in the mass range of $1.2-2.2 \, \Msun$ with mass-radius uncertainties at the two percent level. This scenario is modeled after a best case possibility for the proposed NASA Probe mission \textit{STROBE-X}, where \textit{STROBE-X} performs long targeted observations of the six best candidates. We expect this scenario to deliver uncertainties at the two percent level, which would provide the strongest constraints on the neutron star EoS. The bottom panel of Fig.~\ref{fig_ellipses} shows the uncertainty ellipses corresponding to the \textit{Future-X} scenario.
\begin{figure*}[t!]
\centering
\begin{subfigure}{0.50\textwidth}
   \includegraphics[width=\textwidth]{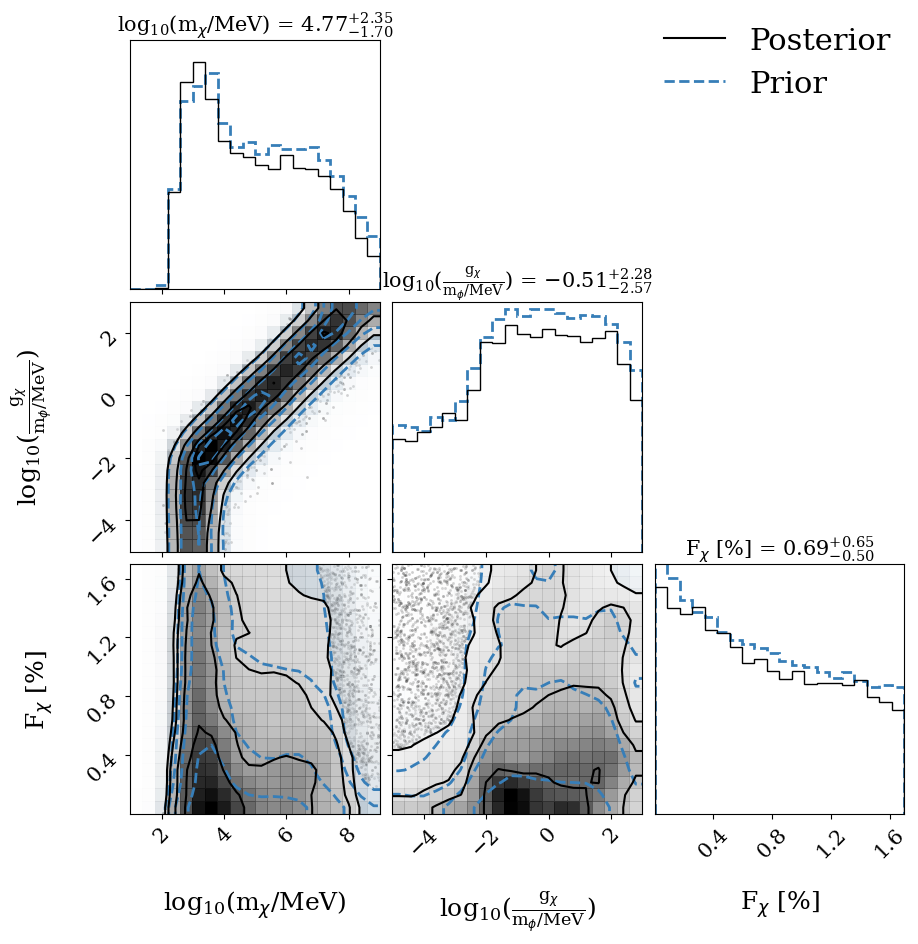}
   \label{fig2} 
\end{subfigure}%
\begin{subfigure}{0.50\textwidth}
   \includegraphics[width=\textwidth]{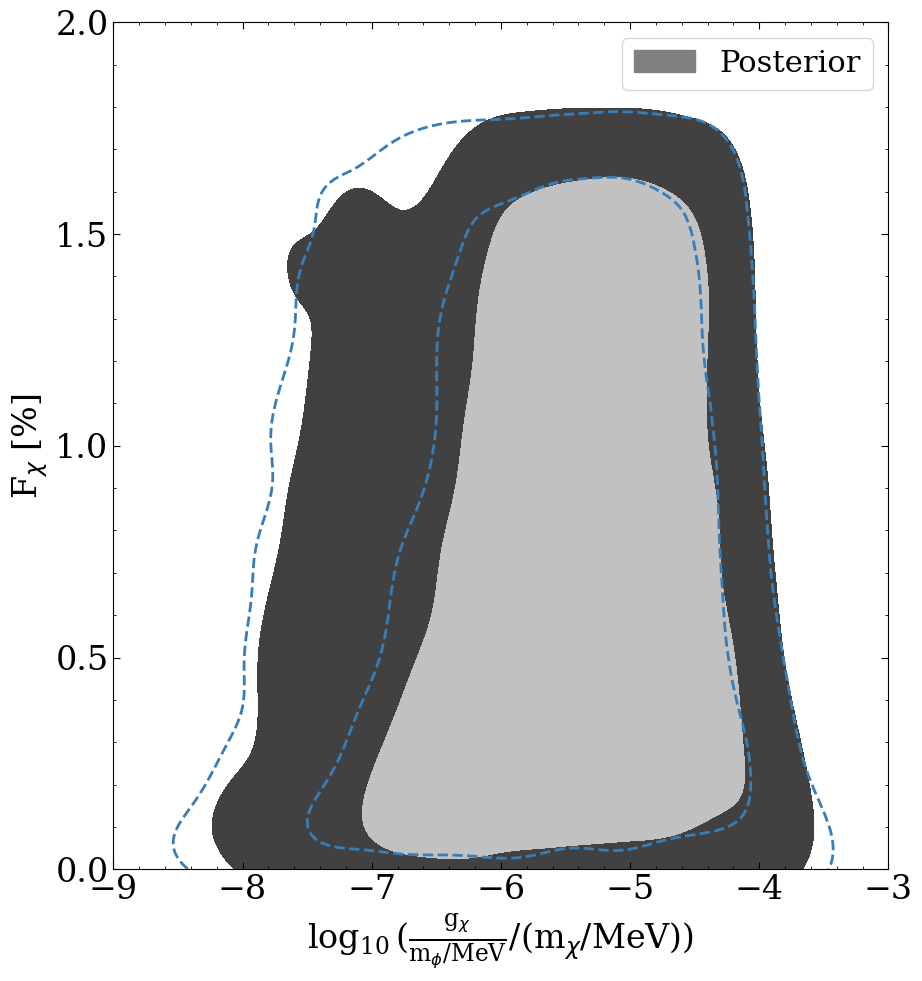}
   \label{fig3}
\end{subfigure}
\caption{Left panel: Posterior distribution of the fermionic ADM EoS
parameters (solid black lines) in which real data is considered. For comparison, we have overlaid the posteriors with their respective priors (dashed blue lines). The contour levels are same as in Fig.~\ref{fig1}; Right panel: Probability density contour plot of the ADM posteriors in the $F_\chi$ vs. $\log_{10} \left(\frac{g_\chi}{(m_\phi/\mathrm{MeV})}/(m_\chi/\mathrm{MeV})\right)$ plane. Note, the contours represent the 1$\sigma$ (light grey) and 2$\sigma$ (dark grey) levels for both the prior and posterior. Here we see that the 1$\sigma$ level posteriors favor slightly higher ratios of $g_\chi/m_\chi$ and $m_\chi$, but the 2$\sigma$ posteriors are almost touching the priors for all $F_\chi$. In the left panel, we find that the priors and posteriors are nearly identical in all panels. However, in the right panel, we find that the 1$\sigma$ and $2\sigma$ level posteriors favor slightly higher ratios of $g_\chi/m_\chi$ and $m_\chi$ than their respective priors.}
\label{fig_2_3}
\end{figure*}

\section{Results and discussion}\label{results}
In all of our Bayesian parameter estimations, we take the most conservative approach of simultaneously varying all EoS parameters. By sampling all parameters in the neutron star EoS model, the most likely combined EoS of baryonic matter and fermionic ADM can be inferred. Additionally, this approach also allows for the constraints on the fermionic ADM EoS to be determined. In this section, we first study the posteriors of the fermionic ADM EoS and baryonic EoS using the mass-radius measurements of PSR J0740$+$6620 \cite{Riley0740} and PSR J0030$+$0451 \cite{Riley0030}. Using the synthetic data of the \textit{Future-X} scenario, we again perform Bayesian inference on the fermionic ADM and baryonic matter EoSs to study the future promise of constraining fermionic ADM cores using neutron star mass-radius measurements.
\subsection{Real data inferences}\label{realdata}
In Fig.~\ref{fig_2_3}, we show the posterior distributions on the fermionic ADM EoS in which we consider the mass-radius measurements of PSR J0740$+$6620 \cite{Riley0740} and PSR J0030$+$0451 \cite{Riley0030}. Here, the corner plot in the left panel of Fig.~\ref{fig_2_3} shows that all of the 1-D histograms and 2-D posterior density contours strongly overlap with their respective priors. The strong overlap of the priors and posteriors on the fermionic ADM EoS parameters is due to the apparent degeneracy between the ADM EoS parameters. That is, varying both the baryonic matter and ADM EoSs allows for scenarios in which neutron stars with a baryonic EoS and an ADM core described by one set of ($m_\chi$, $g_\chi/m_\phi$, $F_\chi$) can be equally well described by a relatively similar baryonic EoS with an ADM core described by a different set ($m_\chi^{'}$, $g_\chi^{'}/m_\phi$, $F_\chi^{'}$), which would give both sets of ADM parameters similar likelihoods. For example, a neutron star with an ADM core defined by ($m_\chi = 45$ GeV, $g_\chi/m_\phi = 0.01 \, \mathrm{MeV^{-1}}$, $F_\chi = 1\%$) radially differs by $< 0.6\%$ from another ADM core with ($m_\chi = 0.5$ GeV, $g_\chi/m_\phi = 0.001 \, \mathrm{MeV^{-1}}$, $F_\chi = 1.7\%$). Therefore, because the radial difference between the two sets of parameters is much smaller than the $\mathcal{O}(10\%)$ measurement uncertainties, both sets will receive similar likelihood evaluations. From the observation that the fermionic ADM priors and posteriors are approximately identical, we conclude that the fermionic ADM EoS parameters cannot be constrained under the chosen priors and current uncertainties of the baryonic EoS.

If the fermionic ADM posteriors and priors are transformed into the $\log_{10}\left(\frac{g_\chi} {(m_\phi/\mathrm{MeV})}/(m_\chi/\mathrm{MeV})\right)$-$F_\chi$ plane (right panel of Fig.~\ref{fig_2_3}), we find that that the lower bound on the ratio of $g_\chi/m_\phi$ and $m_\chi$ can be constrained when compared to the prior. In particular, we find that the prior 68\% (95\%) credible levels on $\log_{10}\left(\frac{g_\chi} {(m_\phi/\mathrm{MeV})}/(m_\chi/\mathrm{MeV})\right)$ is $-5.7^{+1.08}_{-1.15}$ ($-5.7^{+1.62}_{-2.3}$). For the posteriors, we find the 68\%(95\%) credible levels to be $-5.57^{+0.97}_{-1.02}$ ($-5.57^{+1.48}_{-2.2}$). Thus, the lower bound on $\log_{10}\left(\frac{g_\chi} {(m_\phi/\mathrm{MeV})}/(m_\chi/\mathrm{MeV})\right)$ can be constrained to $-6.59$ and $-7.77$ at the 68\% and 95\% credible levels, respectively. The lower bound on the ratio of $g_\chi/m_\phi$ and $m_\chi$ can be constrained while the upper bound cannot because small ratios produce compact ADM cores with ADM central densities that are several orders of magnitude larger than the baryonic central densities for a given $F_\chi$, which significantly reduce the resulting neutron star mass below the $1\Msun$ constraint. While, for the same $F_\chi$, large ratios of $g_\chi/m_\phi$ and $m_\chi$ produce more diffuse fermionic ADM cores with ADM central densities less than baryonic central energy densities, which affect the overall neutron star mass less than the lower ratios. For instance, for $\log_{10}\left(\frac{g_\chi} {(m_\phi/\mathrm{MeV})}/(m_\chi/\mathrm{MeV})\right)$ = $-$8 and $F_\chi = 0.75\%$, the maximum central ADM density is $\approx 10^{20} \, g/cm^3$ and the maximum central baryonic density is $\approx 10^{15} \, g/cm^3$, which results in the maximum neutron star mass of $\approx 0.86 \, \Msun$. However, if we again take $F_\chi = 0.75\%$ and the same maximum baryonic central density, but $\log_{10}\left(\frac{g_\chi} {(m_\phi/\mathrm{MeV})}/(m_\chi/\mathrm{MeV})\right)$ = $-$4, the maximum central ADM density is reduced to $\approx 10^{13.4} \, g/cm^3$ and the maximum neutron star mass increases to $2.38 \, \Msun$

In Fig.~\ref{fig4}, the priors and posteriors of Fig.~\ref{fig_2_3} are converted to the pressure-energy density plane (left) to study the effect that fermionic ADM cores have on the uncertainties of the baryonic EoS\footnote{Note, we have scaled the energy density by a factor of c$^{-2}$ such that it has units of $g/cm^3$.}. In particular, we consider the posteriors that only vary the baryonic EoS parameters (‘Neglecting ADM)’ and the posteriors that vary the combined ADM and baryonic matter EoS parameters (‘Including ADM’). Fig.~\ref{fig4} shows that the 95\% confidence region of the `Including ADM' band (orange dashed dotted band) is marginally wider than the 95\% confidence region of the `Neglecting ADM' band (light green band). Quantitatively, we calculate that the `Including ADM' band is 1.021\% and 1.025\% wider than the `Neglecting ADM' band at $\mathrm{log_{10}}(\varepsilon \, \mathrm{cm^3/g}) = 14.381$ and $\mathrm{log_{10}}(\varepsilon \, \mathrm{cm^3/g}) = 15.010$, respectively. Accounting for the possibility of fermionic ADM cores broadens the uncertainties on the baryonic EoS because ADM cores decrease the neutron star mass and radius, which allows the baryonic EoS to be more stiff and remain in agreement with the source data. However, since including fermionic ADM broadens the 95\% confidence interval on the baryonic EoS by $\mathcal{O}(1\%)$, we conclude that fermionic ADM cores do not significantly impact the uncertainties on the baryonic EoS within the considered ADM priors.

\begin{figure*}[t!]
    \centering
    \includegraphics[width = \textwidth]{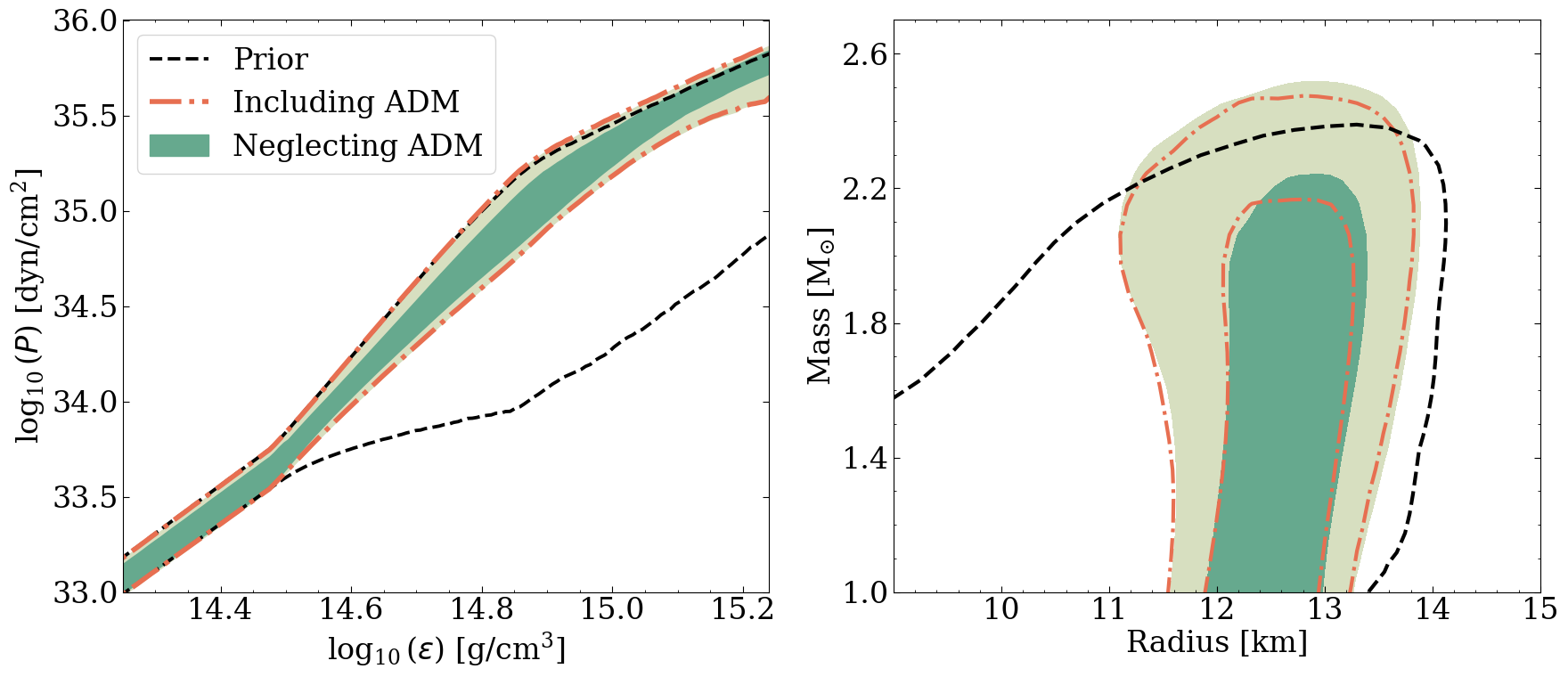}
    \caption{Left panel: Pressure-energy density posterior and prior distributions for the baryonic EoS for when fermionic ADM is included and neglected; Right panel: Mass-radius posterior and prior distributions for the total combined ADM and baryonic equations of state for when fermionic ADM is included and neglected. For both panels, the black dashed line represents the 95\% prior distribution, the orange dashed-dotted lines represent the 68\% and 95\% confidence regions of the posteriors that vary both the baryonic and fermionic ADM EoS parameters, and the light/dark green regions are the 68\% and 95\% confidence regions of the posteriors that only vary the baryonic EoS. Note, in the left panel we only show the 95\% confidence region of the `Including ADM' band. Here we see that the `Including ADM' bands are nearly identical to the `Neglecting ADM' bands in both panels.}
    \label{fig4}
\end{figure*}

 \begin{figure*}
\centering
\begin{subfigure}{.5\textwidth}
  \centering
  \includegraphics[width=\textwidth,scale = 1.5]{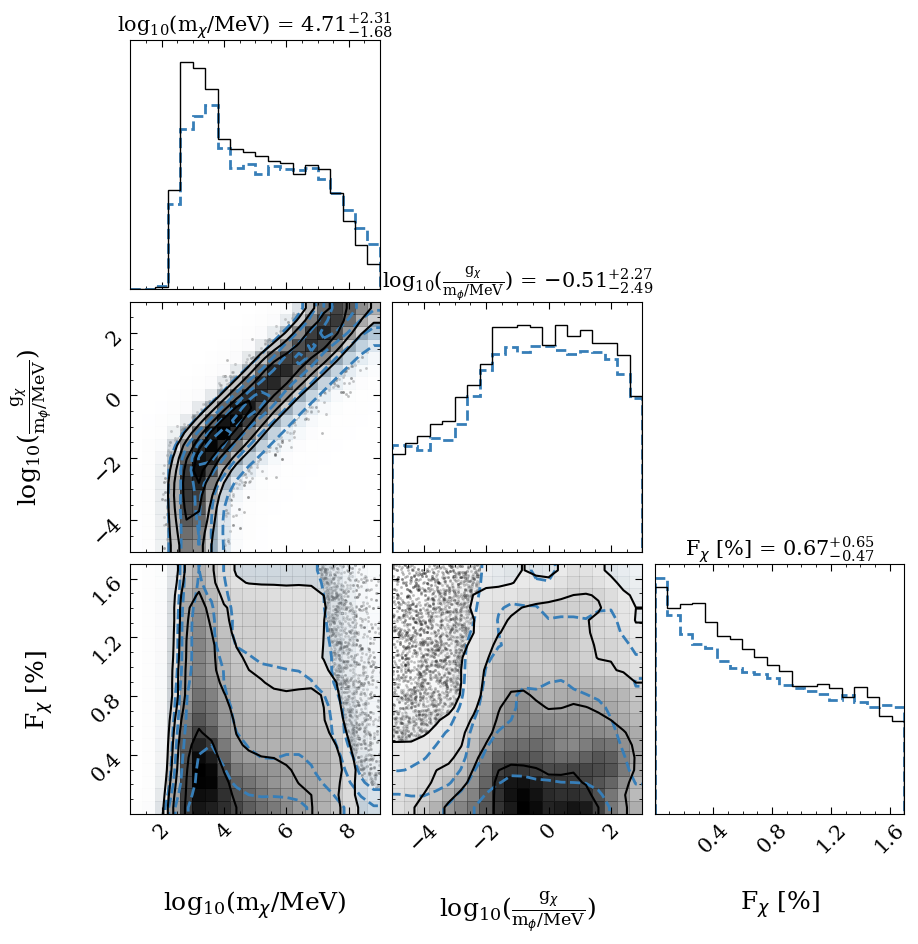}
  \label{fig6:sub1}
\end{subfigure}%
\begin{subfigure}{.5\textwidth}
  \centering
  \includegraphics[width=\textwidth,scale = 1.5]{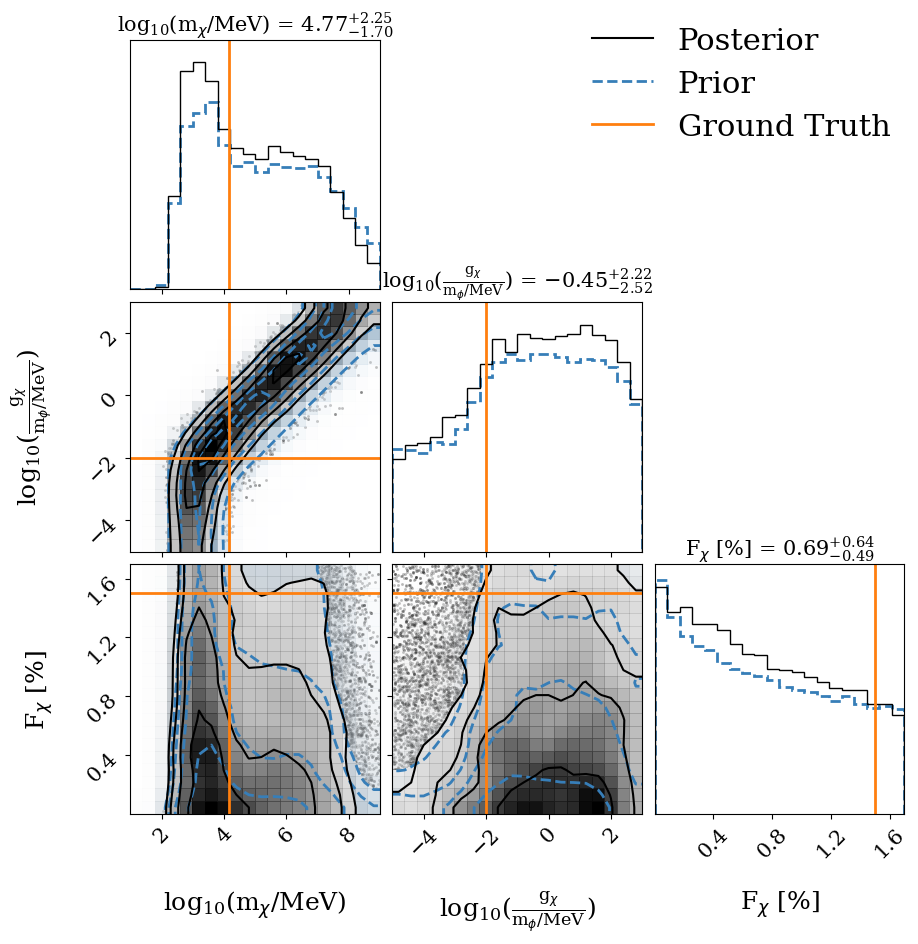}
  \label{fig6:sub2}
\end{subfigure}\\
\begin{subfigure}{.5\textwidth}
  \centering
  \includegraphics[width=\textwidth]{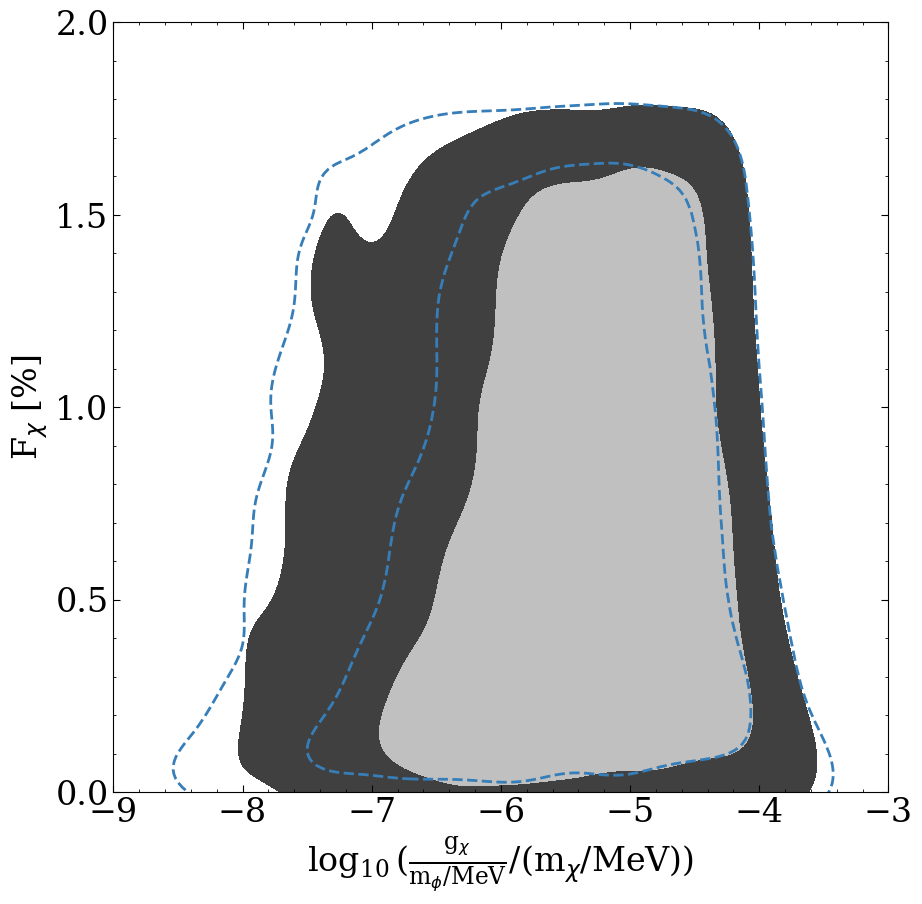}
  \label{fig6:sub3}
\end{subfigure}%
\begin{subfigure}{.5\textwidth}
  \centering
  \includegraphics[width=\textwidth]{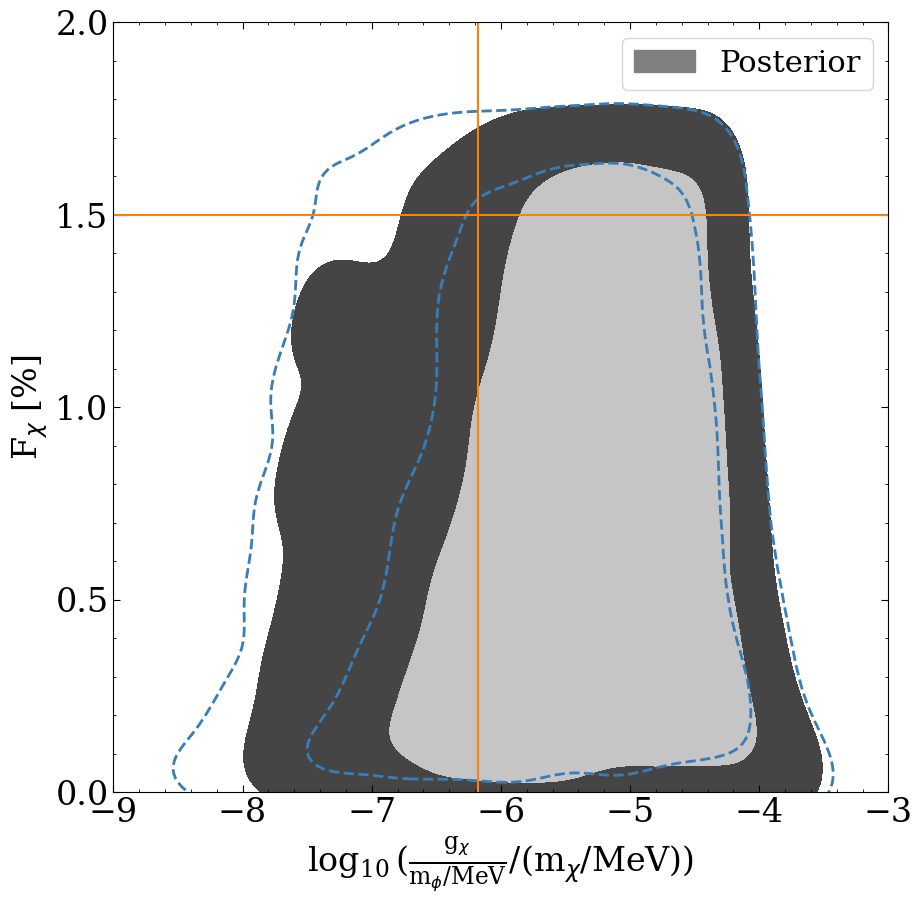}
  \label{fig6:sub4}
\end{subfigure}
\caption{Left two panels: \textit{Future-X} fermionic ADM posteriors for the ``No ADM" model; Right two panels: Same as the left two panels, but for the ``ADM Core'' model. The top two panels are the corner plots of the fermionic ADM EoS posteriors for the ``No ADM" and ``ADM Core'' ground truth models. The posterior and prior contour levels of the upper panels are same as in Fig.~\ref{fig_2_3}. In the bottom panels, we show the fermionic ADM posteriors and priors in the $\log_{10}\left(\frac{g_\chi}{m_\phi/\mathrm{MeV}}/(m_\chi/\mathrm{MeV}\right)$-$F_\chi$ plane. The contour levels of both the priors and posteriors are identical to the levels of Fig.~\ref{fig_2_3}. In all panels, the orange solid lines represent the ground truth values for the ``ADM Core'' model. We find that the corner plots of both ground truth models are approximately identical to one another, while the bottom contour plots differ slightly along the $F_\chi$-axis.}
\label{fig6}
\end{figure*}

 \begin{figure*}
\centering
\begin{subfigure}{.5\textwidth}
  \centering
  \includegraphics[width=\linewidth]{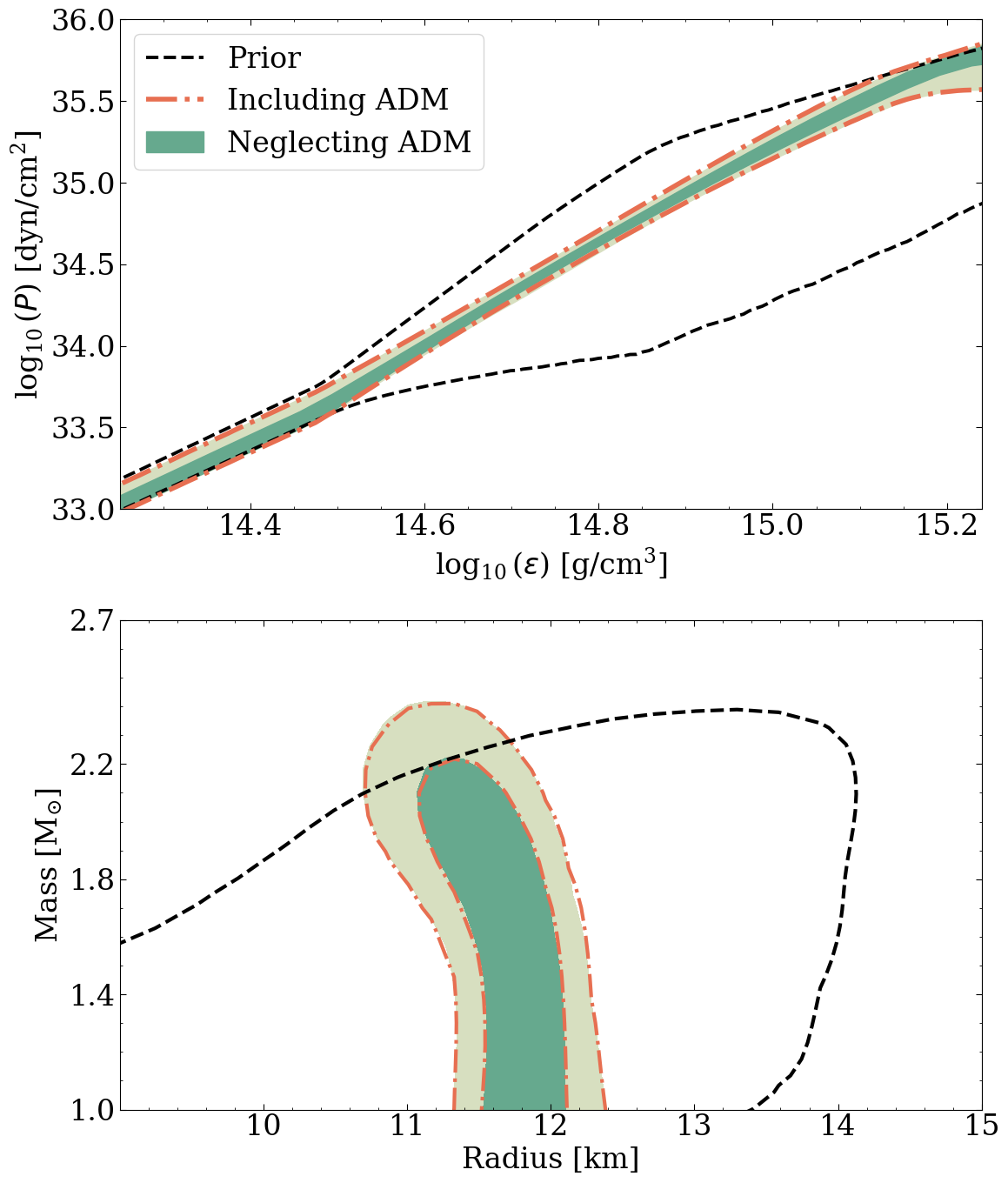}
  \label{fig7:sub1}
\end{subfigure}%
\begin{subfigure}{.5\textwidth}
  \centering
  \includegraphics[width=\linewidth]{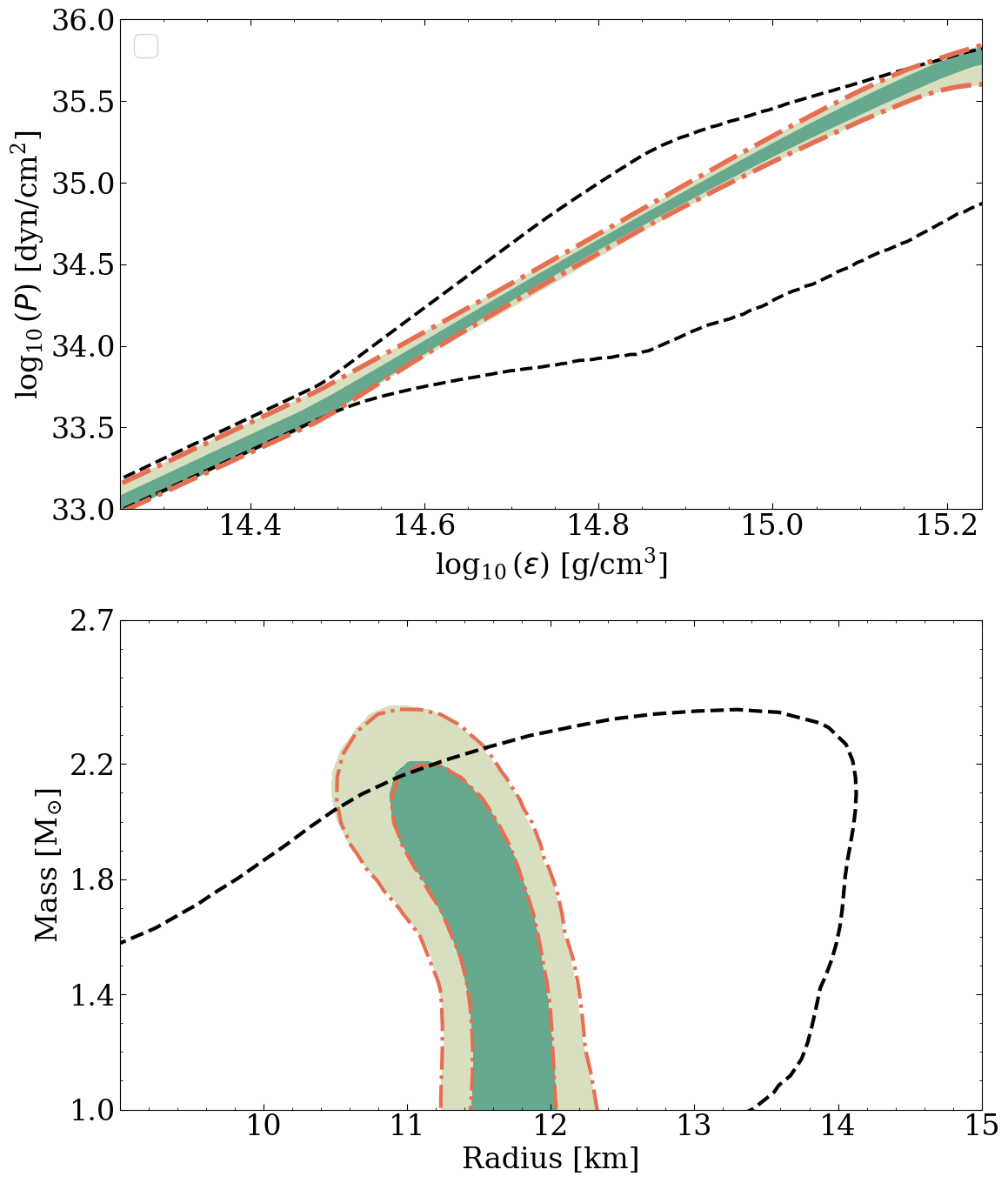}
  \label{fig7:sub2}
\end{subfigure}
\caption{Left two panels: \textit{Future-X} fermionic ADM and baryonic matter EoS posteriors and priors of the ``No ADM" model converted to the pressure-energy density plane (top panel) and mass-radius plane; Right two panels: Same as the left two panels, but for the ``ADM Core'' model. Note, the top two panels following the same legend and contour levels as Fig.~\ref{fig4}, and the bottom two follow the same legend and contour levels as Fig.~\ref{fig6}. Note, the solid orange lines in the bottom two panels are the ``No ADM" and ``ADM Core" model ground truth mass-radius curves, respectively. In all quadrants, we find that the `Including ADM' bands are nearly identical to the `Neglecting ADM' bands.}
\label{fig7}
\end{figure*}
Fig.~\ref{fig4} additionally shows the posterior distributions on the both the fermionic ADM admixed neutron star mass-radius relation (`Including ADM') and the purely baryonic mass-radius relation (`Neglecting ADM'). Along the radial axis, Fig.~\ref{fig4} shows that the `Including ADM' contours predict similar radii to the `Neglecting ADM' contours. Fig.~\ref{fig4} also shows that the `Including ADM' band favor marginally lower maximum masses than the `Neglecting ADM' band. In particular, the 68\% and 95\% confidence regions of the `Including ADM' predict maximum masses of 2.167 $\Msun$ and 2.475 $\Msun$, respectively. While the `Neglecting ADM' band predicts maximum masses of 2.241 $\Msun$ and 2.518 $\Msun$ for the 68\% and 95\% confidence regions, respectively. The `Including ADM' posterior favors lower maximum masses than the `Neglecting ADM' posterior because ADM cores decrease neutron star masses when compared to an identical neutron star with the same baryonic central energy density. This reduction in mass from the presence of ADM cores would push the posteriors to predict lower maximum masses than inferences done with only baryonic matter. Since the `Including ADM' band only marginally favors lower maximum masses and strongly overlaps with the `Neglecting ADM' band, we find that the inclusion of fermionic ADM cores are equally as consistent with the \textit{NICER} data as the posteriors which only account for baryonic matter.

\subsection{Synthetic data inferences}\label{syntheticdata}
We now consider the synthetic mass-radius measurements of a potential \textit{STROBE-X} scenario using the \textit{Future-X} scenario of \cite{Rutherford2023}. To study such a potential scenario, it is useful to define two ground truth models i.e., models in which the synthetic neutron star mass-radius measurements will be computed from: one with an ADM core and one with only baryonic matter. Considering two ground truth models will allow for statements about the ADM EoS, regardless if ADM cores are actually present in neutron stars. The first ground truth model that we consider is the ``ADM Core" model, which is described by the PP model in Sec.~\ref{EoSs} with an ADM core defined by the ADM parameters
\begin{align}
    m_\chi &= 15 \, \mathrm{GeV}\\
    \frac{g_\chi}{m_\phi/\mathrm{MeV}} &= 0.01\\
    F_\chi &= 1.5\%.
\end{align}
The second ground truth model is defined identically to the ``ADM core" model, but $F_\chi = 0\%$ in order to account for the possibility that neutron stars do not accumulate an appreciable total ADM mass, but the possibility of ADM is still considered during sampling. Using the \textit{Future-X} scenario with the ``No ADM" and ``ADM core" models in this way will allow for the best-case future constraints on fermionic ADM to be determined \cite[see][]{Rutherford2023}. In Fig.~\ref{fig_ellipses}, the uncertainty ellipses and ground truth models for the \textit{Future-X} scenario are shown.

In Fig.~\ref{fig6}, we show the fermionic ADM prior and posterior distributions of the ``No ADM" (left) and ``ADM Core'' (right) models for the \textit{Future-X} scenario. In the top two panels, we show the corner plots of the ``No ADM" and ``ADM Core'' models. In both the 2-D density and 1-D histograms plots, the posteriors of both ground truth models are identical to each other, despite having different ground truth ADM mass-fractions. In addition, the posterior distributions of the ``ADM Core'' and ``No ADM'' models are approximately identical to the prior distribution. Since the corner plots of both the ``No ADM" and ``ADM Core'' models are nearly identical to each other as well as the prior, the \textit{Future-X} scenario will not be able to provide any additional constrains on the fermionic ADM particle mass, effective self-repulsion strength, and mass-fraction than the inferences using the neutron star data from \cite{Riley0740,Riley0030}. 

In the bottom two panels of Fig.~\ref{fig6}, the fermionic ADM posteriors and priors are transformed to the $\log_{10}\left(\frac{g_\chi}{m_\phi/\mathrm{MeV}}/(m_\chi/\mathrm{MeV}\right)$-$F_\chi$ plane. Fig.~\ref{fig6} shows that the ``ADM Core'' model posteriors narrow on the left side more than the ``No ADM'' posteriors for increasing $F_\chi$. The posteriors on the ratio of $g_\chi/m_\phi$ and $m_\chi$ differ along the $F_\chi$ axis because the ground truth mass-fractions are 1.5\% and 0\% for the ``ADM Core" and ``No ADM" models, respectively. That is, a given ratio of $g_\chi/m_\phi$ and $m_\chi$ could produce mass-radius curves satisfying the ``No ADM" model data for $F_\chi$ near 0\%, but simultaneously not produce neutron stars satisfying the ``ADM Core" model data for $F_\chi \approx 1.5\%$. Since the posteriors on the ratio of the fermionic ADM self-repulsion and particle mass differ between both ground truth models and the priors, we find that \textit{Future-X} will be able to constrain the lower bound on the ratio of $g_\chi/m_\phi$ and $m_\chi$. In particular, the lower bound on $\log_{10}\left(\frac{g_\chi}{m_\phi/\mathrm{MeV}}/(m_\chi/\mathrm{MeV}\right)$ is constrained to be $\gtrsim -6.49$ and $\gtrsim -7.68$, at the 68\% and 95\% confidence intervals, respectively. However, when compared to the real data posteriors of Fig.~\ref{fig_2_3}, the \textit{Future-X} scenario can only slightly tighten the constraints on $\log_{10}\left(\frac{g_\chi}{m_\phi/\mathrm{MeV}}/(m_\chi/\mathrm{MeV}\right)$ at both the 68\% and 95\% confidence intervals.

Fig.~\ref{fig7} shows the ``No ADM" and ``ADM Core'' model posteriors on the baryonic EoS uncertainty in the pressure-energy density plane (top two panels) and the combined fermionic ADM and baryonic matter EoS in the mass-radius plane (bottom two panels). In the pressure-energy density plane, the 95\% confidence region of the `Including ADM' predicts baryonic EoS uncertainties that are comparable to those of the `Neglecting ADM' band for both the ``No ADM" and ``ADM Core'' models. In the mass-radius plane, the 68\% and 95\% confidence intervals on the `Including ADM' band also do not significantly deviate from their corresponding `Neglecting ADM' bands. The strong overlap between the `Neglecting ADM' and `Including ADM' posteriors is due to an apparent degeneracy between the fermionic ADM and baryonic matter EoSs. Moreover, a strong degeneracy between fermionic ADM and baryonic matter implies that a stiff baryonic EoS with an ADM core can yield an identical mass-radius relation to a relatively softer baryonic EoS without an ADM core, which would result in both neutron star models receiving the same likelihood evaluations. Based on the observations that the `Including ADM' bands favor nearly identical posteriors to the `Neglecting ADM' bands, this figure shows that neutron stars with fermionic ADM cores are indistinguishable from purely baryonic neutron stars. Thus, we conclude that the presence of fermionic ADM cores inside neutron stars are fully consistent with purely baryonic neutron stars down to the 2\% mass-radius uncertainty level.

\section{Summary and conclusions}\label{conclusion}
In this work, we have presented a full Bayesian analysis for fermionic ADM cores in neutron stars using the framework developed in \cite{Rutherford2023}. Here we have modeled the fermionic ADM cores using the \cite{Nelson_2018} ADM model, which describes ADM as spin$-$$1/2$ fermions with repulsive self-interactions. We have considered the mass-radius data of PSR J0740$+$6620 \cite{Riley0740} and PSR J0030$+$0451 \cite{Riley0030} as well as synthetic mass-radius data from a best case scenario of the NASA \textit{STROBE-X} mission. By considering both real and synthetic mass-radius measurements, we inferred the current and possible future constraints on the fermionic ADM particle mass $m_\chi$, effective self-repulsion strength $g_\chi/m_\phi$, and mass-fraction $F_\chi$.

For the inferences which consider the PSR J0740$+$6620 and PSR J0030$+$0451 mass-radius measurements, we find that the 2-D posterior densities of $\log_{10}(m_\chi/\mathrm{MeV}) \, \mathrm{vs.} \, \log_{10}(g_\chi/(m_\phi/\mathrm{MeV}))$, $F_\chi \, \mathrm{vs.} \, \log_{10}(m_\chi/\mathrm{MeV})$, and $F_\chi \, \mathrm{vs.} \log_{10}(g_\chi/(m_\phi/\mathrm{MeV}))$ are nearly identical to their respective prior densities. In addition, the 1-D posterior histograms of each fermionic ADM EoS parameter also strongly coincide with their prior counterparts. The fermionic ADM EoS posteriors are nearly identical to their corresponding priors because the EoS parameters are degenerate with one another such that one set of ($m_\chi$, $g_\chi/m_\phi$, $F_\chi$) produces similar neutron stars as another set of ($m_\chi^{'}$, $g_\chi^{'}/m_\phi$, $F_\chi^{'}$). Thus, we conclude that the fermionic ADM EoS parameters cannot be constrained using the \textit{NICER} mass-radius measurements of PSR J0740$+$6620 and PSR J0030$+$0451. If the ADM posteriors are transformed to the $\log_{10}\left(\frac{g_\chi}{m_\phi/\mathrm{MeV}}/(m_\chi/\mathrm{MeV}\right)$-$F_\chi$ plane, the lower bound on the ratio of the fermionic ADM effective self-repulsion strength to the particle mass can be constrained to $-6.59$ and $-7.77$ at the 68\% and 95\% confidence levels, respectively. These results show that, within the current uncertainties of neutron star mass-radius measurements delivered by \textit{NICER}, the lower bound of the ratio of $g_\chi/m_\phi$ and $m_\chi$ can only be marginally constrained. However, all other combinations of fermionic ADM parameters cannot be constrained.

Converting the fermionic ADM and baryonic matter EoS posteriors to the mass-radius and pressure-energy density planes, we find that the posteriors on the neutron star EoS are largely unaffected by the inclusion of fermionic ADM cores. In the mass-radius plane, we find that the maximum masses of the posteriors which include fermionic ADM differ from the purely baryonic ones at the 95\% percent level by 0.0043 $\Msun$. Moreover, the combined fermionic ADM and baryonic mass-radius posteriors predict similar radii to the purely baryonic posteriors. In the pressure-energy density plane, the baryonic EoS uncertainty slightly broadens when fermionic ADM is accounted for. In particular, at $\log_{10}(\varepsilon \, \mathrm{cm}^3/\mathrm{g}) =$ 14.381 and , the baryonic EoS uncertain widens by 1.021\%. The small differences between the posteriors that include fermionic ADM cores and the ones that do not, show that fermionic ADM cores inside neutron star interiors can be fully consistent with their purely baryonic counterparts.

In order to determine the promise of constraining fermionic ADM cores by missions, like the NASA \textit{STROBE-X} mission, this work has also considered the \textit{Future-X} scenario from \cite{Rutherford2023}. The \textit{Future-X} scenario describes six synthetic neutron star mass-radius measurements with mass and radius uncertainties at the 2\% level. Within the \textit{Future-X} scenario, the fermionic ADM posteriors remain nearly identical to the real data inferences for both the ``ADM core'' and ``No ADM'' models. However, we find that the posteriors on the ratio of $g_\chi/m_\phi$ and $m_\chi$ differ between the ``No ADM" and ``ADM core" models. In particular, the ``ADM core" model infers marginally tighter constraints on the lower bound of $\log_{10}\left(\frac{g_\chi}{m_\phi/\mathrm{MeV}}/(m_\chi/\mathrm{MeV}\right)$ than the ``No ADM" model. The posteriors on the lower bound of $\log_{10}\left(\frac{g_\chi}{m_\phi/\mathrm{MeV}}/(m_\chi/\mathrm{MeV}\right)$ are slightly more narrow in the ``ADM core" model than in the ``No ADM" model because the ground truth mass-fraction of the ``ADM core" model is higher than that of the ``No ADM" model. This allows for ratios of $g_\chi/m_\phi$ and $m_\chi$, which produce neutron stars satisfying the ``No ADM'' data for a given $F_\chi$, to be given a non-zero likelihood. However, the same ratios of $g_\chi/m_\phi$ and $m_\chi$ would be given a zero likelihood because they would not satisfy the data of the ``ADM core'' model for the same given $F_\chi$. Since the posteriors in $\log_{10}\left(\frac{g_\chi}{m_\phi/\mathrm{MeV}}/(m_\chi/\mathrm{MeV}\right)$-$F_\chi$ plane differ between the ``No ADM" and ``ADM core" models, we found that \textit{Future-X} will be able to constrain the lower bound the ratio of $g_\chi/m_\phi$ and $m_\chi$. 

According to the posteriors on the lower bound of the ratio of $g_\chi/m_\phi$ and $m_\chi$, we find that \textit{Future-X} slightly tightens the constraints to $-6.49$ and $-7.68$ at the 68\% and 95\% confidence levels, respectively. It is physically reasonable that the constraints on $\log_{10}\left(\frac{g_\chi}{m_\phi/\mathrm{MeV}}/(m_\chi/\mathrm{MeV}\right)$ improve in the \textit{Future-X} scenario because the mass of an admixed neutron star is sensitive to the compactness of the fermionic ADM core, which is partially controlled by the ratio of $g_\chi/m_\phi$ and $m_\chi$. Therefore, the posteriors on the lower bound of $\log_{10}\left(\frac{g_\chi}{m_\phi/\mathrm{MeV}}/(m_\chi/\mathrm{MeV}\right)$ will slightly improve with the tighter mass and radius uncertainties of the \textit{Future-X} scenario. Note, however, the \textit{Future-X} scenario is a best case scenario for the \textit{STROBE-X} mission and our constraints will relax accordingly for larger mass-radius credible intervals.

In the pressure-energy density and mass-radius plane, we find that the uncertainties on the baryonic matter EoS and the total neutron star mass-radius remain unaffected when the possibility of fermionic ADM cores is considered. That is, similar to the real data inferences, we find that the pressure-energy density and mass-radius of the `Including ADM' bands are identical to their respective `Neglecting ADM' bands in both the ``No ADM" and ``ADM core" models. Our results highlight that neutron star EoS models that additionally allow for fermionic ADM cores are indistinguishable with the baryonic EoS inferences for mass and radius uncertainties down to the $2\%$ level, which implies that ADM admixed neutron stars equally as consistent with neutron star data as purely baryonic neutron stars.

The `Including ADM' posteriors in the pressure-energy density and mass-radius planes of both the real data from \textit{NICER} and the hypothetical data from the \textit{Future-X} scenario (\textit{STROBE-X}) strongly coincide with their corresponding `Neglecting ADM' posteriors is in part because, under the ADM mass-fraction priors considered, the presence of ADM cores do not significantly affect the uncertainties of the baryonic EoS. This is most clearly shown in Fig.~\ref{fig_ellipses}, where the ``ADM core'' and ``NO ADM'' where the ``ADM core'' and ``NO ADM'' mass-radius relations are closely aligned.  The other reason why including fermionic ADM cores are physically consistent with the purely baryonic matter posteriors is because of the strong degeneracy between the ADM and baryonic matter EoSs. That is, the mass-radius relation can equally well described by both a stiff baryonic EoS with an ADM core and a softer baryonic EoS without ADM. Our results show this in the mass-radius posteriors of Figs.~\ref{fig4} and \ref{fig7}, where the purely baryonic posteriors strongly overlap with the fermionic ADM posteriors. Lastly, \cite{Giangrandi2022} pointed out several scenarios that could break this degeneracy, such as a reduction of neutron star masses toward the center of the Galaxy, searching for supplementary peaks in gravitational wave spectra from binary neutron star merger simulations, detecting objects that are in contrast to our understanding of neutron star structure, and by finding a new feature in the binary Love relation. 

Overall, this work shows that the current neutron star measurement of the NASA \textit{NICER} mission, as well as the potential future measurements of the NASA \textit{STROBE-X} mission, can provide constraints on the lower bound of the ratio $g_\chi/m_\phi$ and $m_\chi$, but not the individual quantities of $F_\chi$, $m_\chi$, and $g_\chi/m_\phi$. Moreover, within the uncertainties on the baryonic EoS, neutron stars with ADM cores are equally as consistent with the mass-radius data as stars only made of baryonic matter, which means that neither \textit{NICER} nor \textit{STROBE-X} will be able to distinguish between neutron stars with an ADM core and those without. Although fermionic ADM cores inside neutron stars are indistinguishable from their purely baryonic counterparts to \textit{NICER} and \textit{STROBE-X}, when specific assumptions about the neutron EoS are made, small ADM mass-fractions have been shown to trigger rapid neutron star cooling for low mass neutron stars through the direct Urca process, which could shed light on the presence of ADM in these stars \cite{Avila2024,Giangrandi2024,Scordino2024}. 

In a similar analysis, \cite{Miao_2022} inferred that the ADM particle mass favors masses near 0.6 GeV for ADM cores using a fixed baryonic EoS with the \textit{NICER} PSR J0740$+$6620 and PSR J0030$+$0451 measurements, which is in contrast with the findings of this work. However, when comparing our respective priors on $m_\chi$, we observe that their priors are flat while ours are not. The observed difference between our priors could be due to the fact that our analysis strictly enforces a minimum neutron star mass constraint of 1$\Msun$, while the analysis of \cite{Miao_2022} does not specify one. Imposing a minimum neutron star mass constraint modifies the priors on the ADM parameters, which in-turn can modify the interpretation of the ADM posteriors. Moreover, \cite{Miao_2022} fixes their analysis to one baryonic EoS while our work varies the baryonic EoS, which can also change the ADM EoS posteriors \citep[see e.g.,][]{Rutherford2023}. 

 Another recent work, \cite{Guha2024}, considered ten different choices of baryonic matter EoS and a fermionic ADM model with self-interactions mediated by both a massive scalar and vector mediator. By directly comparing the mass-radius relations of the ten baryonic EoSs for a range of $m_\chi$ and ADM Fermi momenta values with the 95\% contours of the \textit{NICER} and LIGO/VIRGO, measurements, \cite{Guha2024} found that $m_\chi \in [0.1,30]$ GeV for Fermi momenta in the range of [0.01,0.07] GeV are consistent with the observational data. Our results differ from those of \cite{Guha2024} for two key reasons. The first reason is that our analysis fully varies the baryonic EoS and considers the full distribution of the \textit{NICER} PSR J0740$+$6620 and PSR J0030$+$0451 mass-radius posteriors as part of our analysis. \cite{Guha2024}, by contrast, approximates the baryonic EoS uncertainty by considering ten representative choices and uses the \textit{NICER} and LIGO/VIRGO posterior contours as hard-cut offs. Although the \cite{Guha2024} approach has the advantage of determining tight constraints for a handful of fixed baryonic EoSs, our Bayesian analysis is less dependent on the baryonic EoS because it varies both the  ADM and baryonic EoS during inference, allowing the most likely combination of ADM and baryonic matter to be determined. The second is that our analysis directly considers the ADM mass-fraction while \cite{Guha2024} considers the ADM Fermi momentum, which can result in $F_\chi \gtrsim$ 50\% for many combinations of ADM parameters and different $F_\chi$ along the entire admixed mass-radius relation. Therefore, it is physically reasonable that the results of \cite{Guha2024} more tightly constrained $m_\chi$ than this work because, in combination with choosing ten fixed baryonic EoSs, large $F_\chi$ will increasingly significantly affect the resulting mass-radius relation as $m_\chi$ increases, which will preferentially favor lower $m_\chi$. While, our analysis allows for relatively smaller, and physically grounded, values of $F_\chi$, which, of course, affect the resulting mass-radius relation less.

Finally, future work will explore how the proper inclusion of fermionic (as well as bosonic) ADM halos affects our inferences on both the ADM and baryonic matter EoSs. The work of \cite{Shawqi2024} constructed a framework for interpreting neutron star mass-radius measurements in the presence of ADM halos. While many works consider a variety of different plausible ADM mass-fractions \cite[see e.g.,][and references therein]{Shawqi2024,Konstantinou2024,Shakeri2024,Karkevandi2022,Collier2022,Miao_2022, Husain2021,Sen2021,Ellis2018}, an in-depth analysis on the possible accumulation methods of ADM in neutron stars has yet to be done and is left for future work. By appropriately accounting for the possible presence of ADM halos and physically constraining $F_\chi$, full inferences on the neutron star EoS will be able to determine the most general constraints on the ADM EoS.

\section*{Acknowledgments}\label{acknowledge}
We acknowledge Ann Nelson for her pioneering work on dark matter in neutron stars. We thank Michael Lathwood for insightful conversations with N.R. on the consequences of the symmetric nature of the metric tensor. We also thank Nathan Musoke, Yves Kini, Anthony Mirasola, and Melissa Mendes for feedback on the manuscript. Lastly, we thank the anonymous referee for comments that helped improve this work. A.L.W. acknowledges support from ERC Consolidator grant No.865768 AEONS. C.P.W. acknowledges all the administrative and facilities staff at the University of New Hampshire, especially Katie Makem-Boucher and Michelle Mancini. The contributions of C.P.W. and N.R. were supported by NASA grant No.80NSSC22K0092.

\textit{Software}: Python/C language~\cite{python2007}, GNU~Scientific~Library~\cite[GSL;][]{Gough:2009}, NumPy~\cite{Numpy2011}, Cython~\cite{cython2011}, SciPy~\cite{Scipy}, MPI for Python~\cite{mpi4py}, Matplotlib~\cite{Hunter:2007}, Jupyter~\cite{Kluyver:2016aa}, MultiNest~\cite{Feroz13}, \textsc{PyMultiNest}~\cite{Buchner14}, kalepy~\cite{Kelley2021}, corner~\cite{corner}, seaborn~\cite{Waskom2021}, NEoST~\cite{Raaijmakers24}.
\appendix

\section{ Approximating $g_\chi/m_\phi = 0 \, \mathrm{MeV^{-1}}$}\label{AppendixB}
In order to capture the physically allowed parameter space of the effective fermionic self-repulsion strength, $g_\chi/m_\phi = 0\, \mathrm{MeV^{-1}}$ must be considered. However, since we have assumed  $g_B \ll g_\chi$ (see Sec.~\ref{fermionic priors}), a non-zero approximation of $g_\chi/m_\phi = 0\, \mathrm{MeV^{-1}}$ is necessary. Here we approximate zero self-repulsion strength by calculating the average relative radial percent difference (RRPD) between the mass-radius curves of zero self-repulsion and a non-zero self-repulsion strength, which we take to be $10^{-5}\, \mathrm{MeV^{-1}}$, for all neutron star masses $\geq 1 \, \Msun$. We define the RRPD at a fixed neutron star mass as 
\begin{align}\label{RRPD}
    \mathrm{RRPD} = \frac{\abs{R_{-5}-R_0}}{R_0}\cdot 100,
\end{align}
where $R_{-5}$ is the radius of the neutron star produced by $g_\chi/m_\phi = 10^{-5} \, \mathrm{MeV^{-1}}$ and $R_0$ is the radius of the neutron star produced by zero self-repulsion strength.

 \begin{figure*}
\centering
\begin{subfigure}{.5\textwidth}
  \centering
  \includegraphics[width=\textwidth]{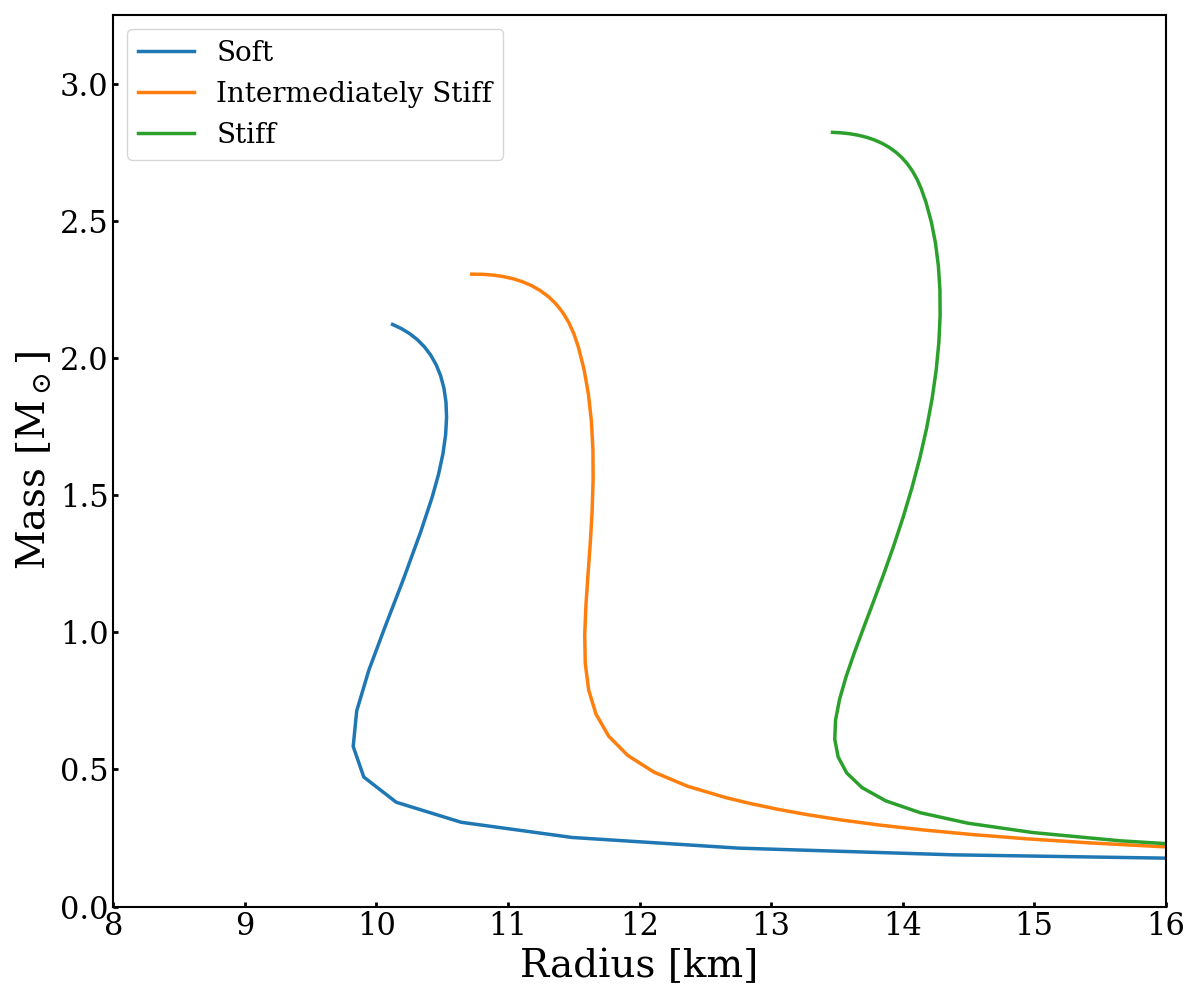}
  \label{app-fig1:sub1}
\end{subfigure}%
\begin{subfigure}{.5\textwidth}
  \centering
  \includegraphics[scale = 0.3325]{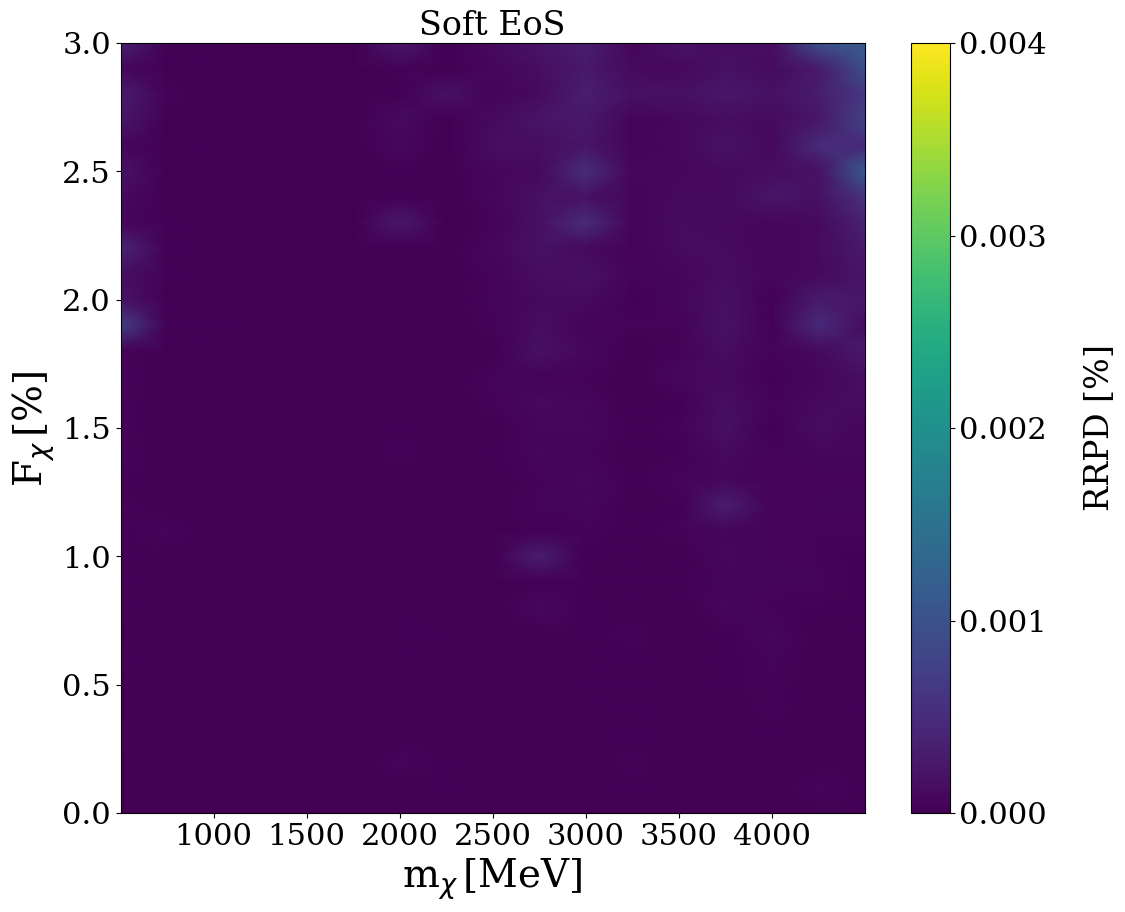}
  \label{app-fig1:sub2}
\end{subfigure}\\
\begin{subfigure}{.5\textwidth}
  \centering
  \includegraphics[scale = 0.3325]{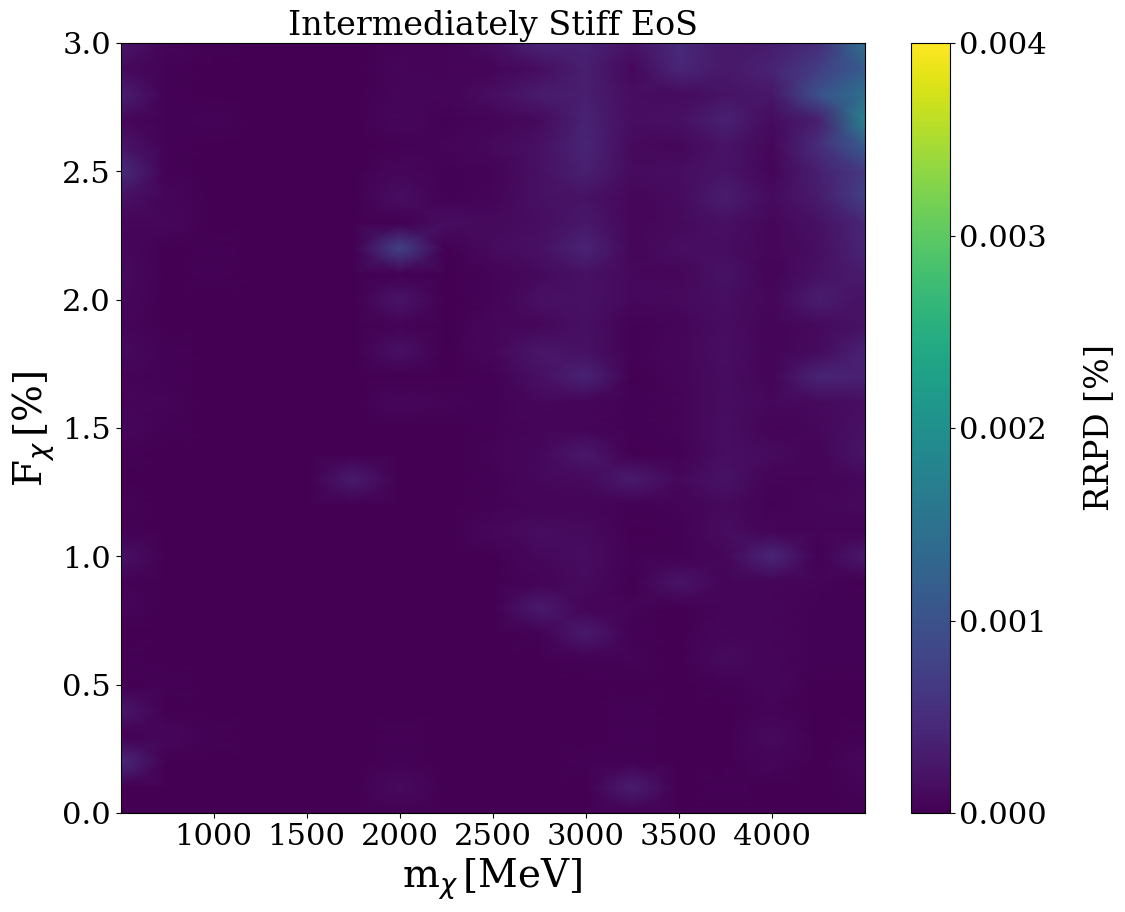}
  \label{app-fig1:sub3}
\end{subfigure}%
\begin{subfigure}{.5\textwidth}
  \centering
  \includegraphics[scale = 0.3325]{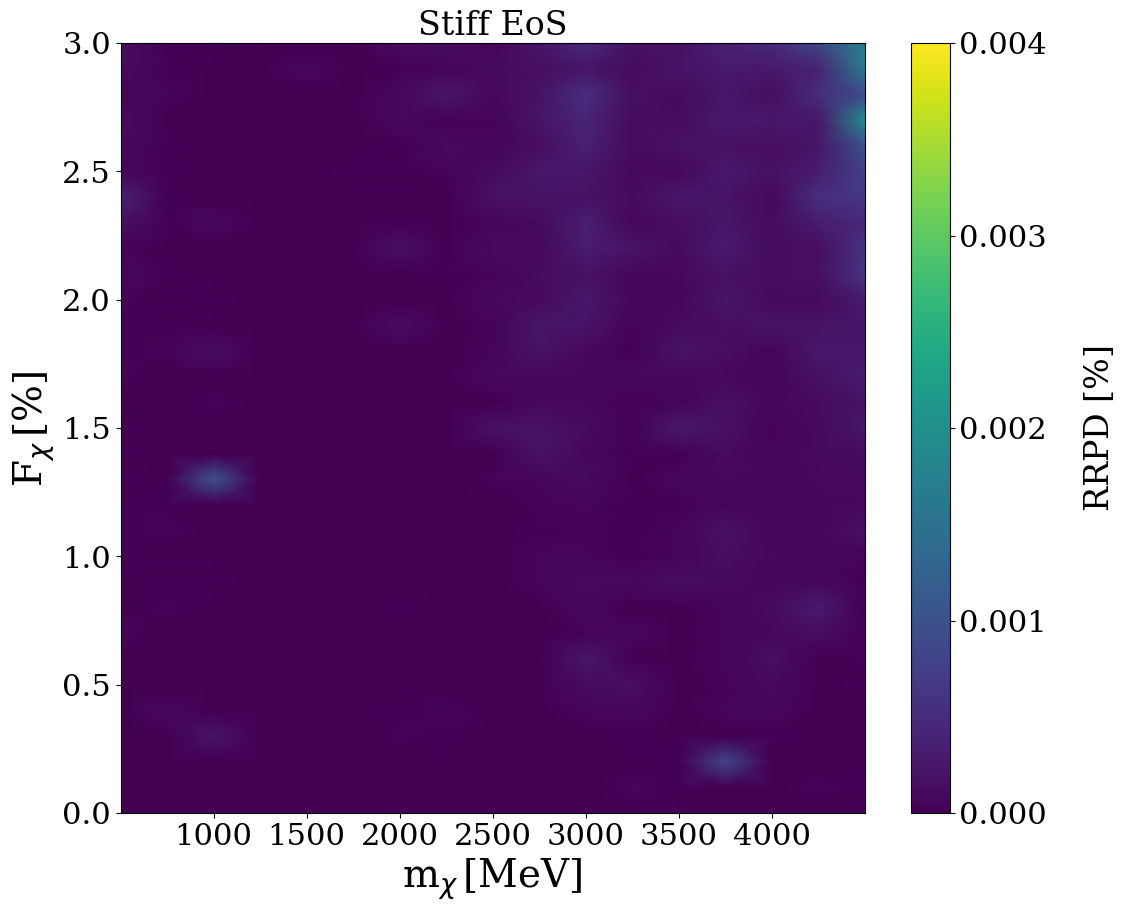}
  \label{app-fig1:sub4}
\end{subfigure}
\caption{Top Left: The three underlying baryonic mass-radius curves of varying stiffness from soft (blue) to intermediately stiff (orange) to stiff (green) used in the remaining three panels, respectively; Top Right: Color plot of the RRPD between $g_\chi/m_\phi = 0  \,\mathrm{MeV^{-1}}$ and $g_\chi/m_\phi = 10^{-5} \, \mathrm{MeV^{-1}}$ in the ADM mass-fraction and particle mass plane for the Soft baryonic EoS; Bottom Left: Same as the top right panel, but for the intermediately stiff baryonic EoS; Bottom right: Same as the top right panel, but for the stiff baryonic EoS. For the top right panel and the bottom two panels, the RRPD values do not exceed 0.004\%. }
\label{app-fig1}
\end{figure*}

 \begin{figure*}
\centering
\begin{subfigure}{.5\textwidth}
  \centering
  \includegraphics[scale = 0.3325]{Percent_diff_intermediate_stiff_plot.png}
  \label{app-fig2:sub1}
\end{subfigure}%
\begin{subfigure}{.5\textwidth}
  \centering
  \includegraphics[scale = 0.3325]{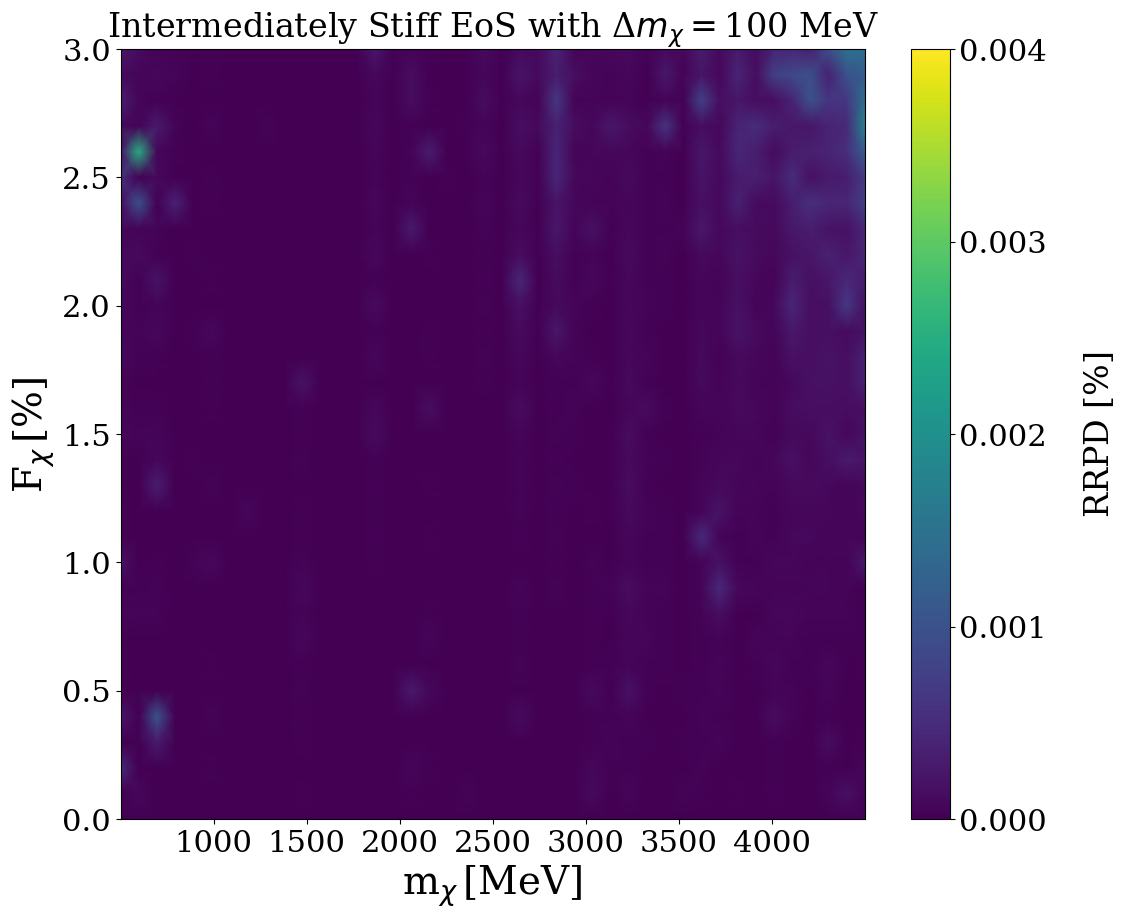}
  \label{app-fig2:sub2}
\end{subfigure}\\
\begin{subfigure}{.5\textwidth}
  \centering
  \includegraphics[scale = 0.3325]{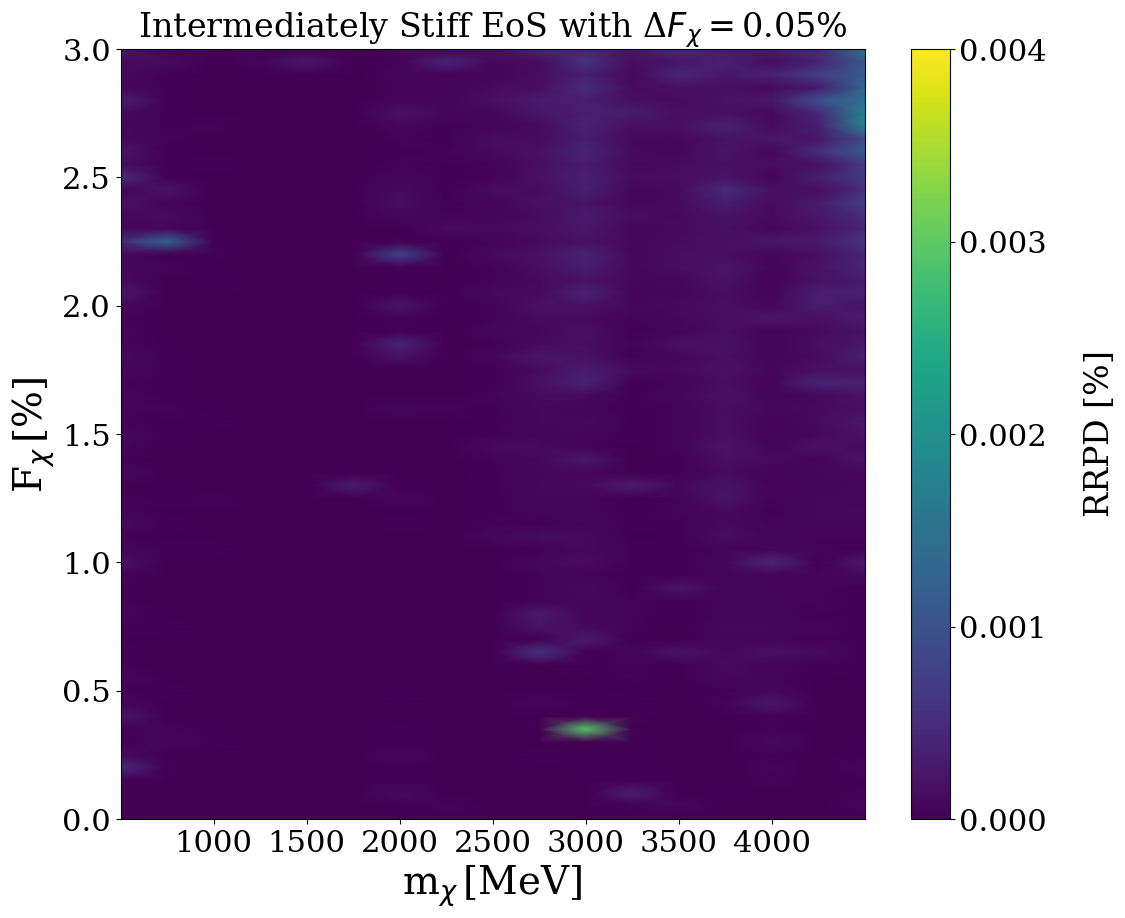}
  \label{app-fig2:sub3}
\end{subfigure}
\caption{Top Left: RRPD color plot in the ADM mass-fraction and particle mass plane for the intermediately stiff EoS; Top Right: Same as the top left panel, but the step size along the $m_\chi$ axis is changed from 250 MeV to 100 MeV; Bottom: Same as the top left panel, but the step size along the $F_\chi$ axis is changed from 0.1\% to 0.05\%. For all three panels, the maximum RRPD value does not exceed 0.004\%.}
\label{app-fig2}
\end{figure*}
To calculate the average RRPD values between the mass-radius relations with zero self-repulsion and $g_\chi/m_\phi = 10^{-5}\, \mathrm{MeV^{-1}}$ in a given interval of $m_\chi$ and $F_\chi$ for a fixed baryonic EoS, we adopt the following procedure. First, we compute the entire mass-radius relation for both $g_\chi/m_\phi = 10^{-5} \, \mathrm{MeV^{-1}}$ and $g_\chi/m_\phi = 0 \, \mathrm{MeV^{-1}}$ for fixed $m_\chi$ and $F_\chi$. Second, we linearly interpolate both mass-radius relations to obtain $R(M)$, i.e, neutron star radius as a function of gravitational mass. With $R(M)$ in-hand, a direct comparison of identical masses between both mass-radius relations can made. Third, we draw 20 evenly spaced masses from 1 $\Msun$ to the maximum mass of the two mass-radius curves and compute the RRPD for all 20 masses. Fourth, we average over the RRPD values of all masses $> 1 \Msun$ and save the average value. Finally, steps 1-4 are repeated until the averaged RRPD value of each combination of $m_\chi$ and $F_\chi$ is obtained.

Using the above procedure to calculate the average RRPD values between $g_\chi/m_\phi = 0\, \mathrm{MeV^{-1}}$ and $g_\chi/m_\phi = 10^{-5}\, \mathrm{MeV^{-1}}$, we compute the average RRPD for ADM particle masses within $m_\chi \in [400,4500]$ MeV and ADM mass-fractions within $F_\chi \in [0,3]$ \%. The interval on $F_\chi$ was chosen such that it extends through and beyond the ADM mass-fraction prior space defined in Sec.~\ref{fermionic priors}. The lower bound on the ADM particle mass interval was determined such that no ADM halo configurations were produced for neutron stars with mass $> 1 \Msun$ with $F_\chi = 3 \%$ and $g_\chi/m_\phi = 0\, \mathrm{MeV^{-1}}$, which ensures only ADM cores will be accounted for all $F_\chi \in [0,3]\%$. Moreover, the upper bound of $m_\chi = $ 4500 MeV was calculated by determining the largest ADM particle mass such that the maximum mass was at least $1 \Msun$ for $F_\chi = 3\%$ and $g_\chi/m_\phi = 0\, \mathrm{MeV^{-1}}$. Adopting the upper bound of the ADM particle mass interval to be $m_\chi = 4500$ MeV captures the physically relevant ADM core configurations because our Bayesian analysis framework assigns all neutron stars with masses $< 1 \Msun$ a zero likelihood evaluation.  

In Fig.~\ref{app-fig1} we show the RRPD distribution between the mass-radius relations of zero self-repulsion and $g_\chi/m_\phi = 10^{-5}\, \mathrm{MeV^{-1}}$ for three representative baryonic EoSs, each of varying stiffness (top left panel), in the $m_\chi-F_\chi$ plane. For each baryonic EoS, we have spaced each ADM particle mass by $\Delta m_\chi = 250$ MeV and each mass-fraction by $\Delta F_\chi = 0.1\%$ because, when all other EoS parameters are held fixed, both $\Delta m_\chi$ and $\Delta F_\chi$ have a small overall effect on the resulting mass-radius relation between each respective step. Although the RRPD distribution is different between each of the baryonic EoSs using $\Delta m_\chi$ and $\Delta F_\chi$, the maximum RRPD value is $4 \cdot 10^{-3}$\% for the soft, intermediately stiff, and stiff baryonic EoSs. Therefore from this observation, we conclude that $g_\chi/m_\phi = 10^{-5}\, \mathrm{MeV^{-1}}$ is a sufficient approximation for $g_\chi/m_\phi = 0\, \mathrm{MeV^{-1}}$, regardless of the baryonic EoS, because the maximum RRPD value is $4 \cdot 10^{-3}$\% which is several orders of magnitude below the observational uncertainties of neutron star radii considered in this work.

Fig.~\ref{app-fig2} shows the RRPD distribution between the mass-radius relations of $g_\chi/m_\phi = 0\, \mathrm{MeV^{-1}}$ and $g_\chi/m_\phi = 10^{-5}\, \mathrm{MeV^{-1}}$ in which the step size between ADM particle mass points ($\Delta m_\chi$) and ADM mass-fractions ($\Delta F_\chi$) is reduced. Note, we fix the underlying baryonic EoS to be the intermediately stiff EoS from Fig.~\ref{app-fig1}. Reducing the ADM particle mass and mass-fraction step sizes will impact the overall mass-radius relation less between each successive step, thus allowing for better interpolations between grid points in the $m_\chi-F_\chi$ plane. An improved interpolation between ($m_\chi,F_\chi$) grid points will allow for the dependency of the RRPD distribution on the grid spacing to be determined. Here we have reduced $\Delta m_\chi$ from 250 MeV to 100 MeV (top right panel) and $\Delta F_\chi$ from 0.1\% to 0.05\% (bottom left panel). Fig.~\ref{app-fig2}, shows that for both cases in which we set $\Delta m_\chi= 100$ MeV and $\Delta F_\chi= 0.05\%$, the maximum RRPD value remains below $4\cdot 10^{-3}$\%. From this observation, we find that the RRPD values are insensitive to variations in $\Delta m_\chi$ and $\Delta F_\chi$. 

Finally, based on all of the previous observations, we conclude that $g_\chi/m_\phi = 10^{-5}\, \mathrm{MeV^{-1}}$ is an adequate approximation to $0\, \mathrm{MeV^{-1}}$ within the interval of $m_\chi \in [400,4500]$ MeV and $F_\chi \in [0,3]$ \%, regardless of the choice of $\Delta m_\chi$, $\Delta F_\chi$, and baryonic EoS.
\clearpage
\bibliography{bibliography}

\begin{thebibliography}{130}
\providecommand{\natexlab}[1]{#1}
\providecommand{\url}[1]{\texttt{#1}}
\expandafter\ifx\csname urlstyle\endcsname\relax
  \providecommand{\doi}[1]{doi: #1}\else
  \providecommand{\doi}{doi: \begingroup \urlstyle{rm}\Url}\fi

\bibitem[{Hebeler} et~al.(2013){Hebeler}, {Lattimer}, {Pethick}, and {Schwenk}]{Hebeler13}
K.~{Hebeler}, J.~M. {Lattimer}, C.~J. {Pethick}, and A.~{Schwenk}.
\newblock {Equation of State and Neutron Star Properties Constrained by Nuclear Physics and Observation}.
\newblock \emph{\apj}, 773\penalty0 (1):\penalty0 11, August 2013.
\newblock \doi{10.1088/0004-637X/773/1/11}.

\bibitem[Oertel et~al.(2017)Oertel, Hempel, Kl\"ahn, and Typel]{Oertel_2017}
M.~Oertel, M.~Hempel, T.~Kl\"ahn, and S.~Typel.
\newblock Equations of state for supernovae and compact stars.
\newblock \emph{Rev. Mod. Phys.}, 89:\penalty0 015007, Mar 2017.
\newblock \doi{10.1103/RevModPhys.89.015007}.

\bibitem[{Caplan} et~al.(2018){Caplan}, {Schneider}, and {Horowitz}]{Caplan2018}
M.~E. {Caplan}, A.~S. {Schneider}, and C.~J. {Horowitz}.
\newblock {Elasticity of Nuclear Pasta}.
\newblock \emph{\prl}, 121\penalty0 (13):\penalty0 132701, September 2018.
\newblock \doi{10.1103/PhysRevLett.121.132701}.

\bibitem[{Tolos} and {Fabbietti}(2020)]{Tolos2020}
L.~{Tolos} and L.~{Fabbietti}.
\newblock {Strangeness in nuclei and neutron stars}.
\newblock \emph{Progress in Particle and Nuclear Physics}, 112:\penalty0 103770, May 2020.
\newblock \doi{10.1016/j.ppnp.2020.103770}.

\bibitem[Burgio et~al.(2021)Burgio, Schulze, Vidaña, and Wei]{Burgio_2021}
G.F. Burgio, H.-J. Schulze, I.~Vidaña, and J.-B. Wei.
\newblock Neutron stars and the nuclear equation of state.
\newblock \emph{Progress in Particle and Nuclear Physics}, 120:\penalty0 103879, 2021.
\newblock ISSN 0146-6410.
\newblock \doi{https://doi.org/10.1016/j.ppnp.2021.103879}.

\bibitem[{Han} et~al.(2023){Han}, {Huang}, {Tang}, and {Fan}]{Han2023}
Ming-Zhe {Han}, Yong-Jia {Huang}, Shao-Peng {Tang}, and Yi-Zhong {Fan}.
\newblock {Plausible presence of new state in neutron stars with masses above 0.98MTOV}.
\newblock \emph{Science Bulletin}, 68\penalty0 (9):\penalty0 913--919, May 2023.
\newblock \doi{10.1016/j.scib.2023.04.007}.

\bibitem[{Keller} et~al.(2024){Keller}, {Hebeler}, {Pethick}, and {Schwenk}]{Keller2024}
J.~{Keller}, K.~{Hebeler}, C.~J. {Pethick}, and A.~{Schwenk}.
\newblock {Neutron Star Matter as a Dilute Solution of Protons in Neutrons}.
\newblock \emph{\prl}, 132\penalty0 (23):\penalty0 232701, June 2024.
\newblock \doi{10.1103/PhysRevLett.132.232701}.

\bibitem[{Lindblom}(1992)]{Lindblom_1992}
Lee {Lindblom}.
\newblock {Determining the Nuclear Equation of State from Neutron-Star Masses and Radii}.
\newblock \emph{\apj}, 398:\penalty0 569, October 1992.
\newblock \doi{10.1086/171882}.

\bibitem[{Watts}(2019)]{Watts2019}
Anna~L. {Watts}.
\newblock {Constraining the neutron star equation of state using pulse profile modeling}.
\newblock In \emph{Xiamen-CUSTIPEN Workshop on the Equation of State of Dense Neutron-Rich Matter in the Era of Gravitational Wave Astronomy}, volume 2127 of \emph{American Institute of Physics Conference Series}, page 020008, July 2019.
\newblock \doi{10.1063/1.5117798}.

\bibitem[{Bogdanov} et~al.(2019){Bogdanov}, {Lamb}, {Mahmoodifar}, {Miller}, {Morsink}, {Riley}, {Strohmayer}, {Tung}, {Watts}, {Dittmann}, {Chakrabarty}, {Guillot}, {Arzoumanian}, and {Gendreau}]{Bogdanov2019}
Slavko {Bogdanov}, Frederick~K. {Lamb}, Simin {Mahmoodifar}, M.~Coleman {Miller}, Sharon~M. {Morsink}, Thomas~E. {Riley}, Tod~E. {Strohmayer}, Albert~K. {Tung}, Anna~L. {Watts}, Alexander~J. {Dittmann}, Deepto {Chakrabarty}, Sebastien {Guillot}, Zaven {Arzoumanian}, and Keith~C. {Gendreau}.
\newblock {Constraining the Neutron Star Mass-Radius Relation and Dense Matter Equation of State with NICER. II. Emission from Hot Spots on a Rapidly Rotating Neutron Star}.
\newblock \emph{\apjl}, 887\penalty0 (1):\penalty0 L26, December 2019.
\newblock \doi{10.3847/2041-8213/ab5968}.

\bibitem[{Bogdanov} et~al.(2021){Bogdanov}, {Dittmann}, {Ho}, {Lamb}, {Mahmoodifar}, {Miller}, {Morsink}, {Riley}, {Strohmayer}, {Watts}, {Choudhury}, {Guillot}, {Harding}, {Ray}, {Wadiasingh}, {Wolff}, {Markwardt}, {Arzoumanian}, and {Gendreau}]{Bogdanov2021}
Slavko {Bogdanov}, Alexander~J. {Dittmann}, Wynn C.~G. {Ho}, Frederick~K. {Lamb}, Simin {Mahmoodifar}, M.~Coleman {Miller}, Sharon~M. {Morsink}, Thomas~E. {Riley}, Tod~E. {Strohmayer}, Anna~L. {Watts}, Devarshi {Choudhury}, Sebastien {Guillot}, Alice~K. {Harding}, Paul~S. {Ray}, Zorawar {Wadiasingh}, Michael~T. {Wolff}, Craig~B. {Markwardt}, Zaven {Arzoumanian}, and Keith~C. {Gendreau}.
\newblock {Constraining the Neutron Star Mass-Radius Relation and Dense Matter Equation of State with NICER. III. Model Description and Verification of Parameter Estimation Codes}.
\newblock \emph{\apjl}, 914\penalty0 (1):\penalty0 L15, June 2021.
\newblock \doi{10.3847/2041-8213/abfb79}.

\bibitem[{Gendreau} et~al.(2016){Gendreau}, {Arzoumanian}, {Adkins}, {Albert}, {Anders}, {Aylward}, {Baker}, {Balsamo}, {Bamford}, {Benegalrao}, {Berry}, {Bhalwani}, {Black}, {Blaurock}, {Bronke}, {Brown}, {Budinoff}, {Cantwell}, {Cazeau}, {Chen}, {Clement}, {Colangelo}, {Coleman}, {Coopersmith}, {Dehaven}, {Doty}, {Egan}, {Enoto}, {Fan}, {Ferro}, {Foster}, {Galassi}, {Gallo}, {Green}, {Grosh}, {Ha}, {Hasouneh}, {Heefner}, {Hestnes}, {Hoge}, {Jacobs}, {J{\o}rgensen}, {Kaiser}, {Kellogg}, {Kenyon}, {Koenecke}, {Kozon}, {LaMarr}, {Lambertson}, {Larson}, {Lentine}, {Lewis}, {Lilly}, {Liu}, {Malonis}, {Manthripragada}, {Markwardt}, {Matonak}, {Mcginnis}, {Miller}, {Mitchell}, {Mitchell}, {Mohammed}, {Monroe}, {Montt de Garcia}, {Mul{\'e}}, {Nagao}, {Ngo}, {Norris}, {Norwood}, {Novotka}, {Okajima}, {Olsen}, {Onyeachu}, {Orosco}, {Peterson}, {Pevear}, {Pham}, {Pollard}, {Pope}, {Powers}, {Powers}, {Price}, {Prigozhin}, {Ramirez}, {Reid}, {Remillard}, {Rogstad}, {Rosecrans}, {Rowe}, {Sager}, {Sanders}, {Savadkin},
  {Saylor}, {Schaeffer}, {Schweiss}, {Semper}, {Serlemitsos}, {Shackelford}, {Soong}, {Struebel}, {Vezie}, {Villasenor}, {Winternitz}, {Wofford}, {Wright}, {Yang}, and {Yu}]{Gendreau2016}
Keith~C. {Gendreau}, Zaven {Arzoumanian}, Phillip~W. {Adkins}, Cheryl~L. {Albert}, John~F. {Anders}, Andrew~T. {Aylward}, Charles~L. {Baker}, Erin~R. {Balsamo}, William~A. {Bamford}, Suyog~S. {Benegalrao}, Daniel~L. {Berry}, Shiraz {Bhalwani}, J.~Kevin {Black}, Carl {Blaurock}, Ginger~M. {Bronke}, Gary~L. {Brown}, Jason~G. {Budinoff}, Jeffrey~D. {Cantwell}, Thoniel {Cazeau}, Philip~T. {Chen}, Thomas~G. {Clement}, Andrew~T. {Colangelo}, Jerry~S. {Coleman}, Jonathan~D. {Coopersmith}, William~E. {Dehaven}, John~P. {Doty}, Mark~D. {Egan}, Teruaki {Enoto}, Terry~W. {Fan}, Deneen~M. {Ferro}, Richard {Foster}, Nicholas~M. {Galassi}, Luis~D. {Gallo}, Chris~M. {Green}, Dave {Grosh}, Kong~Q. {Ha}, Monther~A. {Hasouneh}, Kristofer~B. {Heefner}, Phyllis {Hestnes}, Lisa~J. {Hoge}, Tawanda~M. {Jacobs}, John~L. {J{\o}rgensen}, Michael~A. {Kaiser}, James~W. {Kellogg}, Steven~J. {Kenyon}, Richard~G. {Koenecke}, Robert~P. {Kozon}, Beverly {LaMarr}, Mike~D. {Lambertson}, Anne~M. {Larson}, Steven {Lentine}, Jesse~H. {Lewis},
  Michael~G. {Lilly}, Kuochia~Alice {Liu}, Andrew {Malonis}, Sridhar~S. {Manthripragada}, Craig~B. {Markwardt}, Bryan~D. {Matonak}, Isaac~E. {Mcginnis}, Roger~L. {Miller}, Alissa~L. {Mitchell}, Jason~W. {Mitchell}, Jelila~S. {Mohammed}, Charles~A. {Monroe}, Kristina~M. {Montt de Garcia}, Peter~D. {Mul{\'e}}, Louis~T. {Nagao}, Son~N. {Ngo}, Eric~D. {Norris}, Dwight~A. {Norwood}, Joseph {Novotka}, Takashi {Okajima}, Lawrence~G. {Olsen}, Chimaobi~O. {Onyeachu}, Henry~Y. {Orosco}, Jacqualine~R. {Peterson}, Kristina~N. {Pevear}, Karen~K. {Pham}, Sue~E. {Pollard}, John~S. {Pope}, Daniel~F. {Powers}, Charles~E. {Powers}, Samuel~R. {Price}, Gregory~Y. {Prigozhin}, Julian~B. {Ramirez}, Winston~J. {Reid}, Ronald~A. {Remillard}, Eric~M. {Rogstad}, Glenn~P. {Rosecrans}, John~N. {Rowe}, Jennifer~A. {Sager}, Claude~A. {Sanders}, Bruce {Savadkin}, Maxine~R. {Saylor}, Alexander~F. {Schaeffer}, Nancy~S. {Schweiss}, Sean~R. {Semper}, Peter~J. {Serlemitsos}, Larry~V. {Shackelford}, Yang {Soong}, Jonathan {Struebel}, Michael~L.
  {Vezie}, Joel~S. {Villasenor}, Luke~B. {Winternitz}, George~I. {Wofford}, Michael~R. {Wright}, Mike~Y. {Yang}, and Wayne~H. {Yu}.
\newblock {The Neutron star Interior Composition Explorer (NICER): design and development}.
\newblock In Jan-Willem~A. {den Herder}, Tadayuki {Takahashi}, and Marshall {Bautz}, editors, \emph{Space Telescopes and Instrumentation 2016: Ultraviolet to Gamma Ray}, volume 9905 of \emph{Society of Photo-Optical Instrumentation Engineers (SPIE) Conference Series}, page 99051H, July 2016.
\newblock \doi{10.1117/12.2231304}.

\bibitem[{Fonseca} et~al.(2021){Fonseca}, {Cromartie}, {Pennucci}, {Ray}, {Kirichenko}, {Ransom}, {Demorest}, {Stairs}, {Arzoumanian}, {Guillemot}, {Parthasarathy}, {Kerr}, {Cognard}, {Baker}, {Blumer}, {Brook}, {DeCesar}, {Dolch}, {Dong}, {Ferrara}, {Fiore}, {Garver-Daniels}, {Good}, {Jennings}, {Jones}, {Kaspi}, {Lam}, {Lorimer}, {Luo}, {McEwen}, {McKee}, {McLaughlin}, {McMann}, {Meyers}, {Naidu}, {Ng}, {Nice}, {Pol}, {Radovan}, {Shapiro-Albert}, {Tan}, {Tendulkar}, {Swiggum}, {Wahl}, and {Zhu}]{Fonseca_2021}
E.~{Fonseca}, H.~T. {Cromartie}, T.~T. {Pennucci}, P.~S. {Ray}, A.~Yu. {Kirichenko}, S.~M. {Ransom}, P.~B. {Demorest}, I.~H. {Stairs}, Z.~{Arzoumanian}, L.~{Guillemot}, A.~{Parthasarathy}, M.~{Kerr}, I.~{Cognard}, P.~T. {Baker}, H.~{Blumer}, P.~R. {Brook}, M.~{DeCesar}, T.~{Dolch}, F.~A. {Dong}, E.~C. {Ferrara}, W.~{Fiore}, N.~{Garver-Daniels}, D.~C. {Good}, R.~{Jennings}, M.~L. {Jones}, V.~M. {Kaspi}, M.~T. {Lam}, D.~R. {Lorimer}, J.~{Luo}, A.~{McEwen}, J.~W. {McKee}, M.~A. {McLaughlin}, N.~{McMann}, B.~W. {Meyers}, A.~{Naidu}, C.~{Ng}, D.~J. {Nice}, N.~{Pol}, H.~A. {Radovan}, B.~{Shapiro-Albert}, C.~M. {Tan}, S.~P. {Tendulkar}, J.~K. {Swiggum}, H.~M. {Wahl}, and W.~W. {Zhu}.
\newblock {Refined Mass and Geometric Measurements of the High-mass PSR J0740+6620}.
\newblock \emph{\apjl}, 915\penalty0 (1):\penalty0 L12, July 2021.
\newblock \doi{10.3847/2041-8213/ac03b8}.

\bibitem[{Riley} et~al.(2021){Riley}, {Watts}, {Ray}, {Bogdanov}, {Guillot}, {Morsink}, {Bilous}, {Arzoumanian}, {Choudhury}, {Deneva}, {Gendreau}, {Harding}, {Ho}, {Lattimer}, {Loewenstein}, {Ludlam}, {Markwardt}, {Okajima}, {Prescod-Weinstein}, {Remillard}, {Wolff}, {Fonseca}, {Cromartie}, {Kerr}, {Pennucci}, {Parthasarathy}, {Ransom}, {Stairs}, {Guillemot}, and {Cognard}]{Riley0740}
Thomas~E. {Riley}, Anna~L. {Watts}, Paul~S. {Ray}, Slavko {Bogdanov}, Sebastien {Guillot}, Sharon~M. {Morsink}, Anna~V. {Bilous}, Zaven {Arzoumanian}, Devarshi {Choudhury}, Julia~S. {Deneva}, Keith~C. {Gendreau}, Alice~K. {Harding}, Wynn C.~G. {Ho}, James~M. {Lattimer}, Michael {Loewenstein}, Renee~M. {Ludlam}, Craig~B. {Markwardt}, Takashi {Okajima}, Chanda {Prescod-Weinstein}, Ronald~A. {Remillard}, Michael~T. {Wolff}, Emmanuel {Fonseca}, H.~Thankful {Cromartie}, Matthew {Kerr}, Timothy~T. {Pennucci}, Aditya {Parthasarathy}, Scott {Ransom}, Ingrid {Stairs}, Lucas {Guillemot}, and Ismael {Cognard}.
\newblock {A NICER View of the Massive Pulsar PSR J0740+6620 Informed by Radio Timing and XMM-Newton Spectroscopy}.
\newblock \emph{\apjl}, 918\penalty0 (2):\penalty0 L27, September 2021.
\newblock \doi{10.3847/2041-8213/ac0a81}.

\bibitem[{Miller} et~al.(2021){Miller}, {Lamb}, {Dittmann}, {Bogdanov}, {Arzoumanian}, {Gendreau}, {Guillot}, {Ho}, {Lattimer}, {Loewenstein}, {Morsink}, {Ray}, {Wolff}, {Baker}, {Cazeau}, {Manthripragada}, {Markwardt}, {Okajima}, {Pollard}, {Cognard}, {Cromartie}, {Fonseca}, {Guillemot}, {Kerr}, {Parthasarathy}, {Pennucci}, {Ransom}, and {Stairs}]{Miller2021}
M.~C. {Miller}, F.~K. {Lamb}, A.~J. {Dittmann}, S.~{Bogdanov}, Z.~{Arzoumanian}, K.~C. {Gendreau}, S.~{Guillot}, W.~C.~G. {Ho}, J.~M. {Lattimer}, M.~{Loewenstein}, S.~M. {Morsink}, P.~S. {Ray}, M.~T. {Wolff}, C.~L. {Baker}, T.~{Cazeau}, S.~{Manthripragada}, C.~B. {Markwardt}, T.~{Okajima}, S.~{Pollard}, I.~{Cognard}, H.~T. {Cromartie}, E.~{Fonseca}, L.~{Guillemot}, M.~{Kerr}, A.~{Parthasarathy}, T.~T. {Pennucci}, S.~{Ransom}, and I.~{Stairs}.
\newblock {The Radius of PSR J0740+6620 from NICER and XMM-Newton Data}.
\newblock \emph{\apjl}, 918\penalty0 (2):\penalty0 L28, September 2021.
\newblock \doi{10.3847/2041-8213/ac089b}.

\bibitem[{Salmi} et~al.(2022){Salmi}, {Vinciguerra}, {Choudhury}, {Riley}, {Watts}, {Remillard}, {Ray}, {Bogdanov}, {Guillot}, {Arzoumanian}, {Chirenti}, {Dittmann}, {Gendreau}, {Ho}, {Miller}, {Morsink}, {Wadiasingh}, and {Wolff}]{Salmi2022}
Tuomo {Salmi}, Serena {Vinciguerra}, Devarshi {Choudhury}, Thomas~E. {Riley}, Anna~L. {Watts}, Ronald~A. {Remillard}, Paul~S. {Ray}, Slavko {Bogdanov}, Sebastien {Guillot}, Zaven {Arzoumanian}, Cecilia {Chirenti}, Alexander~J. {Dittmann}, Keith~C. {Gendreau}, Wynn C.~G. {Ho}, M.~Coleman {Miller}, Sharon~M. {Morsink}, Zorawar {Wadiasingh}, and Michael~T. {Wolff}.
\newblock {The Radius of PSR J0740+6620 from NICER with NICER Background Estimates}.
\newblock \emph{\apj}, 941\penalty0 (2):\penalty0 150, December 2022.
\newblock \doi{10.3847/1538-4357/ac983d}.

\bibitem[Salmi et~al.(2024{\natexlab{a}})]{Salmi2024}
Tuomo Salmi et~al.
\newblock {The Radius of the High-mass Pulsar PSR J0740+6620 with 3.6 yr of NICER Data}.
\newblock \emph{Astrophys. J.}, 974\penalty0 (2):\penalty0 294, 2024{\natexlab{a}}.
\newblock \doi{10.3847/1538-4357/ad5f1f}.

\bibitem[Dittmann et~al.(2024)]{Dittmann2024}
Alexander~J. Dittmann et~al.
\newblock {A More Precise Measurement of the Radius of PSR J0740+6620 Using Updated NICER Data}.
\newblock \emph{Astrophys. J.}, 974\penalty0 (2):\penalty0 295, 2024.
\newblock \doi{10.3847/1538-4357/ad5f1e}.

\bibitem[{Riley} et~al.(2019){Riley}, {Watts}, {Bogdanov}, {Ray}, {Ludlam}, {Guillot}, {Arzoumanian}, {Baker}, {Bilous}, {Chakrabarty}, {Gendreau}, {Harding}, {Ho}, {Lattimer}, {Morsink}, and {Strohmayer}]{Riley0030}
T.~E. {Riley}, A.~L. {Watts}, S.~{Bogdanov}, P.~S. {Ray}, R.~M. {Ludlam}, S.~{Guillot}, Z.~{Arzoumanian}, C.~L. {Baker}, A.~V. {Bilous}, D.~{Chakrabarty}, K.~C. {Gendreau}, A.~K. {Harding}, W.~C.~G. {Ho}, J.~M. {Lattimer}, S.~M. {Morsink}, and T.~E. {Strohmayer}.
\newblock {A NICER View of PSR J0030+0451: Millisecond Pulsar Parameter Estimation}.
\newblock \emph{\apjl}, 887\penalty0 (1):\penalty0 L21, December 2019.
\newblock \doi{10.3847/2041-8213/ab481c}.

\bibitem[{Miller} et~al.(2019){Miller}, {Lamb}, {Dittmann}, {Bogdanov}, {Arzoumanian}, {Gendreau}, {Guillot}, {Harding}, {Ho}, {Lattimer}, {Ludlam}, {Mahmoodifar}, {Morsink}, {Ray}, {Strohmayer}, {Wood}, {Enoto}, {Foster}, {Okajima}, {Prigozhin}, and {Soong}]{Miller2019}
M.~C. {Miller}, F.~K. {Lamb}, A.~J. {Dittmann}, S.~{Bogdanov}, Z.~{Arzoumanian}, K.~C. {Gendreau}, S.~{Guillot}, A.~K. {Harding}, W.~C.~G. {Ho}, J.~M. {Lattimer}, R.~M. {Ludlam}, S.~{Mahmoodifar}, S.~M. {Morsink}, P.~S. {Ray}, T.~E. {Strohmayer}, K.~S. {Wood}, T.~{Enoto}, R.~{Foster}, T.~{Okajima}, G.~{Prigozhin}, and Y.~{Soong}.
\newblock {PSR J0030+0451 Mass and Radius from NICER Data and Implications for the Properties of Neutron Star Matter}.
\newblock \emph{\apjl}, 887\penalty0 (1):\penalty0 L24, December 2019.
\newblock \doi{10.3847/2041-8213/ab50c5}.

\bibitem[{Vinciguerra} et~al.(2024){Vinciguerra}, {Salmi}, {Watts}, {Choudhury}, {Riley}, {Ray}, {Bogdanov}, {Kini}, {Guillot}, {Chakrabarty}, {Ho}, {Huppenkothen}, {Morsink}, {Wadiasingh}, and {Wolff}]{Vinciguerra2024}
Serena {Vinciguerra}, Tuomo {Salmi}, Anna~L. {Watts}, Devarshi {Choudhury}, Thomas~E. {Riley}, Paul~S. {Ray}, Slavko {Bogdanov}, Yves {Kini}, Sebastien {Guillot}, Deepto {Chakrabarty}, Wynn C.~G. {Ho}, Daniela {Huppenkothen}, Sharon~M. {Morsink}, Zorawar {Wadiasingh}, and Michael~T. {Wolff}.
\newblock {An Updated Mass-Radius Analysis of the 2017-2018 NICER Data Set of PSR J0030+0451}.
\newblock \emph{\apj}, 961\penalty0 (1):\penalty0 62, January 2024.
\newblock \doi{10.3847/1538-4357/acfb83}.

\bibitem[{Choudhury} et~al.(2024){Choudhury}, {Salmi}, {Vinciguerra}, {Riley}, {Kini}, {Watts}, {Dorsman}, {Bogdanov}, {Guillot}, {Ray}, {Reardon}, {Remillard}, {Bilous}, {Huppenkothen}, {Lattimer}, {Rutherford}, {Arzoumanian}, {Gendreau}, {Morsink}, and {Ho}]{Choudhury2024}
Devarshi {Choudhury}, Tuomo {Salmi}, Serena {Vinciguerra}, Thomas~E. {Riley}, Yves {Kini}, Anna~L. {Watts}, Bas {Dorsman}, Slavko {Bogdanov}, Sebastien {Guillot}, Paul~S. {Ray}, Daniel~J. {Reardon}, Ronald~A. {Remillard}, Anna~V. {Bilous}, Daniela {Huppenkothen}, James~M. {Lattimer}, Nathan {Rutherford}, Zaven {Arzoumanian}, Keith~C. {Gendreau}, Sharon~M. {Morsink}, and Wynn C.~G. {Ho}.
\newblock {A NICER View of the Nearest and Brightest Millisecond Pulsar: PSR J0437{\textendash}4715}.
\newblock \emph{\apjl}, 971\penalty0 (1):\penalty0 L20, August 2024.
\newblock \doi{10.3847/2041-8213/ad5a6f}.

\bibitem[{Reardon} et~al.(2024){Reardon}, {Bailes}, {Shannon}, {Flynn}, {Askew}, {Bhat}, {Chen}, {Cury{\l}o}, {Feng}, {Hobbs}, {Kapur}, {Kerr}, {Liu}, {Manchester}, {Mandow}, {Mishra}, {Russell}, {Shamohammadi}, {Zhang}, and {Zic}]{Reardon2024}
Daniel~J. {Reardon}, Matthew {Bailes}, Ryan~M. {Shannon}, Chris {Flynn}, Jacob {Askew}, N.~D.~Ramesh {Bhat}, Zu-Cheng {Chen}, Ma{\l}gorzata {Cury{\l}o}, Yi~{Feng}, George~B. {Hobbs}, Agastya {Kapur}, Matthew {Kerr}, Xiaojin {Liu}, Richard~N. {Manchester}, Rami {Mandow}, Saurav {Mishra}, Christopher~J. {Russell}, Mohsen {Shamohammadi}, Lei {Zhang}, and Andrew {Zic}.
\newblock {The Neutron Star Mass, Distance, and Inclination from Precision Timing of the Brilliant Millisecond Pulsar J0437-4715}.
\newblock \emph{\apjl}, 971\penalty0 (1):\penalty0 L18, August 2024.
\newblock \doi{10.3847/2041-8213/ad614a}.

\bibitem[Salmi et~al.(2024{\natexlab{b}})]{Salmi:2024bss}
Tuomo Salmi et~al.
\newblock {A NICER View of PSR J1231\ensuremath{-}1411: A Complex Case}.
\newblock \emph{Astrophys. J.}, 976\penalty0 (1):\penalty0 58, 2024{\natexlab{b}}.
\newblock \doi{10.3847/1538-4357/ad81d2}.

\bibitem[{Raaijmakers} et~al.(2019){Raaijmakers}, {Riley}, {Watts}, {Greif}, {Morsink}, {Hebeler}, {Schwenk}, {Hinderer}, {Nissanke}, {Guillot}, {Arzoumanian}, {Bogdanov}, {Chakrabarty}, {Gendreau}, {Ho}, {Lattimer}, {Ludlam}, and {Wolff}]{Raaijmakers2019}
G.~{Raaijmakers}, T.~E. {Riley}, A.~L. {Watts}, S.~K. {Greif}, S.~M. {Morsink}, K.~{Hebeler}, A.~{Schwenk}, T.~{Hinderer}, S.~{Nissanke}, S.~{Guillot}, Z.~{Arzoumanian}, S.~{Bogdanov}, D.~{Chakrabarty}, K.~C. {Gendreau}, W.~C.~G. {Ho}, J.~M. {Lattimer}, R.~M. {Ludlam}, and M.~T. {Wolff}.
\newblock {A Nicer View of PSR J0030+0451: Implications for the Dense Matter Equation of State}.
\newblock \emph{Astrophysical Journal Letters}, 887\penalty0 (1):\penalty0 L22, December 2019.
\newblock \doi{10.3847/2041-8213/ab451a}.

\bibitem[{Raaijmakers} et~al.(2020){Raaijmakers}, {Greif}, {Riley}, {Hinderer}, {Hebeler}, {Schwenk}, {Watts}, {Nissanke}, {Guillot}, {Lattimer}, and {Ludlam}]{Raaijmakers2020}
G.~{Raaijmakers}, S.~K. {Greif}, T.~E. {Riley}, T.~{Hinderer}, K.~{Hebeler}, A.~{Schwenk}, A.~L. {Watts}, S.~{Nissanke}, S.~{Guillot}, J.~M. {Lattimer}, and R.~M. {Ludlam}.
\newblock {Constraining the Dense Matter Equation of State with Joint Analysis of NICER and LIGO/Virgo Measurements}.
\newblock \emph{Astrophysical Journal Letters}, 893\penalty0 (1):\penalty0 L21, April 2020.
\newblock \doi{10.3847/2041-8213/ab822f}.

\bibitem[{Raaijmakers} et~al.(2021){Raaijmakers}, {Greif}, {Hebeler}, {Hinderer}, {Nissanke}, {Schwenk}, {Riley}, {Watts}, {Lattimer}, and {Ho}]{Raaijmakers2021}
G.~{Raaijmakers}, S.~K. {Greif}, K.~{Hebeler}, T.~{Hinderer}, S.~{Nissanke}, A.~{Schwenk}, T.~E. {Riley}, A.~L. {Watts}, J.~M. {Lattimer}, and W.~C.~G. {Ho}.
\newblock {Constraints on the Dense Matter Equation of State and Neutron Star Properties from NICER's Mass-Radius Estimate of PSR J0740+6620 and Multimessenger Observations}.
\newblock \emph{\apjl}, 918\penalty0 (2):\penalty0 L29, September 2021.
\newblock \doi{10.3847/2041-8213/ac089a}.

\bibitem[{Li} et~al.(2021){Li}, {Sedrakian}, and {Alford}]{JieLiJ21}
Jia~Jie {Li}, Armen {Sedrakian}, and Mark {Alford}.
\newblock {Relativistic hybrid stars in light of the NICER PSR J 0740 +6620 radius measurement}.
\newblock \emph{\prd}, 104\penalty0 (12):\penalty0 L121302, December 2021.
\newblock \doi{10.1103/PhysRevD.104.L121302}.

\bibitem[{Legred} et~al.(2021){Legred}, {Chatziioannou}, {Essick}, {Han}, and {Landry}]{Legred21}
Isaac {Legred}, Katerina {Chatziioannou}, Reed {Essick}, Sophia {Han}, and Philippe {Landry}.
\newblock {Impact of the PSR J0740+6620 radius constraint on the properties of high-density matter}.
\newblock \emph{\prd}, 104\penalty0 (6):\penalty0 063003, September 2021.
\newblock \doi{10.1103/PhysRevD.104.063003}.

\bibitem[{Pang} et~al.(2021){Pang}, {Tews}, {Coughlin}, {Bulla}, {Van Den Broeck}, and {Dietrich}]{Pang21}
Peter T.~H. {Pang}, Ingo {Tews}, Michael~W. {Coughlin}, Mattia {Bulla}, Chris {Van Den Broeck}, and Tim {Dietrich}.
\newblock {Nuclear Physics Multimessenger Astrophysics Constraints on the Neutron Star Equation of State: Adding NICER's PSR J0740+6620 Measurement}.
\newblock \emph{\apj}, 922\penalty0 (1):\penalty0 14, November 2021.
\newblock \doi{10.3847/1538-4357/ac19ab}.

\bibitem[{Tang} et~al.(2021){Tang}, {Jiang}, {Han}, {Fan}, and {Wei}]{TangSP21}
Shao-Peng {Tang}, Jin-Liang {Jiang}, Ming-Zhe {Han}, Yi-Zhong {Fan}, and Da-Ming {Wei}.
\newblock {Constraints on the phase transition and nuclear symmetry parameters from PSR J 0740 +6620 and multimessenger data of other neutron stars}.
\newblock \emph{\prd}, 104\penalty0 (6):\penalty0 063032, September 2021.
\newblock \doi{10.1103/PhysRevD.104.063032}.

\bibitem[{Annala} et~al.(2022){Annala}, {Gorda}, {Katerini}, {Kurkela}, {N{\"a}ttil{\"a}}, {Paschalidis}, and {Vuorinen}]{Annala2022}
Eemeli {Annala}, Tyler {Gorda}, Evangelia {Katerini}, Aleksi {Kurkela}, Joonas {N{\"a}ttil{\"a}}, Vasileios {Paschalidis}, and Aleksi {Vuorinen}.
\newblock {Multimessenger Constraints for Ultradense Matter}.
\newblock \emph{Physical Review X}, 12\penalty0 (1):\penalty0 011058, January 2022.
\newblock \doi{10.1103/PhysRevX.12.011058}.

\bibitem[{Biswas}(2022)]{Biswas2022}
Bhaskar {Biswas}.
\newblock {Bayesian Model Selection of Neutron Star Equations of State Using Multi-messenger Observations}.
\newblock \emph{\apj}, 926\penalty0 (1):\penalty0 75, February 2022.
\newblock \doi{10.3847/1538-4357/ac447b}.

\bibitem[{Rutherford} et~al.(2024){Rutherford}, {Mendes}, {Svensson}, {Schwenk}, {Watts}, {Hebeler}, {Keller}, {Prescod-Weinstein}, {Choudhury}, {Raaijmakers}, {Salmi}, {Timmerman}, {Vinciguerra}, {Guillot}, and {Lattimer}]{Rutherford2024}
Nathan {Rutherford}, Melissa {Mendes}, Isak {Svensson}, Achim {Schwenk}, Anna~L. {Watts}, Kai {Hebeler}, Jonas {Keller}, Chanda {Prescod-Weinstein}, Devarshi {Choudhury}, Geert {Raaijmakers}, Tuomo {Salmi}, Patrick {Timmerman}, Serena {Vinciguerra}, Sebastien {Guillot}, and James~M. {Lattimer}.
\newblock {Constraining the Dense Matter Equation of State with New NICER Mass{\textendash}Radius Measurements and New Chiral Effective Field Theory Inputs}.
\newblock \emph{\apjl}, 971\penalty0 (1):\penalty0 L19, August 2024.
\newblock \doi{10.3847/2041-8213/ad5f02}.

\bibitem[{Huang} et~al.(2024){Huang}, {Raaijmakers}, {Watts}, {Tolos}, and {Provid{\^e}ncia}]{Huang2024}
Chun {Huang}, Geert {Raaijmakers}, Anna~L. {Watts}, Laura {Tolos}, and Constan{\c{c}}a {Provid{\^e}ncia}.
\newblock {Constraining a relativistic mean field model using neutron star mass-radius measurements I: nucleonic models}.
\newblock \emph{\mnras}, 529\penalty0 (4):\penalty0 4650--4665, April 2024.
\newblock \doi{10.1093/mnras/stae844}.

\bibitem[Keller et~al.(2023)Keller, Hebeler, and Schwenk]{Keller2022}
J.~Keller, K.~Hebeler, and A.~Schwenk.
\newblock Nuclear equation of state for arbitrary proton fraction and temperature based on chiral effective field theory and a gaussian process emulator.
\newblock \emph{Phys. Rev. Lett.}, 130:\penalty0 072701, Feb 2023.
\newblock \doi{10.1103/PhysRevLett.130.072701}.

\bibitem[{Watts} et~al.(2019){Watts}, {Yu}, {Poutanen}, {Zhang}, {Bhattacharyya}, {Bogdanov}, {Ji}, {Patruno}, {Riley}, {Bakala}, {Baykal}, {Bernardini}, {Bombaci}, {Brown}, {Cavecchi}, {Chakrabarty}, {Chenevez}, {Degenaar}, {Del Santo}, {Di Salvo}, {Doroshenko}, {Falanga}, {Ferdman}, {Feroci}, {Gambino}, {Ge}, {Greif}, {Guillot}, {Gungor}, {Hartmann}, {Hebeler}, {Heger}, {Homan}, {Iaria}, {Zand}, {Kargaltsev}, {Kurkela}, {Lai}, {Li}, {Li}, {Li}, {Linares}, {Lu}, {Mahmoodifar}, {M{\'e}ndez}, {Coleman Miller}, {Morsink}, {N{\"a}ttil{\"a}}, {Possenti}, {Prescod-Weinstein}, {Qu}, {Riggio}, {Salmi}, {Sanna}, {Santangelo}, {Schatz}, {Schwenk}, {Song}, {{\v{S}}r{\'a}mkov{\'a}}, {Stappers}, {Stiele}, {Strohmayer}, {Tews}, {Tolos}, {T{\"o}r{\"o}k}, {Tsang}, {Urbanec}, {Vacchi}, {Xu}, {Xu}, {Zane}, {Zhang}, {Zhang}, {Zhang}, {Zheng}, and {Zhou}]{extp}
Anna~L. {Watts}, WenFei {Yu}, Juri {Poutanen}, Shu {Zhang}, Sudip {Bhattacharyya}, Slavko {Bogdanov}, Long {Ji}, Alessandro {Patruno}, Thomas~E. {Riley}, Pavel {Bakala}, Altan {Baykal}, Federico {Bernardini}, Ignazio {Bombaci}, Edward {Brown}, Yuri {Cavecchi}, Deepto {Chakrabarty}, J{\'e}r{\^o}me {Chenevez}, Nathalie {Degenaar}, Melania {Del Santo}, Tiziana {Di Salvo}, Victor {Doroshenko}, Maurizio {Falanga}, Robert~D. {Ferdman}, Marco {Feroci}, Angelo~F. {Gambino}, MingYu {Ge}, Svenja~K. {Greif}, Sebastien {Guillot}, Can {Gungor}, Dieter~H. {Hartmann}, Kai {Hebeler}, Alexander {Heger}, Jeroen {Homan}, Rosario {Iaria}, Jean~in't. {Zand}, Oleg {Kargaltsev}, Aleksi {Kurkela}, XiaoYu {Lai}, Ang {Li}, XiangDong {Li}, ZhaoSheng {Li}, Manuel {Linares}, FangJun {Lu}, Simin {Mahmoodifar}, Mariano {M{\'e}ndez}, M.~{Coleman Miller}, Sharon {Morsink}, Joonas {N{\"a}ttil{\"a}}, Andrea {Possenti}, Chanda {Prescod-Weinstein}, JinLu {Qu}, Alessandro {Riggio}, Tuomo {Salmi}, Andrea {Sanna}, Andrea {Santangelo}, Hendrik
  {Schatz}, Achim {Schwenk}, LiMing {Song}, Eva {{\v{S}}r{\'a}mkov{\'a}}, Benjamin {Stappers}, Holger {Stiele}, Tod {Strohmayer}, Ingo {Tews}, Laura {Tolos}, Gabriel {T{\"o}r{\"o}k}, David {Tsang}, Martin {Urbanec}, Andrea {Vacchi}, RenXin {Xu}, YuPeng {Xu}, Silvia {Zane}, GuoBao {Zhang}, ShuangNan {Zhang}, WenDa {Zhang}, ShiJie {Zheng}, and Xia {Zhou}.
\newblock {Dense matter with eXTP}.
\newblock \emph{Science China Physics, Mechanics, and Astronomy}, 62\penalty0 (2):\penalty0 29503, February 2019.
\newblock \doi{10.1007/s11433-017-9188-4}.

\bibitem[{Ray} et~al.(2019){Ray}, {Arzoumanian}, {Ballantyne}, {Bozzo}, {Brandt}, {Brenneman}, {Chakrabarty}, {Christophersen}, {DeRosa}, {Feroci}, {Gendreau}, {Goldstein}, {Hartmann}, {Hernanz}, {Jenke}, {Kara}, {Maccarone}, {McDonald}, {Nowak}, {Phlips}, {Remillard}, {Stevens}, {Tomsick}, {Watts}, {Wilson-Hodge}, {Wood}, {Zane}, {Ajello}, {Alston}, {Altamirano}, {Antoniou}, {Arur}, {Ashton}, {Auchettl}, {Ayres}, {Bachetti}, {Balokovic}, {Baring}, {Baykal}, {Begelman}, {Bhat}, {Bogdanov}, {Briggs}, {Bulbul}, {Bult}, {Burns}, {Cackett}, {Campana}, {Caspi}, {Cavecchi}, {Chenevez}, {Cherry}, {Corbet}, {Corcoran}, {Corsi}, {Degenaar}, {Drake}, {Eikenberry}, {Enoto}, {Fragile}, {Fuerst}, {Gandhi}, {Garcia}, {Goldstein}, {Gonzalez}, {Grefenstette}, {Grinberg}, {Grossan}, {Guillot}, {Guver}, {Haggard}, {Heinke}, {Heinz}, {Hemphill}, {Homan}, {Hui}, {Huppenkothen}, {Ingram}, {Irwin}, {Jaisawal}, {Jaodand}, {Kalemci}, {Kaplan}, {Keek}, {Kennea}, {Kerr}, {van der Klis}, {Kocevski}, {Koss}, {Kowalski}, {Lai}, {Lamb},
  {Laycock}, {Lazio}, {Lazzati}, {Longcope}, {Loewenstein}, {Maitra}, {Majid}, {Maksym}, {Malacaria}, {Margutti}, {Martindale}, {McHardy}, {Meyer}, {Middleton}, {Miller}, {Miller}, {Motta}, {Neilsen}, {Nelson}, {Noble}, {O'Brien}, {Osborne}, {Osten}, {Ozel}, {Palliyaguru}, {Pasham}, {Patruno}, {Pelassa}, {Petropoulou}, {Pilia}, {Pohl}, {Pooley}, {Prescod-Weinstein}, {Psaltis}, {Raaijmakers}, {Reynolds}, {Riley}, {Salvesen}, {Santangelo}, {Scaringi}, {Schanne}, {Schnittman}, {Smith}, {Smith}, {Snios}, {Steiner}, {Steiner}, {Stella}, {Strohmayer}, {Sun}, {Tauris}, {Taylor}, {Tohuvavohu}, {Vacchi}, {Vasilopoulos}, {Veledina}, {Walsh}, {Weinberg}, {Wilkins}, {Willingale}, {Wilms}, {Winter}, {Wolff}, {in 't Zand}, {Zezas}, {Zhang}, and {Zoghbi}]{Ray2019}
Paul~S. {Ray}, Zaven {Arzoumanian}, David {Ballantyne}, Enrico {Bozzo}, Soren {Brandt}, Laura {Brenneman}, Deepto {Chakrabarty}, Marc {Christophersen}, Alessand~ra {DeRosa}, Marco {Feroci}, Keith {Gendreau}, Adam {Goldstein}, Dieter {Hartmann}, Margarita {Hernanz}, Peter {Jenke}, Erin {Kara}, Tom {Maccarone}, Michael {McDonald}, Michael {Nowak}, Bernard {Phlips}, Ron {Remillard}, Abigail {Stevens}, John {Tomsick}, Anna {Watts}, Colleen {Wilson-Hodge}, Kent {Wood}, Silvia {Zane}, Marco {Ajello}, Will {Alston}, Diego {Altamirano}, Vallia {Antoniou}, Kavitha {Arur}, Dominic {Ashton}, Katie {Auchettl}, Tom {Ayres}, Matteo {Bachetti}, Mislav {Balokovic}, Matthew {Baring}, Altan {Baykal}, Mitch {Begelman}, Narayana {Bhat}, Slavko {Bogdanov}, Michael {Briggs}, Esra {Bulbul}, Petrus {Bult}, Eric {Burns}, Ed~{Cackett}, Riccardo {Campana}, Amir {Caspi}, Yuri {Cavecchi}, Jerome {Chenevez}, Mike {Cherry}, Robin {Corbet}, Michael {Corcoran}, Alessandra {Corsi}, Nathalie {Degenaar}, Jeremy {Drake}, Steve {Eikenberry},
  Teruaki {Enoto}, Chris {Fragile}, Felix {Fuerst}, Poshak {Gandhi}, Javier {Garcia}, Adam {Goldstein}, Anthony {Gonzalez}, Brian {Grefenstette}, Victoria {Grinberg}, Bruce {Grossan}, Sebastien {Guillot}, Tolga {Guver}, Daryl {Haggard}, Craig {Heinke}, Sebastian {Heinz}, Paul {Hemphill}, Jeroen {Homan}, Michelle {Hui}, Daniela {Huppenkothen}, Adam {Ingram}, Jimmy {Irwin}, Gaurava {Jaisawal}, Amruta {Jaodand}, Emrah {Kalemci}, David {Kaplan}, Laurens {Keek}, Jamie {Kennea}, Matthew {Kerr}, Michiel {van der Klis}, Daniel {Kocevski}, Mike {Koss}, Adam {Kowalski}, Dong {Lai}, Fred {Lamb}, Silas {Laycock}, Joseph {Lazio}, Davide {Lazzati}, Dana {Longcope}, Michael {Loewenstein}, Dipankair {Maitra}, Walid {Majid}, W.~Peter {Maksym}, Christian {Malacaria}, Raffaella {Margutti}, Adrian {Martindale}, Ian {McHardy}, Manuel {Meyer}, Matt {Middleton}, Jon {Miller}, Cole {Miller}, Sara {Motta}, Joey {Neilsen}, Tommy {Nelson}, Scott {Noble}, Paul {O'Brien}, Julian {Osborne}, Rachel {Osten}, Feryal {Ozel}, Nipuni
  {Palliyaguru}, Dheeraj {Pasham}, Alessandro {Patruno}, Vero {Pelassa}, Maria {Petropoulou}, Maura {Pilia}, Martin {Pohl}, David {Pooley}, Chanda {Prescod-Weinstein}, Dimitrios {Psaltis}, Geert {Raaijmakers}, Chris {Reynolds}, Thomas~E. {Riley}, Greg {Salvesen}, Andrea {Santangelo}, Simone {Scaringi}, Stephane {Schanne}, Jeremy {Schnittman}, David {Smith}, Krista~Lynne {Smith}, Bradford {Snios}, Andrew {Steiner}, Jack {Steiner}, Luigi {Stella}, Tod {Strohmayer}, Ming {Sun}, Thomas {Tauris}, Corbin {Taylor}, Aaron {Tohuvavohu}, Andrea {Vacchi}, Georgios {Vasilopoulos}, Alexandra {Veledina}, Jonelle {Walsh}, Nevin {Weinberg}, Dan {Wilkins}, Richard {Willingale}, Joern {Wilms}, Lisa {Winter}, Michael {Wolff}, Jean {in 't Zand}, Andreas {Zezas}, Bing {Zhang}, and Abdu {Zoghbi}.
\newblock {STROBE-X: X-ray Timing and Spectroscopy on Dynamical Timescales from Microseconds to Years}.
\newblock \emph{arXiv e-prints}, art. arXiv:1903.03035, Mar 2019.

\bibitem[{Nandra} et~al.(2013){Nandra}, {Barret}, {Barcons}, {Fabian}, {den Herder}, {Piro}, {Watson}, {Adami}, {Aird}, {Afonso}, {Alexander}, {Argiroffi}, {Amati}, {Arnaud}, {Atteia}, {Audard}, {Badenes}, {Ballet}, {Ballo}, {Bamba}, {Bhardwaj}, {Stefano Battistelli}, {Becker}, {De Becker}, {Behar}, {Bianchi}, {Biffi}, {B{\^\i}rzan}, {Bocchino}, {Bogdanov}, {Boirin}, {Boller}, {Borgani}, {Borm}, {Bouch{\'e}}, {Bourdin}, {Bower}, {Braito}, {Branchini}, {Branduardi-Raymont}, {Bregman}, {Brenneman}, {Brightman}, {Br{\"u}ggen}, {Buchner}, {Bulbul}, {Brusa}, {Bursa}, {Caccianiga}, {Cackett}, {Campana}, {Cappelluti}, {Cappi}, {Carrera}, {Ceballos}, {Christensen}, {Chu}, {Churazov}, {Clerc}, {Corbel}, {Corral}, {Comastri}, {Costantini}, {Croston}, {Dadina}, {D'Ai}, {Decourchelle}, {Della Ceca}, {Dennerl}, {Dolag}, {Done}, {Dovciak}, {Drake}, {Eckert}, {Edge}, {Ettori}, {Ezoe}, {Feigelson}, {Fender}, {Feruglio}, {Finoguenov}, {Fiore}, {Galeazzi}, {Gallagher}, {Gandhi}, {Gaspari}, {Gastaldello}, {Georgakakis},
  {Georgantopoulos}, {Gilfanov}, {Gitti}, {Gladstone}, {Goosmann}, {Gosset}, {Grosso}, {Guedel}, {Guerrero}, {Haberl}, {Hardcastle}, {Heinz}, {Alonso Herrero}, {Herv{\'e}}, {Holmstrom}, {Iwasawa}, {Jonker}, {Kaastra}, {Kara}, {Karas}, {Kastner}, {King}, {Kosenko}, {Koutroumpa}, {Kraft}, {Kreykenbohm}, {Lallement}, {Lanzuisi}, {Lee}, {Lemoine-Goumard}, {Lobban}, {Lodato}, {Lovisari}, {Lotti}, {McCharthy}, {McNamara}, {Maggio}, {Maiolino}, {De Marco}, {de Martino}, {Mateos}, {Matt}, {Maughan}, {Mazzotta}, {Mendez}, {Merloni}, {Micela}, {Miceli}, {Mignani}, {Miller}, {Miniutti}, {Molendi}, {Montez}, {Moretti}, {Motch}, {Naz{\'e}}, {Nevalainen}, {Nicastro}, {Nulsen}, {Ohashi}, {O'Brien}, {Osborne}, {Oskinova}, {Pacaud}, {Paerels}, {Page}, {Papadakis}, {Pareschi}, {Petre}, {Petrucci}, {Piconcelli}, {Pillitteri}, {Pinto}, {de Plaa}, {Pointecouteau}, {Ponman}, {Ponti}, {Porquet}, {Pounds}, {Pratt}, {Predehl}, {Proga}, {Psaltis}, {Rafferty}, {Ramos-Ceja}, {Ranalli}, {Rasia}, {Rau}, {Rauw}, {Rea}, {Read}, {Reeves},
  {Reiprich}, {Renaud}, {Reynolds}, {Risaliti}, {Rodriguez}, {Rodriguez Hidalgo}, {Roncarelli}, {Rosario}, {Rossetti}, {Rozanska}, {Rovilos}, {Salvaterra}, {Salvato}, {Di Salvo}, {Sanders}, {Sanz-Forcada}, {Schawinski}, {Schaye}, {Schwope}, {Sciortino}, {Severgnini}, {Shankar}, {Sijacki}, {Sim}, {Schmid}, {Smith}, {Steiner}, {Stelzer}, {Stewart}, {Strohmayer}, {Str{\"u}der}, {Sun}, {Takei}, {Tatischeff}, {Tiengo}, {Tombesi}, {Trinchieri}, {Tsuru}, {Ud-Doula}, {Ursino}, {Valencic}, {Vanzella}, {Vaughan}, {Vignali}, {Vink}, {Vito}, {Volonteri}, {Wang}, {Webb}, {Willingale}, {Wilms}, {Wise}, {Worrall}, {Young}, {Zampieri}, {In't Zand}, {Zane}, {Zezas}, {Zhang}, and {Zhuravleva}]{Athena13}
Kirpal {Nandra}, Didier {Barret}, Xavier {Barcons}, Andy {Fabian}, Jan-Willem {den Herder}, Luigi {Piro}, Mike {Watson}, Christophe {Adami}, James {Aird}, Jose~Manuel {Afonso}, Dave {Alexander}, Costanza {Argiroffi}, Lorenzo {Amati}, Monique {Arnaud}, Jean-Luc {Atteia}, Marc {Audard}, Carles {Badenes}, Jean {Ballet}, Lucia {Ballo}, Aya {Bamba}, Anil {Bhardwaj}, Elia {Stefano Battistelli}, Werner {Becker}, Micha{\"e}l {De Becker}, Ehud {Behar}, Stefano {Bianchi}, Veronica {Biffi}, Laura {B{\^\i}rzan}, Fabrizio {Bocchino}, Slavko {Bogdanov}, Laurence {Boirin}, Thomas {Boller}, Stefano {Borgani}, Katharina {Borm}, Nicolas {Bouch{\'e}}, Herv{\'e} {Bourdin}, Richard {Bower}, Valentina {Braito}, Enzo {Branchini}, Graziella {Branduardi-Raymont}, Joel {Bregman}, Laura {Brenneman}, Murray {Brightman}, Marcus {Br{\"u}ggen}, Johannes {Buchner}, Esra {Bulbul}, Marcella {Brusa}, Michal {Bursa}, Alessandro {Caccianiga}, Ed~{Cackett}, Sergio {Campana}, Nico {Cappelluti}, Massimo {Cappi}, Francisco {Carrera}, Maite
  {Ceballos}, Finn {Christensen}, You-Hua {Chu}, Eugene {Churazov}, Nicolas {Clerc}, Stephane {Corbel}, Amalia {Corral}, Andrea {Comastri}, Elisa {Costantini}, Judith {Croston}, Mauro {Dadina}, Antonino {D'Ai}, Anne {Decourchelle}, Roberto {Della Ceca}, Konrad {Dennerl}, Klaus {Dolag}, Chris {Done}, Michal {Dovciak}, Jeremy {Drake}, Dominique {Eckert}, Alastair {Edge}, Stefano {Ettori}, Yuichiro {Ezoe}, Eric {Feigelson}, Rob {Fender}, Chiara {Feruglio}, Alexis {Finoguenov}, Fabrizio {Fiore}, Massimiliano {Galeazzi}, Sarah {Gallagher}, Poshak {Gandhi}, Massimo {Gaspari}, Fabio {Gastaldello}, Antonis {Georgakakis}, Ioannis {Georgantopoulos}, Marat {Gilfanov}, Myriam {Gitti}, Randy {Gladstone}, Rene {Goosmann}, Eric {Gosset}, Nicolas {Grosso}, Manuel {Guedel}, Martin {Guerrero}, Frank {Haberl}, Martin {Hardcastle}, Sebastian {Heinz}, Almudena {Alonso Herrero}, Anthony {Herv{\'e}}, Mats {Holmstrom}, Kazushi {Iwasawa}, Peter {Jonker}, Jelle {Kaastra}, Erin {Kara}, Vladimir {Karas}, Joel {Kastner}, Andrew {King},
  Daria {Kosenko}, Dimita {Koutroumpa}, Ralph {Kraft}, Ingo {Kreykenbohm}, Rosine {Lallement}, Giorgio {Lanzuisi}, J.~{Lee}, Marianne {Lemoine-Goumard}, Andrew {Lobban}, Giuseppe {Lodato}, Lorenzo {Lovisari}, Simone {Lotti}, Ian {McCharthy}, Brian {McNamara}, Antonio {Maggio}, Roberto {Maiolino}, Barbara {De Marco}, Domitilla {de Martino}, Silvia {Mateos}, Giorgio {Matt}, Ben {Maughan}, Pasquale {Mazzotta}, Mariano {Mendez}, Andrea {Merloni}, Giuseppina {Micela}, Marco {Miceli}, Robert {Mignani}, Jon {Miller}, Giovanni {Miniutti}, Silvano {Molendi}, Rodolfo {Montez}, Alberto {Moretti}, Christian {Motch}, Ya{\"e}l {Naz{\'e}}, Jukka {Nevalainen}, Fabrizio {Nicastro}, Paul {Nulsen}, Takaya {Ohashi}, Paul {O'Brien}, Julian {Osborne}, Lida {Oskinova}, Florian {Pacaud}, Frederik {Paerels}, Mat {Page}, Iossif {Papadakis}, Giovanni {Pareschi}, Robert {Petre}, Pierre-Olivier {Petrucci}, Enrico {Piconcelli}, Ignazio {Pillitteri}, C.~{Pinto}, Jelle {de Plaa}, Etienne {Pointecouteau}, Trevor {Ponman}, Gabriele {Ponti},
  Delphine {Porquet}, Ken {Pounds}, Gabriel {Pratt}, Peter {Predehl}, Daniel {Proga}, Dimitrios {Psaltis}, David {Rafferty}, Miriam {Ramos-Ceja}, Piero {Ranalli}, Elena {Rasia}, Arne {Rau}, Gregor {Rauw}, Nanda {Rea}, Andy {Read}, James {Reeves}, Thomas {Reiprich}, Matthieu {Renaud}, Chris {Reynolds}, Guido {Risaliti}, Jerome {Rodriguez}, Paola {Rodriguez Hidalgo}, Mauro {Roncarelli}, David {Rosario}, Mariachiara {Rossetti}, Agata {Rozanska}, Emmanouil {Rovilos}, Ruben {Salvaterra}, Mara {Salvato}, Tiziana {Di Salvo}, Jeremy {Sanders}, Jorge {Sanz-Forcada}, Kevin {Schawinski}, Joop {Schaye}, Axel {Schwope}, Salvatore {Sciortino}, Paola {Severgnini}, Francesco {Shankar}, Debora {Sijacki}, Stuart {Sim}, Christian {Schmid}, Randall {Smith}, Andrew {Steiner}, Beate {Stelzer}, Gordon {Stewart}, Tod {Strohmayer}, Lothar {Str{\"u}der}, Ming {Sun}, Yoh {Takei}, V.~{Tatischeff}, Andreas {Tiengo}, Francesco {Tombesi}, Ginevra {Trinchieri}, T.~G. {Tsuru}, Asif {Ud-Doula}, Eugenio {Ursino}, Lynne {Valencic}, Eros
  {Vanzella}, Simon {Vaughan}, Cristian {Vignali}, Jacco {Vink}, Fabio {Vito}, Marta {Volonteri}, Daniel {Wang}, Natalie {Webb}, Richard {Willingale}, Joern {Wilms}, Michael {Wise}, Diana {Worrall}, Andrew {Young}, Luca {Zampieri}, Jean {In't Zand}, Silvia {Zane}, Andreas {Zezas}, Yuying {Zhang}, and Irina {Zhuravleva}.
\newblock {The Hot and Energetic Universe: A White Paper presenting the science theme motivating the Athena+ mission}.
\newblock \emph{arXiv e-prints}, art. arXiv:1306.2307, June 2013.
\newblock \doi{10.48550/arXiv.1306.2307}.

\bibitem[{Nelson} et~al.(2019){Nelson}, {Reddy}, and {Zhou}]{Nelson_2018}
Ann~E. {Nelson}, Sanjay {Reddy}, and Dake {Zhou}.
\newblock {Dark halos around neutron stars and gravitational waves}.
\newblock \emph{Journal of Cosmology and Astroparticle Physics}, 2019\penalty0 (7):\penalty0 012, July 2019.
\newblock \doi{10.1088/1475-7516/2019/07/012}.

\bibitem[Sagun et~al.(2022)Sagun, Giangrandi, Ivanytskyi, Lopes, and Bugaev]{Sagun:2021oml}
V.~Sagun, E.~Giangrandi, O.~Ivanytskyi, I.~Lopes, and K.~A. Bugaev.
\newblock {Constraints on the fermionic dark matter from observations of neutron stars}.
\newblock \emph{PoS}, PANIC2021:\penalty0 313, 2022.
\newblock \doi{10.22323/1.380.0313}.

\bibitem[{Diedrichs} et~al.(2023){Diedrichs}, {Becker}, {Jockel}, {Christian}, {Sagunski}, and {Schaffner-Bielich}]{Diedrichs2023}
Robin~Fynn {Diedrichs}, Niklas {Becker}, C{\'e}dric {Jockel}, Jan-Erik {Christian}, Laura {Sagunski}, and J{\"u}rgen {Schaffner-Bielich}.
\newblock {Tidal deformability of fermion-boson stars: Neutron stars admixed with ultralight dark matter}.
\newblock \emph{\prd}, 108\penalty0 (6):\penalty0 064009, September 2023.
\newblock \doi{10.1103/PhysRevD.108.064009}.

\bibitem[{Bramante} and {Raj}(2024)]{Bramante2024}
Joseph {Bramante} and Nirmal {Raj}.
\newblock {Dark matter in compact stars}.
\newblock \emph{\physrep}, 1052:\penalty0 1--48, February 2024.
\newblock \doi{10.1016/j.physrep.2023.12.001}.

\bibitem[{Buras-Stubbs} and {Lopes}(2024)]{Buras-Stubbs2024}
Zakary {Buras-Stubbs} and Il{\'\i}dio {Lopes}.
\newblock {Bosonic dark matter dynamics in hybrid neutron stars}.
\newblock \emph{\prd}, 109\penalty0 (4):\penalty0 043043, February 2024.
\newblock \doi{10.1103/PhysRevD.109.043043}.

\bibitem[{Guha} and {Sen}(2024)]{Guha2024}
Atanu {Guha} and Debashree {Sen}.
\newblock {Constraining the mass of fermionic dark matter from its feeble interaction with hadronic matter via dark mediators in neutron stars}.
\newblock \emph{\prd}, 109\penalty0 (4):\penalty0 043038, February 2024.
\newblock \doi{10.1103/PhysRevD.109.043038}.

\bibitem[{Jockel} and {Sagunski}(2024)]{Jockel2024}
C{\'e}dric {Jockel} and Laura {Sagunski}.
\newblock {Fermion Proca Stars: Vector-Dark-Matter-Admixed Neutron Stars}.
\newblock \emph{Particles}, 7\penalty0 (1):\penalty0 52--79, January 2024.
\newblock \doi{10.3390/particles7010004}.

\bibitem[{Shawqi} and {Morsink}(2024)]{Shawqi2024}
Shafayat {Shawqi} and Sharon~M. {Morsink}.
\newblock {Interpreting Mass and Radius Measurements of Neutron Stars with Dark Matter Halos}.
\newblock \emph{\apj}, 975\penalty0 (1):\penalty0 123, November 2024.
\newblock \doi{10.3847/1538-4357/ad77c1}.

\bibitem[{Miao} et~al.(2022){Miao}, {Zhu}, {Li}, and {Huang}]{Miao_2022}
Zhiqiang {Miao}, Yaofeng {Zhu}, Ang {Li}, and Feng {Huang}.
\newblock {Dark Matter Admixed Neutron Star Properties in the Light of X-Ray Pulse Profile Observations}.
\newblock \emph{\apj}, 936\penalty0 (1):\penalty0 69, September 2022.
\newblock \doi{10.3847/1538-4357/ac8544}.

\bibitem[{Shakeri} and {Karkevandi}(2024)]{Shakeri2024}
Soroush {Shakeri} and Davood~Rafiei {Karkevandi}.
\newblock {Bosonic dark matter in light of the NICER precise mass-radius measurements}.
\newblock \emph{\prd}, 109\penalty0 (4):\penalty0 043029, February 2024.
\newblock \doi{10.1103/PhysRevD.109.043029}.

\bibitem[Pitz and Schaffner-Bielich(2025)]{Pitz2024}
Sarah~Louisa Pitz and J\"urgen Schaffner-Bielich.
\newblock {Generating ultracompact neutron stars with bosonic dark matter}.
\newblock \emph{Phys. Rev. D}, 111\penalty0 (4):\penalty0 043050, 2025.
\newblock \doi{10.1103/PhysRevD.111.043050}.

\bibitem[{Ellis} et~al.(2018){Ellis}, {H{\"u}tsi}, {Kannike}, {Marzola}, {Raidal}, and {Vaskonen}]{Ellis2018}
John {Ellis}, Gert {H{\"u}tsi}, Kristjan {Kannike}, Luca {Marzola}, Martti {Raidal}, and Ville {Vaskonen}.
\newblock {Dark matter effects on neutron star properties}.
\newblock \emph{\prd}, 97\penalty0 (12):\penalty0 123007, June 2018.
\newblock \doi{10.1103/PhysRevD.97.123007}.

\bibitem[{Ivanytskyi} et~al.(2020){Ivanytskyi}, {Sagun}, and {Lopes}]{Ivanystkyi2020}
O.~{Ivanytskyi}, V.~{Sagun}, and I.~{Lopes}.
\newblock {Neutron stars: New constraints on asymmetric dark matter}.
\newblock \emph{\prd}, 102\penalty0 (6):\penalty0 063028, September 2020.
\newblock \doi{10.1103/PhysRevD.102.063028}.

\bibitem[{Kain}(2021)]{Kain2021}
Ben {Kain}.
\newblock {Dark matter admixed neutron stars}.
\newblock \emph{\prd}, 103\penalty0 (4):\penalty0 043009, February 2021.
\newblock \doi{10.1103/PhysRevD.103.043009}.

\bibitem[{Rafiei Karkevandi} et~al.(2022){Rafiei Karkevandi}, {Shakeri}, {Sagun}, and {Ivanytskyi}]{Karkevandi2022}
Davood {Rafiei Karkevandi}, Soroush {Shakeri}, Violetta {Sagun}, and Oleksii {Ivanytskyi}.
\newblock {Bosonic dark matter in neutron stars and its effect on gravitational wave signal}.
\newblock \emph{\prd}, 105\penalty0 (2):\penalty0 023001, January 2022.
\newblock \doi{10.1103/PhysRevD.105.023001}.

\bibitem[{Rafiei Karkevandi} et~al.(2024){Rafiei Karkevandi}, {Shahrbaf}, {Shakeri}, and {Typel}]{Karkevandi2024}
Davood {Rafiei Karkevandi}, Mahboubeh {Shahrbaf}, Soroush {Shakeri}, and Stefan {Typel}.
\newblock {Exploring the Distribution and Impact of Bosonic Dark Matter in Neutron Stars}.
\newblock \emph{Particles}, 7\penalty0 (1):\penalty0 201--213, March 2024.
\newblock \doi{10.3390/particles7010011}.

\bibitem[Bastero-Gil et~al.(2024)Bastero-Gil, Huertas-Roldan, and Santos]{Bastero-Gil2024}
Mar Bastero-Gil, Teresa Huertas-Roldan, and Daniel Santos.
\newblock {Neutron decay anomaly, neutron stars, and dark matter}.
\newblock \emph{Phys. Rev. D}, 110\penalty0 (8):\penalty0 083003, 2024.
\newblock \doi{10.1103/PhysRevD.110.083003}.

\bibitem[Scordino and Bombaci(2025)]{Scordino2024}
Domenico Scordino and Ignazio Bombaci.
\newblock {Dark matter admixed neutron stars with a realistic nuclear equation of state from chiral nuclear interactions}.
\newblock \emph{JHEAp}, 45:\penalty0 371--381, 2025.
\newblock \doi{10.1016/j.jheap.2025.01.008}.

\bibitem[Konstantinou(2024)]{Konstantinou2024}
Andreas Konstantinou.
\newblock {The Effect of a Dark Matter Core on the Structure of a Rotating Neutron Star}.
\newblock \emph{Astrophys. J.}, 968\penalty0 (2):\penalty0 83, 2024.
\newblock \doi{10.3847/1538-4357/ad4701}.

\bibitem[{Das} et~al.(2022){Das}, {Kumar}, {Kumar}, and {Patra}]{Das_2022}
H.~C. {Das}, Ankit {Kumar}, Bharat {Kumar}, and Suresh~Kumar {Patra}.
\newblock {Dark Matter Effects on the Compact Star Properties}.
\newblock \emph{Galaxies}, 10\penalty0 (1):\penalty0 14, January 2022.
\newblock \doi{10.3390/galaxies10010014}.

\bibitem[Das et~al.(2022)Das, Malik, and Nayak]{Apran2020}
Arpan Das, Tuhin Malik, and Alekha~C. Nayak.
\newblock Dark matter admixed neutron star properties in light of gravitational wave observations: A two fluid approach.
\newblock \emph{Phys. Rev. D}, 105:\penalty0 123034, Jun 2022.
\newblock \doi{10.1103/PhysRevD.105.123034}.

\bibitem[{Sen} and {Guha}(2021)]{Sen2021}
Debashree {Sen} and Atanu {Guha}.
\newblock {Implications of feebly interacting dark sector on neutron star properties and constraints from GW170817}.
\newblock \emph{\mnras}, 504\penalty0 (3):\penalty0 3354--3363, July 2021.
\newblock \doi{10.1093/mnras/stab1056}.

\bibitem[{Guha} and {Sen}(2021)]{Guha2021}
Atanu {Guha} and Debashree {Sen}.
\newblock {Feeble DM-SM interaction via new scalar and vector mediators in rotating neutron stars}.
\newblock \emph{\jcap}, 2021\penalty0 (9):\penalty0 027, September 2021.
\newblock \doi{10.1088/1475-7516/2021/09/027}.

\bibitem[Giangrandi et~al.(2023)Giangrandi, Sagun, Ivanytskyi, Provid\^encia, and Dietrich]{Giangrandi2022}
Edoardo Giangrandi, Violetta Sagun, Oleksii Ivanytskyi, Constan\c{c}a Provid\^encia, and Tim Dietrich.
\newblock {The Effects of Self-interacting Bosonic Dark Matter on Neutron Star Properties}.
\newblock \emph{Astrophys. J.}, 953\penalty0 (1):\penalty0 115, 2023.
\newblock \doi{10.3847/1538-4357/ace104}.

\bibitem[Barbat et~al.(2024)Barbat, Schaffner-Bielich, and Tolos]{Barbat2024}
Mikel~F. Barbat, J\"urgen Schaffner-Bielich, and Laura Tolos.
\newblock {Comprehensive study of compact stars with dark matter}.
\newblock \emph{Phys. Rev. D}, 110\penalty0 (2):\penalty0 023013, 2024.
\newblock \doi{10.1103/PhysRevD.110.023013}.

\bibitem[{Sun} and {Wen}(2024)]{Sun2023}
Hongyi {Sun} and Dehua {Wen}.
\newblock {New criterion for the existence of dark matter in neutron stars}.
\newblock \emph{\prd}, 109\penalty0 (12):\penalty0 123037, June 2024.
\newblock \doi{10.1103/PhysRevD.109.123037}.

\bibitem[{Thakur} et~al.(2024{\natexlab{a}}){Thakur}, {Malik}, and {Jha}]{P.Thakur2024}
Prashant {Thakur}, Tuhin {Malik}, and Tarun~Kumar {Jha}.
\newblock {Towards Uncovering Dark Matter Effects on Neutron Star Properties: A Machine Learning Approach}.
\newblock \emph{Particles}, 7\penalty0 (1):\penalty0 80--95, January 2024{\natexlab{a}}.
\newblock \doi{10.3390/particles7010005}.

\bibitem[{Thakur} et~al.(2024{\natexlab{b}}){Thakur}, {Malik}, {Das}, {Jha}, and {Provid{\^e}ncia}]{P.Thakur2024b}
Prashant {Thakur}, Tuhin {Malik}, Arpan {Das}, T.~K. {Jha}, and Constan{\c{c}}a {Provid{\^e}ncia}.
\newblock {Exploring robust correlations between fermionic dark matter model parameters and neutron star properties: A two-fluid perspective}.
\newblock \emph{\prd}, 109\penalty0 (4):\penalty0 043030, February 2024{\natexlab{b}}.
\newblock \doi{10.1103/PhysRevD.109.043030}.

\bibitem[{Shirke} et~al.(2024){Shirke}, {Keshari Pradhan}, {Chatterjee}, {Sagunski}, and {Schaffner-Bielich}]{Shirke2024}
Swarnim {Shirke}, Bikram {Keshari Pradhan}, Debarati {Chatterjee}, Laura {Sagunski}, and J{\"u}rgen {Schaffner-Bielich}.
\newblock {Effects of Dark Matter on $f$-mode oscillations of Neutron Stars}.
\newblock \emph{arXiv e-prints}, art. arXiv:2403.18740, March 2024.
\newblock \doi{10.48550/arXiv.2403.18740}.

\bibitem[Thakur et~al.(2024)Thakur, Kumar, Thapa, Parmar, and Sinha]{Pratik-Thankur2024}
Pratik Thakur, Anil Kumar, Vivek~Baruah Thapa, Vishal Parmar, and Monika Sinha.
\newblock {Exploring non-radial oscillation modes in dark matter admixed neutron stars}.
\newblock \emph{JCAP}, 12:\penalty0 042, 2024.
\newblock \doi{10.1088/1475-7516/2024/12/042}.

\bibitem[Pal and Chaudhuri(2024)]{Pal2024}
Suman Pal and Gargi Chaudhuri.
\newblock {Effect of dark matter interaction on hybrid star in the light of the recent astrophysical observations}.
\newblock \emph{JCAP}, 10:\penalty0 064, 2024.
\newblock \doi{10.1088/1475-7516/2024/10/064}.

\bibitem[{Mariani} et~al.(2024){Mariani}, {Albertus}, {Alessandroni}, {Orsaria}, {P{\'e}rez-Garc{\'\i}a}, and {Ranea-Sandoval}]{Mariani2024}
Mauro {Mariani}, Conrado {Albertus}, M.~del~Rosario {Alessandroni}, Milva~G. {Orsaria}, M.~{\'A}ngeles {P{\'e}rez-Garc{\'\i}a}, and Ignacio~F. {Ranea-Sandoval}.
\newblock {Constraining self-interacting fermionic dark matter in admixed neutron stars using multimessenger astronomy}.
\newblock \emph{\mnras}, 527\penalty0 (3):\penalty0 6795--6806, January 2024.
\newblock \doi{10.1093/mnras/stad3658}.

\bibitem[{Mahapatra} et~al.(2024){Mahapatra}, {Singha}, {Hazarika}, and {Das}]{Mahapatra2024}
Premachand {Mahapatra}, Chiranjeeb {Singha}, Ayush {Hazarika}, and Prasanta~Kumar {Das}.
\newblock {Implications of Fermionic Dark Matter Interactions on Anisotropic Neutron Stars}.
\newblock \emph{arXiv e-prints}, art. arXiv:2408.14020, August 2024.
\newblock \doi{10.48550/arXiv.2408.14020}.

\bibitem[Kumar and Sotani(2024)]{Kumar2024}
Ankit Kumar and Hajime Sotani.
\newblock {Constraints on the parameter space in dark matter admixed neutron stars}.
\newblock \emph{Phys. Rev. D}, 110\penalty0 (6):\penalty0 063001, 2024.
\newblock \doi{10.1103/PhysRevD.110.063001}.

\bibitem[Routaray et~al.(2023)Routaray, Mohanty, Das, Ghosh, Kalita, Parmar, and Kumar]{Routaray2023}
Pinku Routaray, Sailesh~Ranjan Mohanty, H.C. Das, Sayantan Ghosh, P.J. Kalita, Vishal Parmar, and Bharat Kumar.
\newblock Investigating dark matter-admixed neutron stars with nitr equation of state in light of psr j0952-0607.
\newblock \emph{Journal of Cosmology and Astroparticle Physics}, 2023\penalty0 (10):\penalty0 073, oct 2023.
\newblock \doi{10.1088/1475-7516/2023/10/073}.
\newblock URL \url{https://dx.doi.org/10.1088/1475-7516/2023/10/073}.

\bibitem[{Khlopov} et~al.(1985){Khlopov}, {Malomed}, and {Zeldovich}]{Khlopov1985}
M.~Iu. {Khlopov}, B.~A. {Malomed}, and Ia.~B. {Zeldovich}.
\newblock {Gravitational instability of scalar fields and formation of primordial black holes}.
\newblock \emph{\mnras}, 215:\penalty0 575--589, August 1985.
\newblock \doi{10.1093/mnras/215.4.575}.

\bibitem[{Bertone} and {Fairbairn}(2008)]{Bertone2008}
Gianfranco {Bertone} and Malcolm {Fairbairn}.
\newblock {Compact stars as dark matter probes}.
\newblock \emph{\prd}, 77\penalty0 (4):\penalty0 043515, February 2008.
\newblock \doi{10.1103/PhysRevD.77.043515}.

\bibitem[{Kouvaris} and {Tinyakov}(2011{\natexlab{a}})]{Kouvaris2011}
Chris {Kouvaris} and Peter {Tinyakov}.
\newblock {Constraining asymmetric dark matter through observations of compact stars}.
\newblock \emph{\prd}, 83\penalty0 (8):\penalty0 083512, April 2011{\natexlab{a}}.
\newblock \doi{10.1103/PhysRevD.83.083512}.

\bibitem[{Gresham} and {Zurek}(2019)]{Gresham2018}
Moira~I. {Gresham} and Kathryn~M. {Zurek}.
\newblock {Asymmetric dark stars and neutron star stability}.
\newblock \emph{\prd}, 99\penalty0 (8):\penalty0 083008, April 2019.
\newblock \doi{10.1103/PhysRevD.99.083008}.

\bibitem[{Bauswein} et~al.(2023){Bauswein}, {Guo}, {Lien-Hua}, {Lin}, and {Wu}]{Bauswein2023}
Andreas {Bauswein}, Gang {Guo}, Jr. {Lien-Hua}, Yen-Hsun {Lin}, and Meng-Ru {Wu}.
\newblock {Compact dark objects in neutron star mergers}.
\newblock \emph{\prd}, 107\penalty0 (8):\penalty0 083002, April 2023.
\newblock \doi{10.1103/PhysRevD.107.083002}.

\bibitem[{R{\"u}ter} et~al.(2023){R{\"u}ter}, {Sagun}, {Tichy}, and {Dietrich}]{Ruter2023}
Hannes~R. {R{\"u}ter}, Violetta {Sagun}, Wolfgang {Tichy}, and Tim {Dietrich}.
\newblock {Quasiequilibrium configurations of binary systems of dark matter admixed neutron stars}.
\newblock \emph{\prd}, 108\penalty0 (12):\penalty0 124080, December 2023.
\newblock \doi{10.1103/PhysRevD.108.124080}.

\bibitem[{Emma} et~al.(2022){Emma}, {Schianchi}, {Pannarale}, {Sagun}, and {Dietrich}]{Emma2022}
Mattia {Emma}, Federico {Schianchi}, Francesco {Pannarale}, Violetta {Sagun}, and Tim {Dietrich}.
\newblock {Numerical Simulations of Dark Matter Admixed Neutron Star Binaries}.
\newblock \emph{Particles}, 5\penalty0 (3):\penalty0 273--286, July 2022.
\newblock \doi{10.3390/particles5030024}.

\bibitem[Leane and Tong(2024)]{Tong2024}
Rebecca~K. Leane and Joshua Tong.
\newblock {Optimal celestial bodies for dark matter detection}.
\newblock \emph{JCAP}, 12:\penalty0 031, 2024.
\newblock \doi{10.1088/1475-7516/2024/12/031}.

\bibitem[{Rutherford} et~al.(2023){Rutherford}, {Raaijmakers}, {Prescod-Weinstein}, and {Watts}]{Rutherford2023}
Nathan {Rutherford}, Geert {Raaijmakers}, Chanda {Prescod-Weinstein}, and Anna {Watts}.
\newblock {Constraining bosonic asymmetric dark matter with neutron star mass-radius measurements}.
\newblock \emph{\prd}, 107\penalty0 (10):\penalty0 103051, May 2023.
\newblock \doi{10.1103/PhysRevD.107.103051}.

\bibitem[{Petraki} and {Volkas}(2013)]{Petraki2013}
Kalliopi {Petraki} and Raymond~R. {Volkas}.
\newblock {Review of Asymmetric Dark Matter}.
\newblock \emph{International Journal of Modern Physics A}, 28\penalty0 (19):\penalty0 1330028, July 2013.
\newblock \doi{10.1142/S0217751X13300287}.

\bibitem[{Petraki} et~al.(2014){Petraki}, {Pearce}, and {Kusenko}]{Petraki2014}
Kalliopi {Petraki}, Lauren {Pearce}, and Alexander {Kusenko}.
\newblock {Self-interacting asymmetric dark matter coupled to a light massive dark photon}.
\newblock \emph{\jcap}, 2014\penalty0 (7):\penalty0 039, July 2014.
\newblock \doi{10.1088/1475-7516/2014/07/039}.

\bibitem[{Morsink} et~al.(2007){Morsink}, {Leahy}, {Cadeau}, and {Braga}]{Morsink2007}
Sharon~M. {Morsink}, Denis~A. {Leahy}, Coire {Cadeau}, and John {Braga}.
\newblock {The Oblate Schwarzschild Approximation for Light Curves of Rapidly Rotating Neutron Stars}.
\newblock \emph{\apj}, 663\penalty0 (2):\penalty0 1244--1251, July 2007.
\newblock \doi{10.1086/518648}.

\bibitem[{AlGendy} and {Morsink}(2014)]{AlGendy2014}
Mohammad {AlGendy} and Sharon~M. {Morsink}.
\newblock {Universality of the Acceleration due to Gravity on the Surface of a Rapidly Rotating Neutron Star}.
\newblock \emph{\apj}, 791\penalty0 (2):\penalty0 78, August 2014.
\newblock \doi{10.1088/0004-637X/791/2/78}.

\bibitem[{Ray} et~al.(2018){Ray}, {Arzoumanian}, {Brandt}, {Burns}, {Chakrabarty}, {Feroci}, {Gendreau}, {Gevin}, {Hernanz}, {Jenke}, {Kenyon}, {G{\'a}lvez}, {Maccarone}, {Okajima}, {Remillard}, {Schanne}, {Tenzer}, {Vacchi}, {Wilson-Hodge}, {Winter}, {Zane}, {Ballantyne}, {Bozzo}, {Brenneman}, {Cackett}, {De Rosa}, {Goldstein}, {Hartmann}, {McDonald}, {Stevens}, {Tomsick}, {Watts}, {Wood}, and {Zoghbi}]{STROBEX2}
P.~S. {Ray}, Z.~{Arzoumanian}, S.~{Brandt}, E.~{Burns}, D.~{Chakrabarty}, M.~{Feroci}, K.~C. {Gendreau}, O.~{Gevin}, M.~{Hernanz}, P.~{Jenke}, S.~{Kenyon}, J.~L. {G{\'a}lvez}, T.~J. {Maccarone}, T.~{Okajima}, R.~A. {Remillard}, S.~{Schanne}, C.~{Tenzer}, A.~{Vacchi}, C.~A. {Wilson-Hodge}, B.~{Winter}, S.~{Zane}, D.~R. {Ballantyne}, E.~{Bozzo}, L.~W. {Brenneman}, E.~{Cackett}, A.~{De Rosa}, A.~{Goldstein}, D.~H. {Hartmann}, M.~{McDonald}, A.~L. {Stevens}, J.~A. {Tomsick}, A.~L. {Watts}, K.~S. {Wood}, and A.~{Zoghbi}.
\newblock {STROBE-X: a probe-class mission for x-ray spectroscopy and timing on timescales from microseconds to years}.
\newblock In \emph{Space Telescopes and Instrumentation 2018: Ultraviolet to Gamma Ray}, volume 10699 of \emph{Society of Photo-Optical Instrumentation Engineers (SPIE) Conference Series}, page 1069919, July 2018.
\newblock \doi{10.1117/12.2312257}.

\bibitem[{Tolman}(1939)]{Tolman1939}
Richard~C. {Tolman}.
\newblock {Static Solutions of Einstein's Field Equations for Spheres of Fluid}.
\newblock \emph{Physical Review}, 55\penalty0 (4):\penalty0 364--373, February 1939.
\newblock \doi{10.1103/PhysRev.55.364}.

\bibitem[{Oppenheimer} and {Volkoff}(1939)]{Oppenheimer1939}
J.~R. {Oppenheimer} and G.~M. {Volkoff}.
\newblock {On Massive Neutron Cores}.
\newblock \emph{Physical Review}, 55\penalty0 (4):\penalty0 374--381, February 1939.
\newblock \doi{10.1103/PhysRev.55.374}.

\bibitem[{Sagun} et~al.(2023){Sagun}, {Giangrandi}, {Dietrich}, {Ivanytskyi}, {Negreiros}, and {Provid{\^e}ncia}]{Sagun2023}
Violetta {Sagun}, Edoardo {Giangrandi}, Tim {Dietrich}, Oleksii {Ivanytskyi}, Rodrigo {Negreiros}, and Constan{\c{c}}a {Provid{\^e}ncia}.
\newblock {What Is the Nature of the HESS J1731-347 Compact Object?}
\newblock \emph{\apj}, 958\penalty0 (1):\penalty0 49, November 2023.
\newblock \doi{10.3847/1538-4357/acfc9e}.

\bibitem[{Sandin} and {Ciarcelluti}(2009)]{Sandin2009}
Fredrik {Sandin} and Paolo {Ciarcelluti}.
\newblock {Effects of mirror dark matter on neutron stars}.
\newblock \emph{Astroparticle Physics}, 32\penalty0 (5):\penalty0 278--284, December 2009.
\newblock \doi{10.1016/j.astropartphys.2009.09.005}.

\bibitem[{Marrod{\'a}n Undagoitia} and {Rauch}(2016)]{Marrodan-Undagoitia2016}
Teresa {Marrod{\'a}n Undagoitia} and Ludwig {Rauch}.
\newblock {Dark matter direct-detection experiments}.
\newblock \emph{Journal of Physics G Nuclear Physics}, 43\penalty0 (1):\penalty0 013001, January 2016.
\newblock \doi{10.1088/0954-3899/43/1/013001}.

\bibitem[{Rrapaj} and {Reddy}(2016)]{Reddy2016}
Ermal {Rrapaj} and Sanjay {Reddy}.
\newblock {Nucleon-nucleon bremsstrahlung of dark gauge bosons and revised supernova constraints}.
\newblock \emph{\prc}, 94\penalty0 (4):\penalty0 045805, October 2016.
\newblock \doi{10.1103/PhysRevC.94.045805}.

\bibitem[{Hajkarim} et~al.(2024){Hajkarim}, {Schaffner-Bielich}, and {Tolos}]{Hajkarim24}
Fazlollah {Hajkarim}, J{\"u}rgen {Schaffner-Bielich}, and Laura {Tolos}.
\newblock {Thermodynamic Consistent Description of Compact Stars of Two Interacting Fluids: The Case of Neutron Stars with Higgs Portal Dark Matter}.
\newblock \emph{arXiv e-prints}, art. arXiv:2412.04585, December 2024.
\newblock \doi{10.48550/arXiv.2412.04585}.

\bibitem[{Collier} et~al.(2022){Collier}, {Croon}, and {Leane}]{Collier2022}
Michael {Collier}, Djuna {Croon}, and Rebecca~K. {Leane}.
\newblock {Tidal Love numbers of novel and admixed celestial objects}.
\newblock \emph{\prd}, 106\penalty0 (12):\penalty0 123027, December 2022.
\newblock \doi{10.1103/PhysRevD.106.123027}.

\bibitem[{Greif} et~al.(2019){Greif}, {Raaijmakers}, {Hebeler}, {Schwenk}, and {Watts}]{Greif19}
S.~K. {Greif}, G.~{Raaijmakers}, K.~{Hebeler}, A.~{Schwenk}, and A.~L. {Watts}.
\newblock {Equation of state sensitivities when inferring neutron star and dense matter properties}.
\newblock \emph{\mnras}, 485\penalty0 (4):\penalty0 5363--5376, June 2019.
\newblock \doi{10.1093/mnras/stz654}.

\bibitem[{Read} et~al.(2009){Read}, {Lackey}, {Owen}, and {Friedman}]{Read2009}
Jocelyn~S. {Read}, Benjamin~D. {Lackey}, Benjamin~J. {Owen}, and John~L. {Friedman}.
\newblock {Constraints on a phenomenologically parametrized neutron-star equation of state}.
\newblock \emph{\prd}, 79\penalty0 (12):\penalty0 124032, June 2009.
\newblock \doi{10.1103/PhysRevD.79.124032}.

\bibitem[{Baym} et~al.(1971){Baym}, {Pethick}, and {Sutherland}]{Baym71}
Gordon {Baym}, Christopher {Pethick}, and Peter {Sutherland}.
\newblock {The Ground State of Matter at High Densities: Equation of State and Stellar Models}.
\newblock \emph{\apj}, 170:\penalty0 299, December 1971.
\newblock \doi{10.1086/151216}.

\bibitem[{Drischler} et~al.(2021){Drischler}, {Holt}, and {Wellenhofer}]{Drischler2021}
C.~{Drischler}, J.~W. {Holt}, and C.~{Wellenhofer}.
\newblock {Chiral Effective Field Theory and the High-Density Nuclear Equation of State}.
\newblock \emph{Annual Review of Nuclear and Particle Science}, 71:\penalty0 403--432, September 2021.
\newblock \doi{10.1146/annurev-nucl-102419-041903}.

\bibitem[{Tews} et~al.(2018){Tews}, {Carlson}, {Gandolfi}, and {Reddy}]{Tews2018}
I.~{Tews}, J.~{Carlson}, S.~{Gandolfi}, and S.~{Reddy}.
\newblock {Constraining the Speed of Sound inside Neutron Stars with Chiral Effective Field Theory Interactions and Observations}.
\newblock \emph{\apj}, 860\penalty0 (2):\penalty0 149, June 2018.
\newblock \doi{10.3847/1538-4357/aac267}.

\bibitem[{Hebeler} and {Schwenk}(2010)]{Hebeler10}
K.~{Hebeler} and A.~{Schwenk}.
\newblock {Chiral three-nucleon forces and neutron matter}.
\newblock \emph{\prc}, 82\penalty0 (1):\penalty0 014314, July 2010.
\newblock \doi{10.1103/PhysRevC.82.014314}.

\bibitem[{Raaijmakers} et~al.(2025){Raaijmakers}, {Rutherford}, {Timmerman}, {Salmi}, {Watts}, {Prescod-Weinstein}, {Svensson}, and {Mendes}]{Raaijmakers24}
Geert {Raaijmakers}, Nathan {Rutherford}, Patrick {Timmerman}, Tuomo {Salmi}, Anna~L. {Watts}, Chanda {Prescod-Weinstein}, Isak {Svensson}, and Melissa {Mendes}.
\newblock {NEoST: A Python package for nested sampling of the neutron star equation of state}.
\newblock \emph{Journal of Open Source Software}, 10\penalty0 (105):\penalty0 6003, 2025.
\newblock \doi{10.21105/joss.06003}.
\newblock URL \url{https://doi.org/10.21105/joss.06003}.

\bibitem[{Rutherford} et~al.(2025){Rutherford}, {Prescod-Weinstein}, and {Watts}]{plotdata_fermionicadm}
N.~{Rutherford}, C.~{Prescod-Weinstein}, and A.~L. {Watts}.
\newblock \textit{Probing fermionic asymmetric dark matter cores using global neutron star properties: prior and posterior samples and scripts for generating plots}.
\newblock Zenodo, June 2025.
\newblock \doi{10.5281/zenodo.15237251}.

\bibitem[{Kouvaris} and {Tinyakov}(2011{\natexlab{b}})]{Kouvaris2011_pt2}
Chris {Kouvaris} and Peter {Tinyakov}.
\newblock {Excluding Light Asymmetric Bosonic Dark Matter}.
\newblock \emph{\prl}, 107\penalty0 (9):\penalty0 091301, August 2011{\natexlab{b}}.
\newblock \doi{10.1103/PhysRevLett.107.091301}.

\bibitem[{Husain} and {Thomas}(2023)]{Husain2023}
Wasif {Husain} and Anthony~W. {Thomas}.
\newblock {Novel neutron decay mode inside neutron stars}.
\newblock \emph{Journal of Physics G Nuclear Physics}, 50\penalty0 (1):\penalty0 015202, January 2023.
\newblock \doi{10.1088/1361-6471/aca1d5}.

\bibitem[{Navarro} et~al.(1996){Navarro}, {Frenk}, and {White}]{Nevarro}
Julio~F. {Navarro}, Carlos~S. {Frenk}, and Simon D.~M. {White}.
\newblock {The Structure of Cold Dark Matter Halos}.
\newblock \emph{\apj}, 462:\penalty0 563, May 1996.
\newblock \doi{10.1086/177173}.

\bibitem[{Lin} and {Li}(2019)]{Lin2019}
Hai-Nan {Lin} and Xin {Li}.
\newblock {The dark matter profiles in the Milky Way}.
\newblock \emph{\mnras}, 487\penalty0 (4):\penalty0 5679--5684, August 2019.
\newblock \doi{10.1093/mnras/stz1698}.

\bibitem[{Sofue}(2013)]{Sofue2013}
Yoshiaki {Sofue}.
\newblock {Rotation Curve and Mass Distribution in the Galactic Center - From Black Hole to Entire Galaxy}.
\newblock \emph{\pasj}, 65:\penalty0 118, December 2013.
\newblock \doi{10.1093/pasj/65.6.118}.

\bibitem[{Strobel} et~al.(1999){Strobel}, {Schaab}, and {Weigel}]{Strobel_1999}
K.~{Strobel}, Ch. {Schaab}, and M.~K. {Weigel}.
\newblock {Properties of non-rotating and rapidly rotating protoneutron stars}.
\newblock \emph{\aap}, 350:\penalty0 497--512, October 1999.
\newblock \doi{10.48550/arXiv.astro-ph/9908132}.

\bibitem[{Radice} et~al.(2017){Radice}, {Burrows}, {Vartanyan}, {Skinner}, and {Dolence}]{Radice17}
David {Radice}, Adam {Burrows}, David {Vartanyan}, M.~Aaron {Skinner}, and Joshua~C. {Dolence}.
\newblock {Electron-capture and Low-mass Iron-core-collapse Supernovae: New Neutrino-radiation-hydrodynamics Simulations}.
\newblock \emph{\apj}, 850\penalty0 (1):\penalty0 43, November 2017.
\newblock \doi{10.3847/1538-4357/aa92c5}.

\bibitem[Suwa et~al.(2018)Suwa, Yoshida, Shibata, Umeda, and Takahashi]{Suwa18}
Yudai Suwa, Takashi Yoshida, Masaru Shibata, Hideyuki Umeda, and Koh Takahashi.
\newblock {On the minimum mass of neutron stars}.
\newblock \emph{MNRAS}, 481\penalty0 (3):\penalty0 3305--3312, 09 2018.
\newblock ISSN 0035-8711.
\newblock \doi{10.1093/mnras/sty2460}.

\bibitem[{Doroshenko} et~al.(2022){Doroshenko}, {Suleimanov}, {P{\"u}hlhofer}, and {Santangelo}]{Doroshenko2022}
Victor {Doroshenko}, Valery {Suleimanov}, Gerd {P{\"u}hlhofer}, and Andrea {Santangelo}.
\newblock {A strangely light neutron star within a supernova remnant}.
\newblock \emph{Nature Astronomy}, 6:\penalty0 1444--1451, December 2022.
\newblock \doi{10.1038/s41550-022-01800-1}.

\bibitem[{Alford} and {Halpern}(2023)]{Alford2023}
J.~A.~J. {Alford} and J.~P. {Halpern}.
\newblock {Do Central Compact Objects have Carbon Atmospheres?}
\newblock \emph{\apj}, 944\penalty0 (1):\penalty0 36, February 2023.
\newblock \doi{10.3847/1538-4357/acaf55}.

\bibitem[{{\'A}vila} et~al.(2024){{\'A}vila}, {Giangrandi}, {Sagun}, {Ivanytskyi}, and {Provid{\^e}ncia}]{Avila2024}
Afonso {{\'A}vila}, Edoardo {Giangrandi}, Violetta {Sagun}, Oleksii {Ivanytskyi}, and Constan{\c{c}}a {Provid{\^e}ncia}.
\newblock {Rapid neutron star cooling triggered by dark matter}.
\newblock \emph{\mnras}, 528\penalty0 (4):\penalty0 6319--6328, March 2024.
\newblock \doi{10.1093/mnras/stae337}.

\bibitem[{Giangrandi} et~al.(2024){Giangrandi}, {{\'A}vila}, {Sagun}, {Ivanytskyi}, and {Provid{\^e}ncia}]{Giangrandi2024}
Edoardo {Giangrandi}, Afonso {{\'A}vila}, Violetta {Sagun}, Oleksii {Ivanytskyi}, and Constan{\c{c}}a {Provid{\^e}ncia}.
\newblock {The Impact of Asymmetric Dark Matter on the Thermal Evolution of Nucleonic and Hyperonic Compact Stars}.
\newblock \emph{Particles}, 7\penalty0 (1):\penalty0 179--200, February 2024.
\newblock \doi{10.3390/particles7010010}.

\bibitem[{Husain} and {Thomas}(2021)]{Husain2021}
Wasif {Husain} and Anthony~W. {Thomas}.
\newblock {Possible nature of dark matter}.
\newblock \emph{\jcap}, 2021\penalty0 (10):\penalty0 086, October 2021.
\newblock \doi{10.1088/1475-7516/2021/10/086}.

\bibitem[{Oliphant}(2007)]{python2007}
T.~E. {Oliphant}.
\newblock Python for scientific computing.
\newblock \emph{Computing in Science Engineering}, 9\penalty0 (3):\penalty0 10--20, May 2007.
\newblock ISSN 1521-9615.
\newblock \doi{10.1109/MCSE.2007.58}.

\bibitem[Gough(2009)]{Gough:2009}
Brian Gough.
\newblock \emph{GNU Scientific Library Reference Manual - Third Edition}.
\newblock Network Theory Ltd., 3rd edition, 2009.
\newblock ISBN 0954612078, 9780954612078.

\bibitem[{van der Walt} et~al.(2011){van der Walt}, {Colbert}, and {Varoquaux}]{Numpy2011}
S.~{van der Walt}, S.~C. {Colbert}, and G.~{Varoquaux}.
\newblock The numpy array: A structure for efficient numerical computation.
\newblock \emph{Computing in Science Engineering}, 13\penalty0 (2):\penalty0 22--30, March 2011.
\newblock ISSN 1521-9615.
\newblock \doi{10.1109/MCSE.2011.37}.

\bibitem[{Behnel} et~al.(2011){Behnel}, {Bradshaw}, {Citro}, {Dalcin}, {Seljebotn}, and {Smith}]{cython2011}
S.~{Behnel}, R.~{Bradshaw}, C.~{Citro}, L.~{Dalcin}, D.~S. {Seljebotn}, and K.~{Smith}.
\newblock Cython: The best of both worlds.
\newblock \emph{Computing in Science Engineering}, 13\penalty0 (2):\penalty0 31--39, March 2011.
\newblock ISSN 1521-9615.
\newblock \doi{10.1109/MCSE.2010.118}.

\bibitem[{Virtanen} et~al.(2020){Virtanen}, {Gommers}, {Oliphant}, {Haberland}, {Reddy}, {Cournapeau}, {Burovski}, {Peterson}, {Weckesser}, {Bright}, {van der Walt}, {Brett}, {Wilson}, {Millman}, {Mayorov}, {Nelson}, {Jones}, {Kern}, {Larson}, {Carey}, {Polat}, {Feng}, {Moore}, {VanderPlas}, {Laxalde}, {Perktold}, {Cimrman}, {Henriksen}, {Quintero}, {Harris}, {Archibald}, {Ribeiro}, {Pedregosa}, {van Mulbregt}, and {SciPy 1. 0 Contributors}]{Scipy}
Pauli {Virtanen}, Ralf {Gommers}, Travis~E. {Oliphant}, Matt {Haberland}, Tyler {Reddy}, David {Cournapeau}, Evgeni {Burovski}, Pearu {Peterson}, Warren {Weckesser}, Jonathan {Bright}, St{\'e}fan~J. {van der Walt}, Matthew {Brett}, Joshua {Wilson}, K.~Jarrod {Millman}, Nikolay {Mayorov}, Andrew R.~J. {Nelson}, Eric {Jones}, Robert {Kern}, Eric {Larson}, C.~J. {Carey}, {\.I}lhan {Polat}, Yu~{Feng}, Eric~W. {Moore}, Jake {VanderPlas}, Denis {Laxalde}, Josef {Perktold}, Robert {Cimrman}, Ian {Henriksen}, E.~A. {Quintero}, Charles~R. {Harris}, Anne~M. {Archibald}, Ant{\^o}nio~H. {Ribeiro}, Fabian {Pedregosa}, Paul {van Mulbregt}, and {SciPy 1. 0 Contributors}.
\newblock {SciPy 1.0: fundamental algorithms for scientific computing in Python}.
\newblock \emph{Nature Methods}, 17:\penalty0 261--272, February 2020.
\newblock \doi{10.1038/s41592-019-0686-2}.

\bibitem[Dalc\'{i}n et~al.(2008)Dalc\'{i}n, Paz, Storti, and D'El\'{i}a]{mpi4py}
Lisandro Dalc\'{i}n, Rodrigo Paz, Mario Storti, and Jorge D'El\'{i}a.
\newblock Mpi for python: Performance improvements and mpi-2 extensions.
\newblock \emph{Journal of Parallel and Distributed Computing}, 68\penalty0 (5):\penalty0 655--662, 2008.
\newblock ISSN 0743-7315.
\newblock \doi{https://doi.org/10.1016/j.jpdc.2007.09.005}.

\bibitem[{Hunter}(2007)]{Hunter:2007}
John~D. {Hunter}.
\newblock {Matplotlib: A 2D Graphics Environment}.
\newblock \emph{Computing in Science and Engineering}, 9\penalty0 (3):\penalty0 90--95, May 2007.
\newblock \doi{10.1109/MCSE.2007.55}.

\bibitem[Kluyver et~al.(2016)Kluyver, Ragan-Kelley, P{\'e}rez, Granger, Bussonnier, Frederic, Kelley, Hamrick, Grout, Corlay, Ivanov, Avila, Abdalla, and Willing]{Kluyver:2016aa}
Thomas Kluyver, Benjamin Ragan-Kelley, Fernando P{\'e}rez, Brian Granger, Matthias Bussonnier, Jonathan Frederic, Kyle Kelley, Jessica Hamrick, Jason Grout, Sylvain Corlay, Paul Ivanov, Dami{\'a}n Avila, Safia Abdalla, and Carol Willing.
\newblock Jupyter notebooks -- a publishing format for reproducible computational workflows.
\newblock In F.~Loizides and B.~Schmidt, editors, \emph{Positioning and Power in Academic Publishing: Players, Agents and Agendas}, pages 87 -- 90. IOS Press, 2016.

\bibitem[Feroz et~al.(2019)Feroz, Hobson, Cameron, and Pettitt]{Feroz13}
F.~Feroz, M.~P. Hobson, E.~Cameron, and A.~N. Pettitt.
\newblock {Importance Nested Sampling and the MultiNest Algorithm}.
\newblock \emph{Open J. Astrophys.}, 2\penalty0 (1):\penalty0 10, 2019.
\newblock \doi{10.21105/astro.1306.2144}.

\bibitem[{Buchner} et~al.(2014){Buchner}, {Georgakakis}, {Nandra}, {Hsu}, {Rangel}, {Brightman}, {Merloni}, {Salvato}, {Donley}, and {Kocevski}]{Buchner14}
J.~{Buchner}, A.~{Georgakakis}, K.~{Nandra}, L.~{Hsu}, C.~{Rangel}, M.~{Brightman}, A.~{Merloni}, M.~{Salvato}, J.~{Donley}, and D.~{Kocevski}.
\newblock {X-ray spectral modelling of the AGN obscuring region in the CDFS: Bayesian model selection and catalogue}.
\newblock \emph{\aap}, 564:\penalty0 A125, April 2014.
\newblock \doi{10.1051/0004-6361/201322971}.

\bibitem[Kelley(2021)]{Kelley2021}
Luke~Zoltan Kelley.
\newblock kalepy: a python package for kernel density estimation, sampling and plotting.
\newblock \emph{Journal of Open Source Software}, 6\penalty0 (57):\penalty0 2784, 2021.
\newblock \doi{10.21105/joss.02784}.

\bibitem[Foreman-Mackey(2016)]{corner}
Daniel Foreman-Mackey.
\newblock corner.py: Scatterplot matrices in python.
\newblock \emph{The Journal of Open Source Software}, 1\penalty0 (2):\penalty0 24, jun 2016.
\newblock \doi{10.21105/joss.00024}.

\bibitem[Waskom(2021)]{Waskom2021}
Michael~L. Waskom.
\newblock seaborn: statistical data visualization.
\newblock \emph{Journal of Open Source Software}, 6\penalty0 (60):\penalty0 3021, 2021.
\newblock \doi{10.21105/joss.03021}.

\end{thebibliography}
\end{document}